\documentclass[12pt,twoside]{article}
\usepackage{MSH,amssymb,amsmath,verbatim}
\usepackage[french]{babel}
\usepackage[T1]{fontenc}
\usepackage[ruled,vlined,french,titlenumbered]{algorithm2e}
\usepackage{latexsym}
\usepackage[OT1,T1]{fontenc}
\usepackage{amsfonts,bbm}
\usepackage{graphicx,times}
\usepackage{mathrsfs,oldgerm}
\usepackage{pstricks,pst-node,pst-tree}
\usepackage{graphicx}

\font\xmplbx = cmbx10.8 scaled \magstephalf

\date{}

\font\xmplbx = cmbx10.8 scaled \magstephalf

\newtheorem{theoreme}{\rm \sc th\'eor\`eme}
\newtheorem{demo}{\rm \sc d\'emonstration}

\newtheorem{definitio}{\rm \sc d\'efinition}

\newtheorem{lemme}{\rm \sc lemme}
\newtheorem{exemple}{\sc exemple}
\newtheorem{prop}{\rm \sc proposition}
\newtheorem{rem}{\rm \sc remarque}

\date{}

\usepackage{times}
\usepackage{graphicx}

\font\xmplbx = cmbx10.8 scaled \magstephalf

\begin{document}

\pagestyle{myheadings}
\markboth{\hfill\mbox{\footnotesize \sc t. hamrouni, s. ben yahia, e. mephu nguifo}\hfill}
{\hfill\mbox{\footnotesize \sc construction du treillis de fermés fréquents et extraction de règles génériques}\hfill}
\thispagestyle{empty}

\noindent
\scriptsize\mbox{\it Math.} \& {\it Sci. hum.} / {\it Mathematics and Social Sciences} \textsc{(}{$49^{\rm e}$ ann\'ee, n$^\circ$\,195}, {\it 2011}\textsc{(}3\textsc{)}, p.\,\pageref{hamrouni1}--\pageref{hamrounif}\textsc{)}

\setcounter{page}{5}\label{hamrouni1}                                                                                                                                                                                                                                                                                                    

% pour imposer le numero de la premiere page
\setcounter{page}{5}

\vspace{30mm}
\begin{center}
\normalsize CONSTRUCTION EFFICACE DU TREILLIS DES MOTIFS FERM\'ES FR\'EQUENTS ET EXTRACTION SIMULTAN\'EE DES BASES G\'EN\'ERIQUES DE R\`EGLES

\vspace*{26pt}
Tarek HAMROUNI$^1$, Sadok BEN YAHIA\footnote{\ URPAH, D\'epartement des Sciences de l'Informatique, Facult\'e des Sciences de Tunis, Universit\'e Tunis El Manar, Campus Universitaire, Tunis, Tunisie, tarek.hamrouni@fst.rnu.tn, sadok.benyahia@fst.rnu.tn,}, Engelbert MEPHU NGUIFO\footnote{\ Clermont Université, Université Blaise Pascal, LIMOS, BP 10448, F-63000 Clermont-Ferrand, France, engelbert.mephu\_nguifo@univ-bpclermont.fr,}$^,$\footnote{\ LIMOS, CNRS, UMR 6158, 63173 Aubi\`ere, France.}
 \end{center}

\vspace{8mm}

\footnotesize
\mbox{\normalsize\sc r\'esum\'e} -- {\em Durant ces dernières années, les quantités de données collectées, dans divers domaines d'application de l'informatique, deviennent de plus en plus importantes. Ces quantités suscitent le besoin d'analyse et d'interprétation afin d'en extraire des connaissances utiles. Dans ce travail, nous nous intéressons à la technique d'extraction des règles d'association à partir de larges contextes. Cette dernière est parmi les techniques les plus fréquemment utilisées en fouille de données. Toutefois, le nombre de règles extraites est généralement important avec en outre la présence de règles redondantes. Dans ce papier, nous proposons un nouvel algorithme, appelé \textsc{Prince}, dont la principale originalité est de construire une structure
partiellement ordonnée \textsc{(}nommée treillis d'Iceberg\textsc{)} dans l'objectif d'extraire des ensembles réduits de règles, appelés bases génériques. Ces bases forment un sous-ensemble, sans perte d'information, des règles d'association. Pour réduire le coût de cette construction, le treillis d'Iceberg est calculé grâce aux générateurs minimaux, associés aux motifs fermés fréquents. Ces derniers sont simultanément dérivés avec les bases génériques grâce à un simple parcours ascendant de la structure construite. Les expérimentations que nous avons réalisées sur des contextes de référence et \guillemotleft~pire des cas \guillemotright \ ont montré l'efficacité de l'algorithme proposé, comparativement à des algorithmes tels que \textsc{Close}, \textsc{A-Close} et \textsc{Titanic}}.
\smallskip

\footnotesize
\mbox{\normalsize\sc mots cl\'es} -- Analyse de concepts formels, Générateur minimal, Motif fermé, Règle, Treillis

\vspace{4mm}

\footnotesize
\mbox{\normalsize\sc summary} -- EFFICIENT CONSTRUCTION OF THE LATTICE OF FREQUENT CLOSED PATTERNS AND SIMULTANEOUS EXTRACTION OF GENERIC RULES BASES\\
{\em In the last few years, the amount of collected data, in various computer science applications, becomes increasingly significant. These large volumes of data pointed out the need to analyze them in order to extract useful hidden knowledge. This work focuses on association rule extraction. This
technique is one of the most popular in data mining. Nevertheless, the number of extracted association rules is often very high, a large part of which is redundant. In this paper, we propose a new algorithm, called \textsc{Prince}. Its main feature is the construction of a partially ordered structure towards extracting subsets of association rules, called \textit{generic bases}. These subsets form a lossless representation of the whole association rule set. To reduce the cost of such a construction, the partially ordered structure is built thanks to the minimal generators associated to frequent closed patterns. The closed ones are simultaneously derived with generic bases thanks to a simple bottom-up traversal of the obtained structure. The experimentations we carried out on benchmark and \guillemotleft~worst case \guillemotright \ contexts showed the efficiency of the proposed algorithm, comparatively to algorithms like \textsc{Close}, \textsc{A-Close} and \textsc{Titanic}}.

\smallskip

\footnotesize
\mbox{\normalsize\sc keywords} -- Association rule, Closed pattern, Formal concept analysis, Lattice, Minimal generator

\normalsize

\section{INTRODUCTION ET MOTIVATIONS}

\noindent La technique d'extraction des règles d'association [Agrawal {\em \& al.}, 1993, 1994] est une des techniques exploratoires les plus utilisées pour extraire les connaissances à partir des données collectées dans différentes applications. Elle s'applique lorsque les données se présentent sous forme d'un contexte d'extraction $\mathcal{K}$ représentant un sous-ensemble $\mathcal{M}$ d'un produit cartésien $\mathcal{O}\times \mathcal{I}, \mathcal{O}$ et $\mathcal{I}$ étant des ensembles finis. \'Etant donnés $X$ et $Y$ deux sous-ensembles disjoints de l'ensemble $\mathcal{I}$, si le contexte $\mathcal{K}$ satisfait la formule : $$\forall \ o \in \mathcal{O}, \textsc{(}\forall \ x \in X, \textsc{(}o, x\textsc{)} \in \mathcal{M}\textsc{)} \Rightarrow \textsc{(}\forall \ y \in Y, \textsc{(}o, y\textsc{)} \in \mathcal{M}\textsc{)}$$ nous disons que le couple\textsc{(}$X, Y$\textsc{)} est une règle d'association \textsc{(}extraite\textsc{)} du contexte $\mathcal{K}$. Nous écrivons dans la suite $X \Rightarrow_{\mathcal{K}} Y$ ou plus simplement $X \Rightarrow Y$. Les applications de cette notion sont ainsi multiples. Par exemple, si $\mathcal{O}$ est un ensemble de patients atteints de la même maladie, $\mathcal{I}$ l'ensemble des symptômes de ladite maladie et le contexte $\mathcal{K}$ l'ensemble des couples \textsc{(}$p, s$\textsc{)} pour lesquels le patient $p$ présente le symptôme $s$, alors une règle d'association $X \Rightarrow Y$ signifie que, pour la population concernée, l'occurrence des symptômes regroupés dans $X$ s'accompagne toujours de celle des symptômes dans $Y$. La notion de règle d'association s'étend naturellement à des règles pondérées par un indice de confiance. Les règles d'association exactes ou pondérées se révèlent donc utiles dans une variété de domaines.

\vspace{2mm}

Toutefois, le nombre de règles extraites peut être si grand qu'il affaiblit l'intérêt pratique de la technique. Or, beaucoup sont redondantes. De nombreux travaux, s'inspirant de l'analyse des concepts formels [Ganter, Wille, 1999], ont alors montrer comment se limiter, sans perte d'information, à des sous-ensembles de règles appelés \textit{base générique}.

\vspace{2mm}

Une base générique est sous-ensemble de l'ensemble total des règles d'association tel que toute règle non retenue dans la base peut être déduite à partir de celles retenues par une méthode appropriée. La plupart des bases génériques, qui ont été proposées dans la littérature [Ashrafi {\em \& al.}, 2007 ; Bastide, 2000; Ceglar, Roddick, 2006 ; Kryszkiewicz, 2002], véhiculent des règles d'association qui représentent des implications entre des ensembles d'attributs \textsc{(}ou motifs\textsc{)} particuliers à savoir les \textit{générateurs minimaux} [Bastide {\em \& al.}, 2000\textsc{(}a\textsc{)}] \textsc{(}appelé aussi motifs libres dans [Boulicaut {\em \& al.}, 2003] et motifs clés dans [Stumme {\em \& al.}, 2002]\textsc{)} et les \textit{motifs fermés} [Pasquier {\em \& al.}, 1999\textsc{(}b\textsc{)}]. En effet, il a été montré que ce type de règles d'association offre le maximum d'informations [Ashrafi {\em \& al.}, 2007 ; Bastide, 2000 ; Kryszkiewicz, 2002 ; Pasquier, 2000]. Néanmoins, pour extraire ces bases de règles, trois composantes primordiales doivent être calculées, à savoir :
\begin{description}
\item \textsc{(}i\textsc{)} l'ensemble des motifs fermés fréquents, c'est-à-dire ceux présentant une fréquence d'apparition jugée satisfaisante par l'utilisateur,
\item \textsc{(}ii\textsc{)} la liste des générateurs minimaux fréquents associés à chaque motif fermé fréquent,
\item \textsc{(}iii\textsc{)} la relation d'ordre partiel ordonnant les motifs fermés fréquents.
\end{description}

Une étude critique des algorithmes d'extraction des motifs fermés fréquents de la littérature a été
menée dans [Ben Yahia {\em \& al.}, 2006]. Le constat essentiel peut être résumé comme suit :
\begin{enumerate}
\item Ces algorithmes se sont concentrés sur l'énumération des motifs fermés fréquents. En effet, leur principal but est de réduire le temps d'extraction de ces motifs moyennant l'utilisation de structures de données évoluées. Leurs performances sont intéressantes sur des contextes qualifiés de denses [Bayardo, 1998]. Cependant, ils présentent de modestes performances sur des contextes épars. En effet, calculer les fermetures des motifs dans pareils contextes réduit leurs performances puisque l'espace de recherche des motifs fermés fréquents tend à se superposer avec celui des motifs fréquents.
\item Les algorithmes d'extraction des motifs fermés fréquents ne tiennent pas compte de l'extraction des bases génériques de règles. En effet, d'une part, seule une partie permet d'extraire les générateurs minimaux et nécessite un surcoût pour les associer à leurs fermés sans redondance\footnote{\ Un motif fermé peut être extrait plusieurs fois puisqu'il peut avoir plusieurs générateurs minimaux associés.}. D'autre part, aucun de ces algorithmes ne détermine la relation d'ordre. Ceci est argumenté par le fait qu'ils se sont seulement focalisés sur le calcul efficace des motifs fermés fréquents.
\end{enumerate}

Pour pallier ces limites, nous proposons dans ce papier un nouvel algorithme, appelé \textsc{Prince}, dédié à l'extraction des bases génériques de règles d'association. \textsc{Prince} effectue une exploration par niveau de l'espace de recherche, c'est-à-dire en partant de l'ensemble vide vers les motifs de taille {\xmplbx1}, ensuite ceux de taille {\xmplbx2}, et ainsi de suite. Sa principale caractéristique est qu'il détermine la relation d'ordre partiel entre les motifs fermés fréquents afin d'extraire les bases génériques de règles d'association. Son originalité est que le treillis d'Iceberg est obtenu grâce à des comparaisons entre les générateurs minimaux fréquents seulement, et non plus entre les motifs fermés. Ainsi, les bases génériques de règles sont dérivées par un simple balayage de la structure ordonnée obtenue sans avoir à calculer les fermetures. De plus, grâce à une gestion efficace des classes d'équivalence, \textsc{Prince} optimise la construction de la relation d'ordre et évite aussi la redondance dans la dérivation des motifs fermés et des règles d'association. Par ailleurs, en adoptant une optimisation issue des caractéristiques des générateurs minimaux, la relation d'ordre peut être facilement déduite. Les résultats des expérimentations, menées sur des contextes de référence et \guillemotleft~pire des cas \guillemotright, sont très encourageants. En effet, les performances de \textsc{Prince}, comparées aux algorithmes de référence d'exploration par niveau, \textit{e.g.} \textsc{Close}, \textsc{A-Close} et \textsc{Titanic}, sont largement supérieures.

\vspace{2mm}

Nous n'avons pas comparé les performances de \textsc{Prince} à celles d'algorithmes plus récents, \textit{e.g.} \textsc{DCI-Closed} [Lucchese {\em \& al.}, 2006] et \textsc{LCM} [Uno {\em \& al.}, 2004], considérés dans l'étude menée dans [Ben Yahia {\em \& al.}, 2006]. Ceci est argumenté par le fait que les algorithmes \textsc{Close}, \textsc{A-Close} et \textsc{Titanic} déterminent, en outre des fermés, l'information clé fournie par l'ensemble des générateurs minimaux, ce qui n'est pas le cas pour le reste des algorithmes.

\vspace{2mm}

La suite de l'article est organisée comme suit. La Section suivante rappelle brièvement les fondements mathématiques de l'analyse de concepts formels ainsi que son lien avec la dérivation des règles d'association génériques. La Section \ref{section_EdeA_IFFs} discute les travaux liés à notre problématique. La quatrième Section est dédiée à une présentation détaillée de l'algorithme \textsc{Prince}. La Section \ref{section_proprietes_Prince} détaille différentes propriétés de l'algorithme proposé. Les résultats des expérimentations montrant l'utilité de l'approche proposée sont présentés dans la sixième Section.

\section{FONDEMENTS MATH\'EMATIQUES}

\noindent Dans ce qui suit, nous présentons brièvement quelques résultats clés provenant de l'analyse de concepts formels et leur lien avec le contexte de dérivation des règles d'association génériques.

\subsection{notions de base} \label{sous_section_notion_de_base}

\noindent Dans la suite du papier, nous utilisons le cadre théorique de l'analyse de concepts formels présenté dans [Ganter, Wille, 1999]. Dans cette partie, nous rappelons d'une manière succincte les notions de base de ce cadre théorique.

\vspace{1.5mm}

\noindent {\em Contexte d'extraction} : un contexte d'extraction est un triplet $\mathcal{K} = \textsc{(}\mathcal{O}, \mathcal{I}, \mathcal{M}$\textsc{)} décrivant un ensemble fini $\mathcal{O}$ d'objets \textsc{(}ou transactions\textsc{)}, un ensemble fini $\mathcal{I}$\ d'items \textsc{(}ou attributs\textsc{)} et une relation \textsc{(}d'incidence\textsc{)} binaire $\mathcal{M}$
\textsc{(}c'est-à-dire $\mathcal{M} \subseteq \mathcal{O}\times \mathcal{I}$\textsc{)}. L'appartenance du couple
$\textsc{(}o,i$\textsc{)} à $\mathcal{M}$ désigne le fait que l'objet $o\in \mathcal{O}$ contient l'item $i\in \mathcal{I}$.

\vspace{1.5mm}

\noindent {\em Correspondance de Galois} : soit le contexte d'extraction $\mathcal{K} = \textsc{(}\mathcal{O}, \mathcal{I}, \mathcal{M}$\textsc{)}. Soit l'application $\Phi$ de l'ensemble des parties de $\mathcal{O}$ \textsc{(}c'est-à-dire l'ensemble de tous les sous-ensembles de $\mathcal{O}$\textsc{)}, noté par $\mathcal{P} \textsc{(}\mathcal{O}$\textsc{)}, dans l'ensemble des parties de $\mathcal{I}$, noté par $\mathcal{P} \textsc{(}\mathcal{I}$\textsc{)}. L'application $\Phi$ associe à un ensemble d'objets \textit{O} $\subseteq \mathcal{O}$, l'ensemble des items \textit{i} $\in \mathcal{I}$ communs à tous les objets \textit{o} $\in$ \textit{O}.
$$\Phi : \mathcal{P}\textsc{(}\mathcal{O}\textsc{)} \rightarrow \mathcal{P}\textsc{(}\mathcal{I}\textsc{)}$$
$$\Phi\textsc{(}O\textsc{)} = \{i \in \mathcal{I} | \forall o \in O, \textsc{(}o, i\textsc{)} \in \mathcal{M}\}$$

Soit l'application $\Psi$ de l'ensemble des parties de $\mathcal{I}$ dans l'ensemble des parties de $\mathcal{O}$. Elle associe à un ensemble d'items \textit{I} $\subseteq \mathcal{I}$, l'ensemble d'objets \textit{o} $\in \mathcal{O}$ communs à tous les items \textsl{i} $\in$ \textit{I} :
$$\Psi : \mathcal{P}\textsc{(}\mathcal{I}\textsc{)} \rightarrow \mathcal{P}\textsc{(}\mathcal{O}\textsc{)}$$
$$\Psi\textsc{(}I\textsc{)} = \{o \in \mathcal{O} | \forall i \in I, \textsc{(}o, i\textsc{)} \in \mathcal{M}\}$$

Le couple d'applications \textsc{(}$\Phi, \Psi$\textsc{)} définit une correspondance de Galois entre l'ensemble des parties de $\mathcal{O}$ et l'ensemble des parties de $\mathcal{I}$. Les applications $\gamma = \Phi \circ \Psi$ et $\omega = \Psi \circ \Phi$ sont appelées les opérateurs de fermeture de la correspondance de Galois [Ganter, Wille, 1999]. L'opérateur de fermeture $\gamma$, tout comme $\omega$, est caractérisé par le fait qu'il est :
\begin{description}
\item{-} Isotone : $I_1 \subseteq I_2 \Rightarrow \gamma\textsc{(}I_1\textsc{)} \subseteq \gamma\textsc{(}I_2$\textsc{)} ;
\item{-} Extensive : $I \subseteq \gamma\textsc{(}I$\textsc{)} ;
\item{-} Idempotent : $\gamma\textsc{(}\gamma\textsc{(}I\textsc{)}\textsc{)} = \gamma\textsc{(}I$\textsc{)}.
\end{description}

Nous allons maintenant introduire la notion de motif fréquent et celle de motif fermé.

\vspace{2mm}

\noindent{\em Motif fréquent} : un motif $I \subseteq \mathcal{I}$ est dit \textit{fréquent} si son support
relatif, \textit{Supp}$\textsc{(}I\textsc{)} = \displaystyle{\frac{\vert \Psi\textsc{(}I\textsc{)} \vert}{\vert \mathcal{O}\vert}}$, dépasse un seuil minimum fixé par l'utilisateur noté \textit{minsupp}. Notons que $\vert \Psi\textsc{(}I\textsc{)} \vert $ est appelé \textit{support absolu} de $I$.

Il est à noter que le support des motifs est anti-monotone par rapport à l'inclusion ensembliste, c'est-à-dire que si $I_1 \subseteq I_2$, alors $\textit{Supp}\textsc{(}I_1\textsc{)} \geq \textit{Supp}\textsc{(}I_2$\textsc{)}. Dans la suite et pour simplifier l'explication, \textit{Supp}\textsc{(}$I$\textsc{)} désignera le support \textit{absolu} du motif $I$.

\vspace{2mm}

\noindent {\em Motif fermé} [Pasquier {\em \& al.}, 1999\textsc{(}b\textsc{)}] : un motif $I \subseteq \mathcal{I}$ est dit \textit{fermé} si $I = \gamma\textsc{(}I$\textsc{)}. Le motif $I$ est un ensemble maximal d'items communs à un ensemble d'objets.

\vspace{2mm}

\noindent {\em Concept formel} : un concept formel est une paire $c = \textsc{(}O,I\textsc{)} \in \mathcal{O}\times\mathcal{I}$, où $O$ et $I$ sont reliés par les opérateurs de la correspondance de Galois, c'est-à-dire
$\Phi\textsc{(}O\textsc{)} = I$ et $\Psi\textsc{(}I\textsc{)} = O$. Les ensembles $O$ et $I$ sont appelés respectivement \textit{extension} et \textit{intension} de $c$.

Un motif fermé est l'intension d'un concept formel alors que son support est la cardinalité de l'extension du concept.

\vspace{2mm}

\noindent {\em Classe d'équivalence} [Bastide {\em \& al.}, 2000\textsc{(}b\textsc{)}] : l'opérateur de fermeture $\gamma$ induit une relation d'équivalence sur l'ensemble des parties de $\mathcal{I}$, c'est-à-dire l'ensemble de parties est subdivisé en des sous-ensembles disjoints, appelés aussi \textit{classes d'équivalence}. Dans chaque classe, tous les éléments possèdent la même fermeture : soit $I \subseteq \mathcal{I}$, la classe d'équivalence de $I$, dénotée [$I$], est : [$I$] = $\{I_1 \subseteq \mathcal{I} | \gamma \textsc{(}I\textsc{)} = \gamma\textsc{(}I_1\textsc{)}\}$. Les éléments de [$I$] ont ainsi la même valeur de support.

Deux classes d'équivalence sont dites \textit{comparables} si leurs motifs fermés associés peuvent être ordonnés par inclusion ensembliste, sinon elles sont dites \textit{incomparables}.\\

La définition d'une classe d'équivalence nous amène à celle d'un générateur minimal.

\vspace{2mm}

\noindent {\em Générateur minimal} [Bastide {\em \& al.}, 2000\textsc{(}b\textsc{)} ; Stumme {\em \& al.}, 2002] : soit $f$ un
motif fermé et [$f$] sa classe d'équivalence. L'ensemble $GM_{f}$ des générateurs {\em minimaux} de $f$ est défini comme suit : $GM_{f} = {\{}g \in [f] | \nexists \ g_{1} \subset g$ tel que $g_{1} \in [f]{\}}$.

\vspace{2mm}

Les générateurs minimaux d'une classe sont les éléments incomparables les plus petits \textsc{(}par rapport à la relation d'inclusion ensembliste\textsc{)}, tandis que le motif fermé est l'élément le plus grand de cette classe. Ainsi, tout motif est nécessairement compris entre un générateur minimal et un motif fermé. D'après la définition d'une classe d'équivalence, un générateur minimal a un support strictement inférieur à celui de ses sous-ensembles. D'une manière duale, un motif fermé admet un support strictement supérieur que celui de ses sur-ensembles.

\vspace{2mm}

Nous allons maintenant nous focaliser sur des propriétés structurelles importantes associées à l'ensemble des motifs fermés et à l'ensemble des générateurs minimaux.

\vspace{2mm}

\noindent {\em Treillis de concepts formels \textsc{(}de Galois\textsc{)}} :  étant donné un contexte d'extraction $\mathcal{K}$, l'ensemble de concepts formels $\mathcal{C}_{\mathcal{K}}$, extrait à partir de $\mathcal{K}$, est un treillis complet $\mathcal{L}_{\mathcal{C}_{\mathcal{K}}} = \textsc{(}\mathcal{C}_{\mathcal{K}}, \leq$\textsc{)}, appelé \textit{treillis de concepts \textsc{(}ou treillis de Galois\textsc{)}}, quand l'ensemble $\mathcal{C}_{\mathcal{K}}$ est considéré avec la relation d'inclusion ensembliste entre les motifs [Barbut, Monjardet, 1970 ; Ganter, Wille, 1999] : soient $c_1 = \textsc{(}O_1,I_1\textsc{)}$ et $c_2 = \textsc{(}O_2,I_2\textsc{)}$ deux concepts formels, $c_1 \leq c_2$ si $I_1 \subseteq I_2$.

Dans le treillis de Galois, chaque élément $c \in \mathcal{C}_{\mathcal{K}}$ est connecté aussi bien à l'ensemble de ses {\em successeurs immédiats}, appelé {\em couverture supérieure} \textsc{(}$Couv^s$\textsc{)}, qu'à l'ensemble de ses {\em prédécesseurs immédiats}, appelé {\em couverture inférieure} \textsc{(}$Couv_i$\textsc{)}.

\vspace{2mm}

\noindent {\em Treillis d'Iceberg} \label{def_treillis_iceberg} quand nous considérons seulement
l'ensemble $\mathcal{IFF}_{\mathcal{K}}$ des motifs fermés fréquents extrait à partir de $\mathcal{K}$ et ordonnés par la relation d'inclusion ensembliste, la structure obtenue \textsc{(}$\hat{\mathcal{L}}, \subseteq$\textsc{)} préserve seulement l'opérateur \textit{Sup} [Ganter, Wille, 1999]. Cette structure forme un sup demi-treillis [Mephu Nguifo, 1994] que, par abus volontaire [Stumme {\em \& al.}, 2002] appelle tout de même treillis, le \textit{treillis d'Iceberg}.

\vspace{2mm}

\begin{exemple} {\em Considérons le contexte d'extraction $\mathcal{K}$ donné par la Figure \ref{runfig} \textsc{(}gauche\textsc{)}. Quelques exemples \ de classes \ d'équivalence extraites \ de ce contexte \ sont donnés \ par \ la \ Figure \ \ref{runfig} \textsc{(}centre\textsc{)}. \ Le \ treillis \ d'Iceberg, \ pour \ \textit{minsupp} = {\xmplbx2}, \ est donné \ par \ la \ Figure~\ref{runfig} \textsc{(}droite\textsc{)}. Chaque n\oe ud dans le treillis d'Iceberg est formé par le motif fermé fréquent et le support correspondant, et est étiqueté par la liste de ses générateurs minimaux associés.}

\begin{figure}[h]
\parbox{3.cm}{\footnotesize
\begin{tabular}{|l|l|l|l|l|l|}
  \hline
   & \texttt{A} & \texttt{B} & \texttt{C} & \texttt{D} & \texttt{E} \\\hline
  1 & $\times$ &  & $\times$ & $\times$ &  \\\hline
  2 &  & $\times$ & $\times$ & & $\times$ \\\hline
  3 & $\times$ & $\times$ & $\times$ &  & $\times$ \\\hline
  4 &  & $\times$ &  &  & $\times$ \\\hline
  5 & $\times$ & $\times$ & $\times$ &  & $\times$ \\\hline
\end{tabular}
}
\hspace{1cm}
\parbox{4.7cm}{\includegraphics[scale = .45]{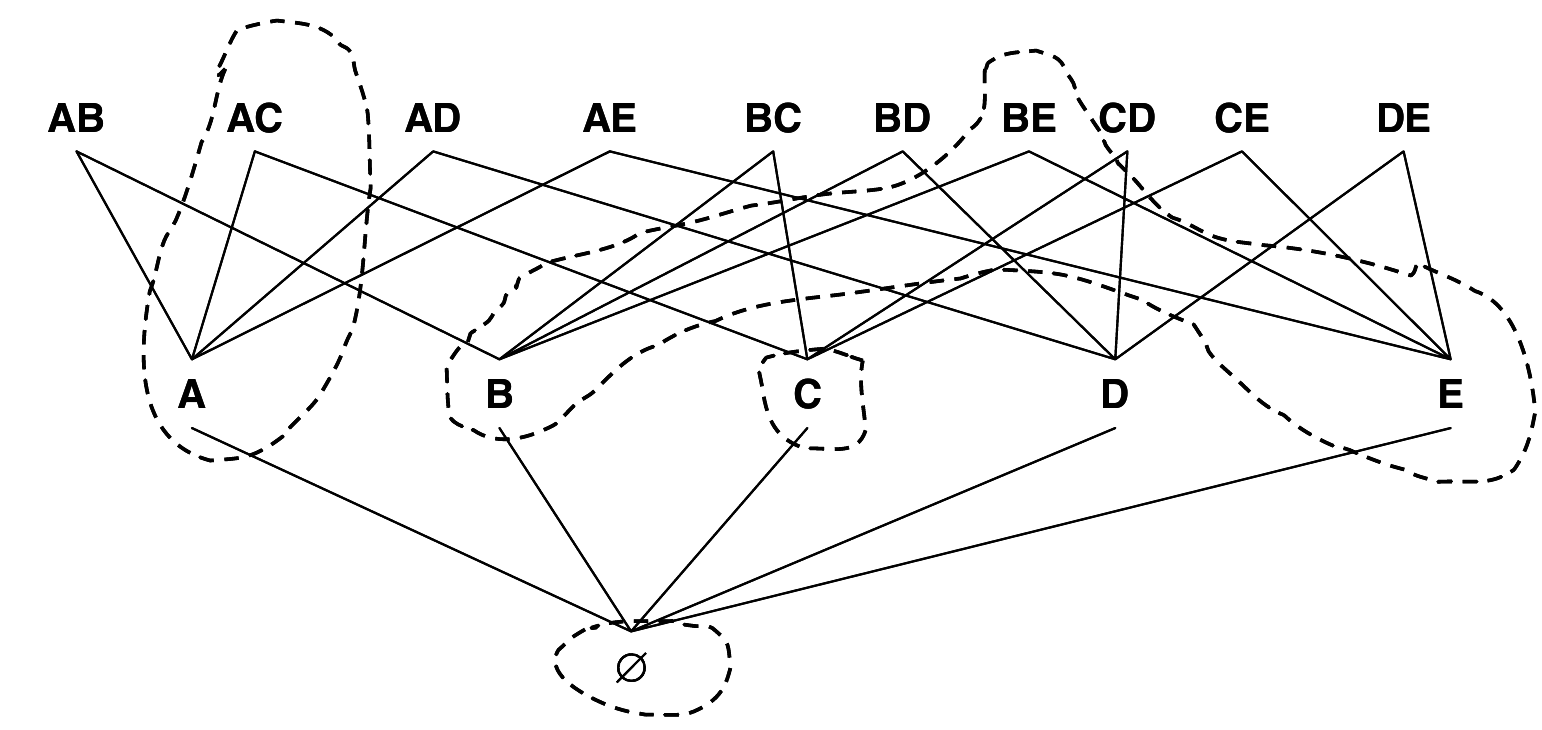}}
\hspace{2.2cm}
\parbox{6cm}{\includegraphics[scale = .5]{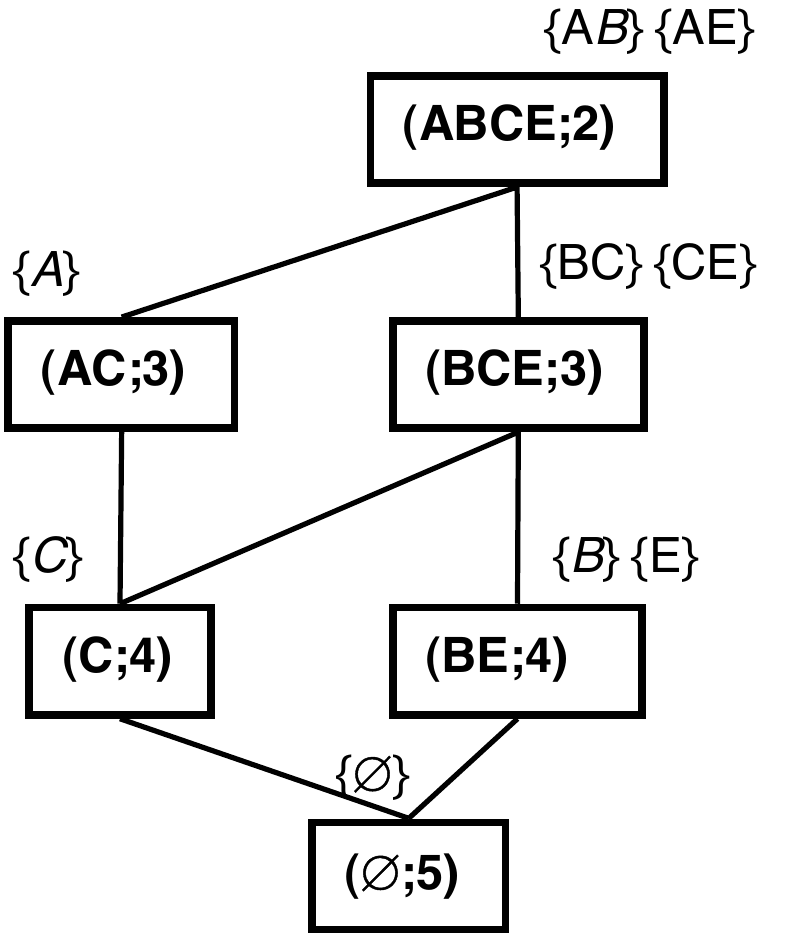}}
\caption{{\sc Gauche} : Contexte d'extraction $\mathcal{K}$. {\sc Centre} : \'{E}chantillon des classes de la relation d'équivalence induite. {\sc Droite} : Le treillis d'Iceberg associé pour \textit{minsupp} = ~2. Chaque n\oe ud dans ce treillis représente une classe d'équivalence. Il contient un motif fermé fréquent ainsi que son support et est étiqueté par les générateurs minimaux associés.} \label{runfig}
\end{figure}
\end{exemple}

\vspace{3mm}

Dans ce qui suit, nous introduisons la notion d'idéal d'ordre [Ganter, Wille, 1999] définie sur l'ensemble des parties d'un ensemble $\mathcal{E}$, c'est-à-dire $\mathcal{P}\textsc{(}\mathcal{E}\textsc{)}$.

\begin{definitio} Un sous-ensemble $\mathcal{S}$ de $\mathcal{P}\textsc{(}\mathcal{E}\textsc{)}$ est un idéal d'ordre s'il vérifie les propriétés suivantes :
\begin{description}
\item {-} Si $x \in \mathcal{S}$, \mbox{\rm alors} $\forall \ x_1 \subseteq x : x_1 \in \mathcal{S}$.
\item {-} Si $x \notin \mathcal{S}$, \mbox{\rm alors} $\forall \ x \subseteq x_1 : x_1 \notin \mathcal{S}$.
\end{description}
\end{definitio}

Le lemme suivant énonce, dans le cas général, une propriété intéressante de l'ensemble des générateurs minimaux \textsc{(}ou clés\textsc{)} associés à un opérateur de fermeture.

\begin{lemme}\label{lemme_ideal_ordre_cas_general} L'ensemble des générateurs minimaux d'un opérateur de fermeture $\theta$ est un idéal de l'ordre d'inclusion ensembliste.
\end{lemme}

\begin{demo} {\em Soit $\theta$ un opérateur de fermeture sur un ensemble $\mathcal{E}$. Soit $Y$ un
sous-ensemble générateur minimal de $\mathcal{E}$, c'est-à-dire pour tout $Y' \subset Y$, alors $\theta\textsc{(}Y'\textsc{)} \subset \theta\textsc{(}Y\textsc{)}$. L'objectif est de montrer que tout sous-ensemble propre de $Y$ est un sous-ensemble générateur minimal. Cette propriété est trivialement vraie si $Y$ est vide ou bien si $Y$ est un singleton. On suppose donc que $Y$ contient au moins une paire. Soit $X$ un
sous-ensemble propre de $Y$. On montre par l'absurde que $X$ est un sous-ensemble
générateur minimal. On suppose le contraire. Il existe alors un sous-ensemble propre $Z$ de $X$ tel que $\theta\textsc{(}X\textsc{)} = \theta\textsc{(}Z\textsc{)}$. On pose $\Delta = X \backslash Z$. Par hypothèse, $\Delta$ n'est pas vide et $X = Z \cup \Delta$. Puisque $X \subseteq \theta\textsc{(}X\textsc{)}$, on a $\Delta \subseteq \theta\textsc{(}X\textsc{)}$, donc  $\Delta \subseteq \theta\textsc{(}Z\textsc{)} \quad \textsc{(}1\textsc{)}$.}

{\em On s'appuie dans la suite de la preuve sur la propriété suivante des opérateurs de fermeture : une application extensive $\theta$ de $\mathcal{P}\textsc{(}\mathcal{E}\textsc{)}$ dans $\mathcal{P}\textsc{(}\mathcal{E}\textsc{)}$ est un opérateur de fermeture \textit{ssi}} :

{\em \textsc{(}2\textsc{)} \quad pour tous sous-ensembles $X, Y$ de $\mathcal{E}, \theta\textsc{(}X \cup Y\textsc{)} = \theta\textsc{(}\theta\textsc{(}X$\textsc{)} $\cup Y\textsc{)}$.}

\vspace{1.5mm}

{\em Il résulte par une double application, de la propriété \textsc{(}2\textsc{)}, appelée \textit{propriété des chemins indépendants}, que si $\theta$ est un opérateur de fermeture, alors :

\vspace{2mm}

\textsc{(}3\textsc{)} \quad $\theta\textsc{(}X \cup Y\textsc{)} = \theta\textsc{(}\theta\textsc{(}X\textsc{)} \cup \theta\textsc{(}Y\textsc{)}\textsc{)}$.

\vspace{1.5mm}

\noindent On a :

\vspace{2mm}

%\parbox{12.cm}{
\begin{tabular}{l}
$\theta\textsc{(}Y\textsc{)} = \theta\textsc{(}\textsc{(}Y \backslash X\textsc{)} \cup X\textsc{)}$\\
\hspace{0.9cm} $= \theta\textsc{(}\theta\textsc{(}Y \backslash X\textsc{)} \cup \theta\textsc{(}X\textsc{)}\textsc{)} \quad \mbox{\rm \textsc{(}propriété \textsc{(}3\textsc{)}\textsc{)}}$\\
\hspace{0.9cm} $= \theta\textsc{(}\theta\textsc{(}Y \backslash X\textsc{)} \cup \theta\textsc{(}Z \cup \Delta\textsc{)}\textsc{)}$\\
\hspace{0.9cm} $= \theta\textsc{(}\theta\textsc{(}Y \backslash X\textsc{)} \cup \theta\textsc{(}\theta\textsc{(}Z\textsc{)} \cup \Delta\textsc{)}\textsc{)} \quad \mbox{\rm \textsc{(}propriété \textsc{(}2\textsc{)}\textsc{)}}$\\
\hspace{0.9cm} $= \theta\textsc{(}\theta\textsc{(}Y \backslash X\textsc{)} \cup \theta\textsc{(}\theta\textsc{(}Z\textsc{)}\textsc{)}\textsc{)} \quad \mbox{\rm \textsc{(}propriété \textsc{(}1\textsc{)} ci-dessus}, \ \Delta \subseteq \theta \textsc{(}Z\textsc{)}\textsc{)}$\\
\hspace{0.9cm} $= \theta\textsc{(}\theta\textsc{(}Y \backslash X\textsc{)} \cup \theta\textsc{(}Z\textsc{)}\textsc{)} \quad \mbox{\rm \textsc{(}idempotence de} \ \theta\textsc{)}$\\
\hspace{0.9cm} $= \theta\textsc{(}\theta\textsc{(}Y \backslash X\textsc{)} \cup Z\textsc{)} \quad \mbox{\rm \textsc{(}propriété \textsc{(}3\textsc{)}\textsc{)}}$\\
\hspace{0.9cm} $= \theta\textsc{(}Y \backslash \Delta\textsc{)} \quad \mbox{\rm \textsc{(}car} \ Z \mbox{\rm est inclus dans} Y\textsc{)}$\\
\end{tabular}

\vspace{1.5mm}

\noindent qui contredit que $Y$ est un sous-ensemble générateur minimal de $\theta$. \quad $\diamondsuit$}
\end{demo}

Il découle du Lemme \ref{lemme_ideal_ordre_cas_general} la proposition suivante :

\begin{prop} \label{propOI} L'ensemble des générateurs minimaux fréquents d'un contexte, associés à l'opérateur de fermeture de Galois $\gamma$, est un idéal d'ordre pour l'inclusion ensembliste, autrement dit, tout sous-ensemble d'un générateur minimal fréquent est un générateur minimal fréquent.
\end{prop}

La Proposition \ref{propOI} permet de rejeter tout motif, dont un des sous-ensembles
n'est pas un générateur minimal fréquent. Un motif est dit \textit{candidat} si tous ses
sous-ensembles sont des générateurs minimaux fréquents.

\vspace{2mm}

Dans la conception de l'algorithme \textsc{Prince}, nous visons à construire le treillis d'Iceberg sans avoir à accéder au contexte d'extraction. Ceci nécessite d'extraire dans une étape préalable une \textit{représentation concise exacte des motifs fréquents} basée sur les générateurs minimaux. Cette représentation aura pour but de déterminer si un motif quelconque est fréquent ou non et de déterminer son support s'il est fréquent. \'Etant donné que l'ensemble des générateurs minimaux fréquents, dénoté $\mathcal{GMF}_{\mathcal{K}}$, ne constitue pas par lui-même une représentation exacte des motifs fréquents, il doit être augmenté par une bordure. Dans cet article, nous l'augmentons par la \textit{bordure négative non fréquente}. Nous appelons bordure négative non fréquente, notée $\mathcal{GB}$d$^{-}$, l'ensemble des plus petits \textsc{(}au sens de l'inclusion\textsc{)} motifs qui ne sont pas fréquents et dont tous les sous-ensembles sont des générateurs minimaux fréquents. Le lemme suivant indique que les éléments de $\mathcal{GB}$d$^{-}$ sont des générateurs minimaux.

\begin{lemme} Les éléments de la bordure négative non fréquente $\mathcal{GB}$d$^{-}$ sont des générateurs minimaux.
\end{lemme}

\begin{demo} {\em Soit $X \in \mathcal{GB} d^{-}$. Par définition, $X$ est non fréquent alors que tous ses sous-ensembles sont fréquents. Ainsi, le support de $X$ est strictement inférieur à celui de ses sous-ensembles stricts. D'où, $X$ est un générateur minimal. \quad $\diamondsuit$}
\end{demo}

L'utilisation de la bordure $\mathcal{GB}d^{-}$ s'explique donc par le fait que son union avec l'ensemble des générateurs minimaux fréquents $\mathcal{GMF}_{\mathcal{K}}$ forme une représentation concise exacte de l'ensemble des motifs fréquents [Calders {\em \& al.}, 2005]. Ainsi, en utilisant l'ensemble résultat de cette union, nous pouvons déterminer le support de tout motif sans effectuer un accès au contexte d'extraction. Le support d'un motif quelconque sera dérivé s'il est fréquent, sinon il sera détecté comme étant non fréquent. Ceci est explicité à travers la proposition suivante :

\begin{prop} \label{proptitanic} Soit $X$ un motif. Si $\exists \ Z \in \mathcal{GB} d^{-}$ et $Z \subseteq
X$ alors $X$ est non fréquent. Sinon, $X$ est fréquent et $\textit{Supp}\textsc{(}X\textsc{)} = \min \{Supp\textsc{(}g\textsc{)} \vert g \in \mathcal{GMF}_{\mathcal{K}}$ et $g \subseteq X \}$.
\end{prop}

\vspace{1.5mm}

Les notions de \textit{bloqueur minimal} et de \textit{face} nous seront aussi utiles dans la suite :

\vspace{1.5mm}

\noindent -- {\em Bloqueur minimal} : Soit $G = \{G_{1},G_{2},\cdots,G_{_{n}}\}$ une famille d'ensembles tel que $\forall \ i = 1\cdots n, G_{i} \neq \emptyset$. Un \textit{bloqueur} $B$ de la famille $G$ est un ensemble dont l'intersection avec tout ensemble $G_{i} \in G$ est non vide. Le bloqueur $B$ est dit {\em minimal} s'il n'existe aucun bloqueur $B_{1}$ de $G$ inclus dans $B$ [Pfaltz, Taylor, 2002].

\vspace{1.5mm}
Il est à noter que l'union des $G_{i}$ est un bloqueur.

\begin{exemple} {\em Considérons le contexte d'extraction $\mathcal{K}$ donné par la Figure~\ref{runfig} \textsc{(}Gauche\textsc{)}. Soit la famille d'ensemble $G = \{\{\texttt{B}, \texttt{C}\}, \{\texttt{C}, \texttt{E}\}\}$, composée par les générateurs minimaux du motif fermé $\texttt{BCE}$ \textsc{(}cf. Figure \ref{runfig} \textsc{(}Droite\textsc{)}\textsc{)}. Ainsi,  $G_{1} = \{\texttt{B}, \texttt{C}\}$ et $G_{2} = \{\texttt{C}, \texttt{E}\}$. L'union de $G_{1}$ et de $G_{2}$, égale à $\{\texttt{B}, \texttt{C}, \texttt{E}\}$, est un bloqueur de $G$. Il en est de même pour $\{\texttt{B}, \texttt{C}\}$. Par ailleurs, les ensembles $\{\texttt{C}\}$ et $\{\texttt{B}, \texttt{E}\}$ sont des bloqueurs {\em minimaux} de la famille $G$.}
\end{exemple}

\noindent -- {\em Face} : Soient $f, f_{1} \in \mathcal{IFF}_{\mathcal{K}}$. Si $f$ couvre $f_{1}$ dans le treillis d'Iceberg $\textsc{(}\hat{\mathcal{L}}, \subseteq$\textsc{)} \textsc{(}c'est-à-dire $f \in Couv^s\textsc{(}f_{1}\textsc{)}\textsc{)}$, \ alors la {\em face} de $f$ par rapport à $f_{1}$, notée {\em face} $\textsc{(}f | f_{1}\textsc{)}$, est égale à : face $\textsc{(}f | f_{1}\textsc{)} = f \backslash f_{1}$ [Pfaltz, Taylor, 2002].

\begin{exemple} {\em Considérons la Figure \ref{runfig}. Soient les deux motifs fermés $f = \texttt{ABCE}$ et $f_1 = \texttt{BCE}$. Le fermé $f$ couvre le fermé $f_1$ dans le treillis d'Iceberg \textsc{(}cf. Figure \ref{runfig} \textsc{(}Droite\textsc{)}\textsc{)}. La face de $f$ par rapport à $f_{1}$ est égale à $f \backslash f_{1} = \texttt{A}.$}
\end{exemple}

Dans ce qui suit, nous présentons le cadre général pour la dérivation des règles d'association, puis nous établissons son lien avec la théorie des concepts formels.

\subsection{dérivation des règles d'association}

\noindent Une règle d'association $R$ est une relation entre motifs de la forme $R : X \Rightarrow \textsc{(}Y\backslash X\textsc{)}$, dans laquelle $X$ et $Y$ sont des motifs fréquents, tels que $X\subset Y$. Les motifs $X$ et $\textsc{(}Y \backslash X\textsc{)}$ sont appelés, respectivement, {\em prémisse} et {\em conclusion} de la règle. La mesure de support de $R$, dénotée $\textit{Supp}\textsc{(}R\textsc{)}$, est égale à $\textit{Supp}\textsc{(}Y\textsc{)}$. Une règle d'association $R$ est dite {\em valide} [Agrawal {\em \& al.}, 1993] si sa mesure de confiance $Conf\textsc{(}R\textsc{)} = \displaystyle{\frac{Supp\textsc{(}Y\textsc{)}}{Supp\textsc{(}X\textsc{)}}}$ est supérieure ou égale à un seuil minimal \textit{minconf} de confiance. Une règle d'association $R$ est dite {\em exacte} si $Conf\textsc{(}R\textsc{)} = {\xmplbx1}$ sinon elle est dite {\em approximative} [Pasquier, 2000].

\vspace{2mm}

Le problème de l'extraction des règles d'association est résolu par un algorithme fondamental, à savoir \textsc{Apriori} [Agrawal, Srikant, 1994]. Cependant, cette approche d'extraction des règles, fondée sur les motifs fréquents, présente deux problèmes majeurs :
\begin{enumerate}
\item le coût de l'extraction des motifs fréquents notamment pour des contextes denses ;
\item en général, le nombre de règles d'association générées peut être excessivement grand, dont une grande partie est redondante [Ashrafi {\em \& al.}, 2007].
\end{enumerate}

Ainsi, une nouvelle approche fondée sur l'extraction des motifs fermés fréquents [Pasquier, 1999], a vu le jour et a pour ambition de :
\begin{enumerate}
\item réduire le coût de l'extraction des motifs fréquents en se basant sur le fait que l'ensemble des motifs fermés fréquents est un ensemble {\em générateur} de l'ensemble des motifs fréquents [Pasquier, 1999] ;
\item permettre, sans perte d'information, la sélection d'un sous-ensemble de toutes les règles d'association, appelé {\em base générique}, à partir duquel le reste des règles pourra être dérivé. Ceci donne la possibilité de présenter le minimum possible de règles à l'utilisateur afin de lui permettre de mieux les visualiser et les exploiter.
\end{enumerate}

Depuis, plusieurs bases génériques ont été introduites, dont celles de Bastide {\em \& al.} [2000\textsc{(}a\textsc{)}] et qui sont définies dans ce qui suit.\\

\noindent 1. La {\em base générique de règles d'association exactes} est définie comme suit :

\begin{definitio}\label{defbase-exact} Soit $\mathcal{IFF}_{\mathcal{K}}$ l'ensemble des motifs fermés fréquents extrait d'un contexte d'extraction $\mathcal{K}$. Pour chaque motif fermé fréquent $f \in \mathcal{IFF}_{\mathcal{K}}$, nous désignons par $GMF_{f}$ l'ensemble de ses générateurs minimaux. La base générique de règles d'association exactes $\mathcal{BG}$ est donnée par : $\mathcal{BG} = \{R : g  \Rightarrow  \textsc{(}f \backslash g\textsc{)} \vert  f \in \mathcal{IFF}_{\mathcal{K}}$ et $g \in GMF_{f}$ et $g \ne f\}$\footnote{\ La condition $g \ne f$ permet de ne pas retenir les règles de la forme $g \Rightarrow \emptyset $.}.
\end{definitio}

\begin{rem} Tous les éléments de $GMF_{f}$ sont fréquents. Par ailleurs, une règle exacte est toujours valide puisque sa confiance, égale à {\xmplbx1}, est toujours supérieure ou égale à \textit{minconf}.
\end{rem}

\noindent 2. La base générique de règles d'association approximatives appelée \textit{base informative de règles d'association approximatives} est définie comme suit :

\begin{definitio}\label{defbase-infor} Soit $\mathcal{GMF}_{\mathcal{K}}$ l'ensemble des générateurs
minimaux fréquents extrait d'un contexte d'extraction $\mathcal{K}$. La base informative de règles d'association approximatives $\mathcal{IB}$ est donnée par : $\mathcal{IB} = \{R : g \overset{c} \Rightarrow \textsc{(}f \backslash g\textsc{)} \vert  f  \in  \mathcal{IFF}_{\mathcal{K}}$ et $g  \in \mathcal{GMF}_{\mathcal{K}}$ et $\gamma\textsc{(}g\textsc{)} \subset f$ et $c = Conf\textsc{(}R\textsc{)} \ge \textit{minconf}\}$.
\end{definitio}

\vspace{1.5mm}

Afin de réduire encore plus le nombre de règles approximatives, Bastide \textit{et al.} proposent une réduction transitive de la base informative [Bastide, 2000 ; Bastide {\em \& al.}, 2000\textsc{(}a\textsc{)}] qui est elle-même une base pour toutes les règles approximatives. La réduction transitive est définie comme suit :

\begin{definitio}\label{defbase-trans} La réduction transitive $\mathcal{RI}$ est donnée par : $\mathcal{RI} = \{R : g \ \overset{c} \Rightarrow \textsc{(}f \backslash g\textsc{)} \vert f \in \mathcal{IFF}_{\mathcal{K}}$ et $g \in \mathcal{GMF}_{\mathcal{K}}$ et $\gamma\textsc{(}g\textsc{)} \in Couv_i\textsc{(}f\textsc{)}$ et $c = Conf\textsc{(}R\textsc{)} \ge \textit{minconf}\}$.
\end{definitio}

\vspace{1.5mm}

Dans la suite, nous allons considérer les règles d'association génériques formées par l'union de la
base générique de règles exactes $\mathcal{BG}$ et la réduction transitive de la base informative $\mathcal{RI}$. Ces règles seront désignées par {\em règles d'association informatives}.

\begin{exemple} {\em Considérons le treillis d'Iceberg donné par la Figure~\ref{runfig} \textsc{(}Droite\textsc{)}. \`{A} partir de la classe d'équivalence dont le motif fermé fréquent est \texttt{ABCE}, deux règles exactes sont obtenues : $\texttt{AB} \Rightarrow \texttt{CE}$ et $\texttt{AE} \Rightarrow \texttt{BC}$. D'autre part, la règle approximative $\texttt{C} \overset{0,75}{\Rightarrow} \texttt{A}$ est générée à partir des deux classes d'équivalence dont les sommets respectifs sont les motifs fermés $\texttt{C}$ et $\texttt{AC}$}.
\end{exemple}

\vspace{1.5mm}

Il faut noter que les bases considérées présentent plusieurs avantages à savoir le fait qu'elles sont formées d'implications à prémisse minimale et à conclusion maximale, ce qui \ donne les règles \  les plus \ informatives \ pour \ l'utilisateur \ [Bastide {\em \& al.}, 2000\textsc{(}a\textsc{)} ; \ Kryszkiewicz, 2002 ; Pasquier, 2000]. En plus, elles satisfont les conditions suivantes [Kryszkiewicz, 2002] :
\begin{enumerate}
\item {\em Dérivabilité} : Le système axiomatique proposé dans [Ben Yahia, Mephu Nguifo, 2004] afin de dériver toutes les règles valides \textsc{(}redondantes\textsc{)} à partir de ces bases est {\em correct}  \textsc{(}c'est-à-dire, le système ne permet de dériver que les règles d'association {\em valides}\textsc{)} et {\em complet} \textsc{(}c'est-à-dire, l'ensemble de {\em toutes} les règles valides est dérivé\textsc{)}.
\item {\em Informativité} : Ces bases génériques des règles d'association permettent de déterminer avec exactitude le support et la confiance des règles dérivées.
\end{enumerate}

Par ailleurs, la réduction transitive regroupe les règles approximatives minimales ayant des valeurs de confiance élevées et sauf rares exceptions, les règles les plus intéressantes sont celles de support et confiance élevés [Bastide, 2000]. En outre, un nombre important de travaux de la littérature témoigne de son utilité dans des cas pratiques.

\section{EXTRACTION DES MOTIFS FERM\'{E}S FR\'{E}QUENTS}\label{section_EdeA_IFFs}

\noindent Il est bien connu que l'étape la plus complexe et la plus consommatrice en temps d'exécution est celle du calcul des motifs fréquents. Cette étape peut aussi extraire un nombre important de motifs fréquents, desquels un nombre prohibitif de règles d'association sera dérivé, ce qui rend leur usage très difficile. Les algorithmes basés sur l'extraction des motifs fermés fréquents sont alors une nouvelle alternative avec la promesse claire de réduire considérablement la taille de l'ensemble des règles d'association. Ainsi, seules les règles d'association informatives devaient être maintenues étant donné qu'elles permettent une réduction de l'ensemble de toutes les règles valides, tout en convoyant le maximum d'information. Une étude critique de la littérature dédiée nous a permis de dégager que :
\begin{enumerate}
\item Beaucoup d'algorithmes orientés fouille de données [Lucchese {\em \& al.}, 2006 ; Pasquier {\em \& al.}, 1999\textsc{(}b\textsc{)} ; Pei {\em \& al.}, 2000 ; Stumme {\em \& al.}, 2002 ; Uno {\em \& al.}, 2004 ; Zaki, Hsiao, 2002] permettent l'extraction des motifs fermés fréquents. Cependant, seuls certains [Pasquier {\em \& al.}, 1999\textsc{(}a\textsc{)}, 1999\textsc{(}b\textsc{)} ; Stumme {\em \& al.}, 2002] se basent sur la notion de générateur minimal\footnote{\ En réalité, ces algorithmes utilisent les générateurs minimaux comme étape intermédiaire pour extraire les motifs fermés.}. Toutefois, ces algorithmes ne construisent pas la relation d'ordre partiel. Ils nécessitent alors l'exécution en aval d'un autre algorithme tel que celui proposé par [Valtchev {\em \& al.}, 2000].

\vspace{2mm}

Les algorithmes \textsc{(}{\em e.g.} [Bastide, 2000 ; Pasquier, 2000 ; Pasquier {\em \& al.}, 1999\textsc{(}b\textsc{)}]\textsc{)} permettant
d'extraire les règles formant le couple $\textsc{(}\mathcal{BG}, \mathcal{RI}\textsc{)}$ supposent {\em l'existence} des motifs fermés fréquents ainsi que leurs générateurs minimaux respectifs. Ceci nécessite un {\em autre} algorithme tel que \textsc{Close} [Pasquier, {\em \& al.}, 1999\textsc{(}b\textsc{)}], \textsc{A-Close} [Pasquier {\em \& al.}, 1999\textsc{(}a\textsc{)}], \textsc{Gc-Growth} [Li {\em \& al.}, 2005] ou une modification de \textsc{Pascal} [Szathmary {\em \& al.}, 2007], etc. Un tel algorithme doit alors associer pour chaque motif fermé fréquent ses générateurs minimaux étant donné qu'il peut être calculé plusieurs fois. La génération de la base générique des règles exactes $\mathcal{BG}$ se fait d'une manière directe. Cependant pour les règles approximatives formant $\mathcal{RI}$, des tests d'inclusion coûteux, mettant en jeu les motifs fermés fréquents, sont aussi réalisés pour déterminer les successeurs {\em immédiats} de chaque motif fermé fréquent.
\item Les algorithmes orientés concepts formels permettent de générer l'ensemble des concepts formels ainsi que la relation d'ordre [Kuznetsov, Obiedkov, 2002]. Toutefois, ils ne génèrent pas l'ensemble des générateurs minimaux associés. Ils nécessitent alors l'application d'un autre algorithme, tel que \textsc{JEN} [Le Floc'h {\em \& al.}, 2003] permettant de déterminer les générateurs minimaux étant donné que la relation d'ordre est déjà construite. Par ailleurs, leur performance reste limitée dans le cas des contextes réels [Stumme {\em \& al.}, 2002] \textsc{(}cf. aussi [Kuznetsov, Obiedkov, 2002] où les algorithmes n'ont pu être testés que sur de petits contextes aléatoirement produits\textsc{)}. De plus, leur consommation en espace mémoire est élevée vu que les intensions des concepts sont maintenues. Notons que l'algorithme proposé dans [Zaki, Hsiao, 2005] permet d'extraire l'ensemble des motifs fermés fréquents munis de la relation d'ordre partiel. Ainsi, l'algorithme  \textsc{JEN} est aussi applicable dans ce cas pour dériver les générateurs minimaux.
\end{enumerate}

Les principaux algorithmes permettant l'extraction des motifs fermés fréquents et leurs générateurs minimaux associés sont \textsc{Close} [Pasquier {\em \& al.}, 1999\textsc{(}b\textsc{)}], \textsc{A-Close} [Pasquier {\em \& al.}, 1999\textsc{(}a\textsc{)}] et \textsc{Titanic} [Stumme {\em \& al.}, 2002]. Ces algorithmes, reposant sur la technique {\em Générer-et-tester}, explorent l'espace de recherche par niveau, c'est-à-dire en partant de l'ensemble vide vers les motifs de taille {\xmplbx1}, ensuite ceux de taille {\xmplbx2}, et ainsi de suite. En plus de l'élagage basé sur la mesure statistique \textit{minsupp}, ces algorithmes mettent en oeuvre un élagage efficace, basé sur la propriété d'idéal d'ordre de l'ensemble des générateurs minimaux fréquents.

\vspace{1.5mm}

Dans la suite, nous allons passer en revue les principaux algorithmes permettant l'extraction des motifs fermés fréquents ainsi que leurs générateurs minimaux associés. Une étude des algorithmes dédiés seulement à l'énumération des motifs fermés fréquents se trouve dans [Ben Yahia {\em \& al.}, 2006].

\subsection{algorithme Close}

\noindent L'algorithme \textsc{Close} est proposé par Pasquier {\em \& al.} [1999\textsc{(}b\textsc{)}]. \`{A} chaque itération, \textsc{Close} génère un ensemble de candidats en joignant les générateurs minimaux retenus durant l'itération précédente. \textsc{Close} calcule alors leurs supports et leurs fermetures dans une même étape par le biais d'un accès au contexte d'extraction. La fermeture d'un candidat générateur minimal $g$ est calculée en exécutant des intersections de l'ensemble des objets auxquelles appartient $g$. Afin de réduire l'espace de recherche, c'est-à-dire le nombre de candidats à tester, \textsc{Close} utilise des stratégies d'élagage. Ces dernières sont basées sur une mesure statistique à savoir \textit{minsupp} et sur l'idéal d'ordre des générateurs minimaux ainsi que sur le
fait qu'un candidat générateur minimal $g$ de taille $k$ ne doit pas être couvert par la fermeture d'un de ses sous-ensembles de taille \textsc{(}$k$ - $1$\textsc{)}.

\subsection{algorithme A-Close}

\noindent  L'algorithme \textsc{A-Close} est aussi proposé par Pasquier {\em \& al.} [1999\textsc{(}a\textsc{)}]. Il opère en deux étapes successives. D'abord, il détermine tous les générateurs minimaux fréquents des différentes classes d'équivalence à l'aide des accès itératifs au contexte d'extraction. Ensuite, il calcule leurs fermetures de la même façon que dans \textsc{Close}. \textsc{A-Close} utilise trois stratégies d'élagages à savoir \textit{minsupp}, l'idéal d'ordre des générateurs minimaux et le fait qu'un candidat générateur minimal $g$ de taille $k$ ne doit pas avoir le même support qu'un de ses sous-ensembles de taille \textsc{(}$k$ - $1$\textsc{)}. Afin de vérifier cette dernière condition, \textsc{A-Close} effectue un balayage des générateurs minimaux retenus de taille \textsc{(}$k$ - $1$\textsc{)}. Pour alléger le calcul des fermetures, l'algorithme \textsc{A-Close} mémorise le numéro de la première itération durant laquelle un des candidats s'avère fréquent mais non générateur minimal \textsc{(}c'est-à-dire ayant un support égal à celui d'un de ses sous-ensembles\textsc{)}. Le numéro de cette itération correspond à la taille $k$ de ce candidat. Il n'est alors pas nécessaire de calculer la fermeture des générateurs minimaux fréquents de tailles inférieures à \textsc{(}$k$ - $1$\textsc{)}, puisqu'ils sont tous égaux à leurs fermetures.

\subsection{algorithme Titanic}

\noindent L'algorithme \textsc{Titanic} est proposé par Stumme {\em \& al.} [2002]. \textsc{Titanic} détermine dans chaque itération les générateurs minimaux fréquents associés, moyennant un accès au contexte d'extraction. Il utilise pour cela les mêmes stratégies d'élagage que \textsc{A-Close}. Cependant, \textsc{Titanic} évite le balayage coûteux effectué par \textsc{A-Close} pour vérifier la dernière stratégie d'élagage. Pour cela, \textsc{Titanic} utilise pour chaque candidat $g$ de taille $k$ une variable où il stocke son support estimé, c'est-à-dire le minimum du support de ses
sous-ensembles de taille \textsc{(}$k$ - $1$\textsc{)}, et qui doit être différent de son support réel, sinon $g$ n'est pas minimal. Ceci est basé sur le lemme suivant :

\begin{lemme}\label{lemmeTitanics-support} Soient $X, Y \subseteq  \mathcal{I}$. Si $X \subseteq Y$ et \textit{Supp}$\textsc{(}X\textsc{)}$ = \textit{Supp}$\textsc{(}Y\textsc{)}$, alors $\gamma\textsc{(}X\textsc{)} = \gamma\textsc{(}Y\textsc{)}$.
\end{lemme}

\textsc{Titanic} évite aussi l'accès au contexte d'extraction pour calculer les fermetures des générateurs minimaux fréquents. Ceci est réalisé moyennant un mécanisme de comptage par inférence \textsc{(}utilisé aussi dans l'algorithme \textsc{Pascal} [Bastide {\em \& al.}, 2000\textsc{(}b\textsc{)}], dédié à l'extraction des motifs fréquents\textsc{)}. Le mécanisme employé est fondé sur le fait qu'on peut déterminer le support de tout motif {\em fréquent} en utilisant la Proposition \ref{proptitanic}. \textsc{Titanic} cherche alors à étendre tout générateur minimal fréquent par les items adéquats appartenant à sa fermeture.

\vspace{2mm}

Les algorithmes décrits dans cette section présentent un inconvénient majeur à savoir le calcul redondant des fermetures. En effet, un motif fermé fréquent peut admettre plusieurs générateurs minimaux et sera donc calculé plusieurs fois, spécialement dans le cas de contextes denses. Par ailleurs, dans le cas des contextes épars, le calcul des fermetures est généralement inutile car les générateurs s'avèrent aussi fermés. Il est aussi important de noter que contrairement aux affirmations des auteurs dans [Stumme {\em \& al.}, 2002], \textsc{Titanic} ne construit pas le treillis d'Iceberg. En effet, il ne génère aucun lien de précédence entre motifs fermés fréquents et se limite simplement à leur extraction ainsi que celle des générateurs minimaux fréquents, tout comme \textsc{Close} et \textsc{A-Close}. Toutefois, ces algorithmes sont aptes à dériver les règles d'association informatives moyennant l'utilisation de la procédure de génération de ces règles, décrite dans [Pasquier, 2000] et qui prend comme entrée l'ensemble des motifs fermés fréquents associés à leurs générateurs minimaux.

\section{DESCRIPTION DE L'ALGORITHME \textsc{Prince}}\label{section_description_Prince}

\noindent Dans cette section, nous allons introduire un nouvel algorithme, appelé \textsc{Prince}, dont l'objectif principal est de pallier les principales lacunes de ces algorithmes dédiés à l'extraction des motifs fermés fréquents, c'est-à-dire le coût du calcul des fermetures ainsi que le fait de ne pas construire la relation d'ordre partiel. La principale originalité de \textsc{Prince} réside dans le fait qu'il construit une structure isomorphe au treillis des motifs fermés. Dans cette structure, le treillis d'Iceberg est construit non plus grâce aux motifs fermés fréquents mais moyennant des comparaisons entre générateurs minimaux fréquents. Rappelons que le treillis d'Iceberg est une structure qui ordonne
partiellement les classes d'équivalence. Une première variante de ce treillis a été définie dans la sous-section \ref{sous_section_notion_de_base} \textsc{(}page \pageref{def_treillis_iceberg}\textsc{)} où chaque classe d'équivalence est représentée par son motif fermé fréquent, c'est-à-dire par le plus grand élément correspondant. Nous proposons ici une nouvelle variante du treillis d'Iceberg, appelée {\em treillis des générateurs minimaux}, où chaque classe d'équivalence est représentée par ses éléments minimaux. Cette variante est définie comme suit :

\begin{definitio}
Le {\em treillis des générateurs minimaux} est une variante du treillis d'Iceberg, où chaque classe d'équivalence est représentée par les générateurs minimaux qu'elle contient.
\end{definitio}

L'algorithme \textsc{Prince} met alors en place un mécanisme de gestion des classes d'équivalence permettant de générer la liste intégrale des motifs fermés fréquents sans duplication et sans recours aux tests de couvertures. Il permet aussi de réduire d'une manière notable le coût du calcul des fermetures, en les dérivant simplement grâce aux notions de bloqueur minimal et de face [Pfaltz, Taylor, 2002]. De plus et grâce à cette structure partiellement ordonnée, \textsc{Prince} permet d'extraire les bases génériques de règles sans avoir à l'associer avec un autre algorithme. La construction des liens de précédence est optimisée grâce à une gestion efficace des classes d'équivalence ainsi que la détection d'information pouvant rendre partielle cette construction.

\vspace{2mm}

\textsc{Prince} prend en entrée un contexte d'extraction $\mathcal{K}$, le seuil minimum de support \textit{minsupp} et le seuil minimum de confiance \textit{minconf}. Il donne en sortie la liste des motifs fermés fréquents et leurs générateurs minimaux respectifs ainsi que les bases génériques de règles. \textsc{Prince} opère en trois étapes successives :
\begin{enumerate}
\item Détermination des générateurs minimaux ;
\item Construction du {\em treillis des générateurs minimaux} ;
\item Extraction des bases génériques de règles.
\end{enumerate}

Ces étapes sont décrites dans ce qui suit. Leur déroulement à partir de l'exemple de la Figure~\ref{runfig} est détaillé dans l'exemple \ref{exemple_deroulement_algo}, page \pageref {exemple_deroulement_algo}.

\subsection{détermination des générateurs minimaux}

\noindent En explorant l'espace de recherche par niveau, l'algorithme \textsc{Prince} détermine lors de sa première étape l'ensemble des générateurs minimaux fréquents $\mathcal{GMF}_{\mathcal{K}}$ ainsi que la bordure négative non fréquente $\mathcal{GB} d^{-}$.

\subsubsection{Stratégies d'élagage adoptées}

\noindent Afin d'optimiser l'extraction, \textsc{Prince} adopte les stratégies d'élagage suivantes : la contrainte de fréquence \textit{minsupp}, l'idéal d'ordre associé à l'ensemble des générateurs minimaux fréquents \textsc{(}cf. Proposition \ref{propOI}\textsc{)}, ainsi que le support estimé \textsc{(}cf. Lemme \ref{lemmeTitanics-support}\textsc{)}.

\subsubsection{Pseudo-code de la première étape de l'algorithme \textsc{Prince}}

\noindent Les notations qui seront utilisées dans l'algorithme \textsc{Prince} sont résumées dans le Tableau \ref{notationsPrince}. Le pseudo-code relatif à cette étape est donné par la procédure \textsc{Gen-GMs} \textsc{(}cf. Algorithme \ref{algoGen-GMs}\textsc{)}. Cette procédure prend en entrée un contexte d'extraction $\mathcal{K}$ et le support minimal \textit{minsupp}. Elle donne en sortie l'ensemble des générateurs minimaux fréquents $\mathcal{GMF}_{\mathcal{K}}$ de façon à pouvoir les
parcourir par ordre décroissant de leurs supports respectifs lors de la deuxième étape de l'algorithme \textsc{Prince}. L'ensemble $\mathcal{GMF}_{\mathcal{K}}$ est alors considéré comme étant divisé en plusieurs sous-ensembles. Chaque sous-ensemble est caractérisé par la même valeur du support. Ainsi, chaque fois qu'un générateur minimal fréquent est déterminé, il est ajouté à l'ensemble représentant son support. La procédure \textsc{Gen-GMs} garde aussi la trace des générateurs minimaux non fréquents, formant la bordure $\mathcal{GB} d^{-}$, afin de les utiliser lors de la deuxième étape.

\begin{table}[htbp]
\begin{center}
\footnotesize{\begin{tabular}{|p{420pt}|}
 \hline
\begin{description}
\item[$k$] : un compteur qui indique l'itération courante. Durant la $k^{i\grave{e}me}$ itération, tous les générateurs minimaux de taille $k$ sont déterminés.
\item[$\mathcal{P}re\mathcal{CGM}_{k}$] : ensemble des motifs résultats de l'application de \textsc{Apriori-Gen}.
\item[$\mathcal{CGM}_{k}$] : ensemble des candidats générateurs minimaux de taille $k$.
\item[$\mathcal{GMF}_{k}$] : ensemble des générateurs minimaux fréquents de taille $k$.
\item[$\mathcal{GMF}_\mathcal{K}$] : ensemble des générateurs minimaux fréquents triés par ordre décroissant du support.
\item[$\mathcal{GB}d^{-}$] : bordure négative non fréquente des générateurs minimaux fréquents.
\end{description}\\

-- Chaque élément $c$ de $\mathcal{P}re\mathcal{CGM}_{k}$, $\mathcal{CGM}_{k}$ ou de
$\mathcal{GMF}_{k}$ est caractérisé par les champs suivants :
\begin{enumerate}
    \item \texttt{\textit{support-réel}} : support réel de $c$, initialisé à {\xmplbx0}.
    \item \texttt{\textit{sous-ens-directs}} : liste des sous-ensembles de $c$ de
    taille \textsc{(}$k$ - $1$\textsc{)}, initialisé à l'ensemble vide.
\end{enumerate}

-- Chaque élément $c$ de $\mathcal{CGM}_{k}$ est aussi caractérisé par un support estimé \textsc{(}le champ \texttt{\textit{support-estimé}}\textsc{)} et qui sera utilisé pour éliminer les candidats non générateurs minimaux.

\mbox{}\\

-- Chaque élément $c$ de $\mathcal{GMF}_{k}$ est aussi caractérisé par les champs suivants :
\begin{enumerate}
\item \texttt{\textit{succs-immédiats}} : liste des successeurs immédiats de [$g$], initialisé à l'ensemble
vide.
\item \texttt{\textit{iff}} :  motif fermé fréquent correspondant, initialisé à l'ensemble vide.
\end{enumerate}\\
\hline
\end{tabular}}
\end{center} \caption{Notations utilisées dans l'algorithme \textsc{Prince}}\label{notationsPrince}
\end{table}

Dans cette procédure, l'ensemble des candidats générateurs minimaux $\mathcal{CGM}_{1}$ est initialisé par l'ensemble des items du contexte d'extraction \textsc{(}ligne 2\textsc{)}. Le support des items est alors calculé via un accès au contexte d'extraction \textsc{(}ligne 3\textsc{)}. Le support de l'ensemble vide est égal au nombre d'objets du contexte d'extraction, c'est-à-dire $|\mathcal{O}|$ \textsc{(}ligne 4\textsc{)}. L'ensemble vide, étant le générateur minimal fréquent de taille {\xmplbx0}, est inséré dans $\mathcal{GMF}_{0}$ \textsc{(}ligne 5\textsc{)}. Pour tout item $c$, nous distinguons les deux cas suivants \textsc{(}lignes 6-14\textsc{)} :
\begin{enumerate}
\item si $\textit{Supp}\textsc{(}c\textsc{)} = \textit{Supp}\textsc{(}\emptyset\textsc{)}$, alors $c$ n'est pas un générateur minimal \textsc{(}ligne 8\textsc{)} ;
\item sinon $c$ est un générateur minimal. Il est ajouté à $\mathcal{GMF}_{1}$ si $\textit{Supp}\textsc{(}c\textsc{)} \geq \textit{minsupp}$ \textsc{(}ligne 12\textsc{)}, sinon il est ajouté à $\mathcal{GB}d^{-}$ \textsc{(}ligne 14\textsc{)}\footnote{\ Il est à noter que l'item $c$ peut ne pas être ajouté à $\mathcal{GB}d^{-}$ car il ne sera plus utilisé dans la suite. Son ajout a pour seul intérêt que d'avoir une bordure complète.}.
\end{enumerate}

Ensuite, le calcul est effectué par niveau. Pour cela, nous utilisons la procédure \textsc{Gen-GMs-suivants} \textsc{(}lignes 15-16\textsc{)}, dont le pseudo-code est donné par l'algorithme \ref{algoGen-GMs-suivants}. La procédure \textsc{Gen-GMs-suivants} prend en entrée l'ensemble des générateurs minimaux fréquents de taille $k$ et retourne l'ensemble des générateurs minimaux fréquents de taille \textsc{(}$k$ + $1$\textsc{)}.

\incmargin{1em}
\restylealgo{algoruled}
\linesnumbered
\begin{algorithm}[h]\footnotesize{
 \caption{\textsc{Gen-GMs}}}  \label{algoGen-GMs}
   \SetVline
   \setnlskip{-3pt}
\Donnees {- un contexte d'extraction $\mathcal{K}$ = ($\mathcal{O}$, $\mathcal{I}$, $\mathcal{M}$), et le seuil \textit{minsupp}.}

\Res{  \begin{enumerate}
    \item L'ensemble $\mathcal{GMF}_{\mathcal{K}}$ des générateurs minimaux fréquents.
    \item L'ensemble $\mathcal{GB}$d$^{-}$ contenant la bordure négative
non fréquente des générateurs minimaux fréquents.
\item La fermeture de l'ensemble vide.
\end{enumerate}    }

    \Deb {
$\mathcal{CGM}$$_{1}$ = $\mathcal{I}$; // $\mathcal{I}$ est l'ensemble des items\\
\textsc{Calcul-Support}
\textsc{(}$\mathcal{CGM}$$_{1}$\textsc{)}\;
$\emptyset$.\texttt{\textit{support-réel}} = $|$$\mathcal{O}$|\;
$\mathcal{GMF}$$_{0}$ = {\{}$\emptyset${\}}\;

\PourCh{ \textsc{(}$c$ $\in$
$\mathcal{CGM}$$_{1}$\textsc{)}}{\eSi{\textsc{(}$c$.\texttt{\textit{support-réel}}
 = $|$$\mathcal{O}$|\textsc{)}}{$\emptyset$.\texttt{\textit{iff}} =
$\emptyset$.\texttt{\textit{iff}} $\cup$ $\{$$c$$\}$\;}{\eSi{\textsc{(}c.\texttt{\textit{support-réel}} $\geq$
\textit{minsupp}\textsc{)}}{$c$.\texttt{\textit{sous-ens-directs}} = {\{}$\emptyset${\}}\;
$\mathcal{GMF}_{1}$ = $\mathcal{GMF}_{1}$ $\cup$
$\{$$c$$\}$\;}{$\mathcal{GB}d^{-}$ = $\mathcal{GB}d^{-}$ $\cup$ $\{$$c$$\}$\;}}}

\Pour{\textsc{(}$k$ = {\xmplbx1} ; $\mathcal{GMF}$$_{k}$ $ \ne
\emptyset $ ; $k$++\textsc{)}}
{$\mathcal{GMF}$$_{\textsc{(}k+1\textsc{)}}$ = \textsc{Gen-GMs-suivants}\textsc{(}$\mathcal{GMF}$$_{k}$\textsc{)}\;}

$\mathcal{GMF}_\mathcal{K}$ = \ $\cup$
\textsc{\{}$\mathcal{GMF}_{i}$ $|$ $i$ = {\xmplbx0} \ldots $k$
\textsc{\}}\;

}
\end{algorithm}
\decmargin{1em}

La première étape de l'algorithme \textsc{Prince} prend alors fin lorsque l'ensemble des candidats est vide. Cette étape retourne ainsi l'ensemble des générateurs minimaux fréquents $\mathcal{GMF}_{\mathcal{K}}$ trié par ordre décroissant du support, et pour un support donné, par ordre lexicographique \textsc{(}ligne 17\textsc{)}, ainsi que la bordure négative non fréquente $\mathcal{GB}d^{-}$ et la fermeture de l'ensemble vide. Il est important de noter que  la considération de la fermeture de l'ensemble vide dans le cas de l'algorithme, que nous proposons, est réalisée afin d'obtenir un treillis complet des motifs fermés \textsc{(}c'est-à-dire incluant aussi la fermeture de l'ensemble vide\textsc{)}. Toutefois, les arcs de succession immédiate, auxquelles la procédure de construction \textsc{Gen-Ordre} dédiée sera décrite dans la sous-section suivante, peuvent être déterminés sans avoir à insérer au préalable l'ensemble vide dans le treillis.

\incmargin{1em}
\restylealgo{algoruled}
\linesnumbered
\begin{algorithm}[h]\small{
 \caption{\textsc{Gen-GMs-suivants}}}  \label{algoGen-GMs-suivants}
   \SetVline
   \setnlskip{-3pt}
\Donnees{       - $\mathcal{GMF}$$_{k}$.}

\Res{       - $\mathcal{GMF}$$_{\textsc{(}k+1\textsc{)}}$.}

    \Deb {
/* Phase {\xmplbx1} : \textsc{Apriori-Gen} */\\
$\mathcal{P}re\mathcal{CGM}_{\textsc{(}k+1\textsc{)}}$ = \textsc{Apriori-Gen}\textsc{(}$\mathcal{GMF}$$_{k}$\textsc{)}

/* Phase {\xmplbx2} : Vérification de l'idéal d'ordre des
générateurs minimaux fréquents */\\

\PourCh {\textsc{(}$c$ $\in$
$\mathcal{P}re\mathcal{CGM}_{\textsc{(}k+1\textsc{)}}$\textsc{)}}{candidat = \textit{vrai}\;
$c$.\texttt{\textit{support-estimé}} = $|$$\mathcal{O}$$|$; /*
support maximal possible */\\\PourCh {\textsc{(}$c_{1}$ tel que
$\vert c_{1} \vert$ = $k$ et $c_{1}$ $\subset$
$c$\textsc{)}}{ \Si {\textsc{(}$c_1$ $\notin$
$\mathcal{GMF}$$_{k}$\textsc{)}}{candidat = \textit{faux}\; \textbf{sortie}; // arrêt de l'énumération de la ligne 8\\}

$c$.\texttt{\textit{support-estimé}} =
min\textsc{(}$c$.\texttt{\textit{support-estimé}},
$c_{1}$.\texttt{\textit{support-réel}}\textsc{)}\;
$c$.\texttt{\textit{sous-ens-directs}} =
$c$.\texttt{\textit{sous-ens-directs}} $\cup $ $\{$$c_{1}$$\}$\;}\lSi{\textsc{(}candidat = \textit{vrai}\textsc{)}} $\mathcal{CGM}$$_{\textsc{(}k+1\textsc{)}}$
=  $\mathcal{CGM}$$_{\textsc{(}k+1\textsc{)}}$ $\cup$ $\{$$c$$\}$\;}

/* Phase {\xmplbx3} : Calcul des supports des candidats et
élagage des non fréquents */

\textsc{Calcul-Support}
\textsc{(}$\mathcal{CGM}$$_{\textsc{(}k+1\textsc{)}}$\textsc{)}\;

\PourCh {\textsc{(}$c$ $\in$
$\mathcal{CGM}$$_{\textsc{(}k+1\textsc{)}}$\textsc{)}}{\eSi
{\textsc{(}$c$.\texttt{\textit{support-réel}} $\neq$
$c$.\texttt{\textit{support-estimé}} et
c.\texttt{\textit{support-réel}} $\geq$
\textit{minsupp}\textsc{)}}
{$\mathcal{GMF}_{\textsc{(}k+1\textsc{)}}$ = $\mathcal{GMF}_{\textsc{(}k+1\textsc{)}}$
$\cup$ $\{$$c$$\}$\;}{ {\Si{\textsc{(}c.\texttt{\textit{support-réel}}
< \textit{minsupp}\textsc{)}}{$\mathcal{GB}d^{-}$ =
$\mathcal{GB}d^{-}$ $\cup $ $\{$$c$$\}$\;}}}}

\Retour{$\mathcal{GMF}_{\textsc{(}k+1\textsc{)}}$}
   }
\end{algorithm}
\decmargin{1em}

La première phase de la procédure \textsc{Gen-GMs-suivants} consiste à déterminer l'ensemble $\mathcal{P}re\mathcal{CGM}_{\textsc{(}k+1\textsc{)}}$ qui est un sur-ensemble de l'ensemble $\mathcal{CGM}_{\textsc{(}k+1\textsc{)}}$ des candidats de taille \textsc{(}$k$ + $1$\textsc{)} \textsc{(}lignes 2-3\textsc{)}. Chaque élément de l'ensemble $\mathcal{P}re\mathcal{CGM}_{\textsc{(}k+1\textsc{)}}$ est dérivé à partir de deux générateurs minimaux fréquents de taille $k$ ayant \textsc{(}$k$ - $1$\textsc{)} items en commun. Lors de la deuxième phase et pour chaque élément $c$ de $\mathcal{P}re\mathcal{CGM}_{\textsc{(}k+1\textsc{)}}$, nous testons s'il vérifie l'idéal d'ordre des générateurs minimaux fréquents \textsc{(}lignes 4-14\textsc{)}. En même temps, nous calculons le support estimé de $c$ et qui est égal au minimum des supports de ses sous-ensembles de taille $k$ \textsc{(}ligne 12\textsc{)}. Des liens vers ces derniers sont stockés dans le champ \texttt{{\em sous-ens-directs}} et qui seront utilisés dans la seconde étape de l'algorithme \textsc{Prince} \textsc{(}ligne 13\textsc{)}. Si $c$ ne vérifie pas l'idéal d'ordre, alors $c$ est éliminé \textsc{(}lignes 9-11\textsc{)}, sinon il est ajouté à $\mathcal{CGM}_{\textsc{(}k+1\textsc{)}}$ \textsc{(}ligne 14\textsc{)}. Une fois le test de l'idéal d'ordre effectué, nous entamons la troisième phase \textsc{(}lignes 15-22\textsc{)}. Ainsi, un accès au contexte d'extraction permettra de calculer les supports réels des candidats retenus dans $\mathcal{CGM}_{\textsc{(}k+1\textsc{)}}$ \textsc{(}ligne 16\textsc{)}. Une fois cet accès effectué, le support réel de chaque candidat $c$ de $\mathcal{CGM}_{\textsc{(}k+1\textsc{)}}$, est comparé à son support estimé \textsc{(}lignes 17-22\textsc{)}. Si ces derniers sont égaux, alors $c$ n'est pas considéré comme un générateur minimal. Sinon, $c$ est un générateur minimal et la comparaison de son support réel avec \textit{minsupp} permettra de le classer parmi les générateurs minimaux fréquents ou parmi ceux non fréquents \textsc{(}lignes 18-21\textsc{)}. Après l'exécution de ces trois phases, la procédure \textsc{Gen-GMs-suivants} retourne l'ensemble des générateurs minimaux fréquents de taille $\textsc{(}k+1\textsc{)}$ \textsc{(}ligne 23\textsc{)}.

\subsubsection{Structure de données utilisée}

\noindent Du point de vue structure de données, nous avons utilisé un unique arbre lexicographique [Bodon, R\'onyai, 2003 ; Knuth, 1968] pour stocker les générateurs minimaux afin d'accélérer l'extraction des informations qui seront utilisées lors des prochaines étapes. L'ensemble des générateurs minimaux étant un idéal pour l'ordre d'inclusion, le chemin de la racine à un n\oe ud quelconque de l'arbre représente un générateur minimal. Ceci a pour avantage de réduire la nécessité en espace mémoire comparée à l'utilisation d'un arbre lexicographique pour {\em chaque} ensemble de générateurs minimaux de taille $k$ donnée, comme c'est le cas pour les algorithmes \textsc{Close}, \textsc{A-Close} et \textsc{Titanic}. Cette structure de données a été privilégiée dans plusieurs travaux, tels que [Bastide, 2000 ; Goethals, 2004]. Il est aussi à noter qu'afin de rendre efficace le calcul des supports des motifs candidats, le contexte d'extraction a été lui aussi stocké dans une structure dédiée permettant d'optimiser l'espace mémoire nécessaire au stockage des objets du contexte partageant les mêmes items.

\vspace{2mm}

Dans la suite, nous allons noter par \texttt{\em support}, au lieu de \texttt{\em support-réel}, le champ contenant le support réel de chaque générateur minimal étant donné que nous n'avons plus à distinguer le support réel et le support estimé d'un motif.

\subsection{construction du treillis des générateurs minimaux}

\noindent L'objectif de cette étape est d'organiser les générateurs minimaux fréquents sous forme d'un
{\em treillis des générateurs minimaux}. Pour construire le treillis, l'ensemble trié $\mathcal{GMF}_{\mathcal{K}}$ est alors parcouru en introduisant un par un ses éléments dans le {\em treillis des générateurs minimaux} partiellement construit. La couverture supérieure de chaque classe d'équivalence sera alors déterminée au fur et à mesure. Dans la suite de cette sous-section et étant donné que nous allons construire le treillis d'Iceberg en comparant seulement des générateurs minimaux fréquents, chaque classe d'équivalence sera caractérisée par un générateur minimal {\em représentant}. Ce représentant remplace donc le motif fermé associé dans la variante classique du treillis d'Iceberg \textsc{(}cf. la sous-section \ref{sous_section_notion_de_base}, page \pageref{def_treillis_iceberg}\textsc{)}.

\vspace{2mm}

Dans la suite, nous dénotons par $\approx$ la relation d'équivalence entre générateurs minimaux dans le sens que $g \approx g_1$ {\em ssi} $g$ et $g_1$ sont deux générateurs minimaux qui appartiennent à la même classe d'équivalence \textsc{(}ou d'une manière équivalente, admettent la même fermeture de Galois\textsc{)}. Les générateurs $g$ et $g_1$ sont dits {\em équivalents}.

\vspace{2mm}

Soit un générateur minimal fréquent $A$. Nous noterons dans la suite $\rho\textsc{(}A\textsc{)}$ le plus petit élément, au sens de l'ordre lexicographique $\prec$, de la classe d'équivalence de $A$. $\rho\textsc{(}A\textsc{)}$ est égal au générateur minimal fréquent qui sera retenu comme représentant de la classe de $A$ par l'algorithme proposé dans cette étape dédié à la construction du {\em treillis des générateurs minimaux}. \`A cet égard, dans la suite, par le terme {\em successeur immédiat}, nous entendons un {\em générateur minimal fréquent représentant} sauf indication contraire. Le {\em treillis des générateurs minimaux} est ainsi une relation d'ordre sur l'ensemble $\{\rho\textsc{(}g\textsc{)} | g$ est un générateur minimal fréquent\}.

\subsubsection{Détermination des liens de précédence}

\noindent D'une manière générale, afin de déterminer les liens de précédence entre les classes d'équivalence de deux motifs $X$ et $Y$, nous comparons le support de ces motifs avec celui de leur union. \`A cet effet, la représentation concise des motifs fréquents basée sur les générateurs minimaux fréquents et la bordure négative non fréquente, extraite lors de l'étape précédente, sera utilisée pour déterminer le support de l'union moyennant la Proposition \ref{proptitanic} \textsc{(}cf. page \pageref{proptitanic}\textsc{)}. Ceci permet de construire le treillis d'Iceberg {\em sans effectuer un accès supplémentaire au contexte d'extraction}.

\vspace{2mm}

La Proposition \ref{proprelation} regroupe les différents cas possibles résultants de la comparaison des supports de $X$ et de $Y$ et pour lesquels les classes d'équivalence associées sont dites {\em comparables}. Sa preuve utilise le Lemme \ref{lemmeTitanics-support}.

\begin{prop}\label{proprelation} Soient $X$ et $Y$ deux motifs distincts tels que $\textit{Supp}\textsc{(}X\textsc{)} \leq  \textit{Supp}\textsc{(}Y\textsc{)}, [X]$ et $[Y]$ leurs classes d'équivalence respectives.

\vspace{1.5mm}

\noindent 1. $[X] = [Y]$ {\em ssi} $\textit{Supp}\textsc{(}X\textsc{)} = \textit{Supp}\textsc{(}Y\textsc{)} = \textit{Supp}\textsc{(}X \cup Y\textsc{)}$.

\noindent 2. $[X]$ \textsc{(}resp. $[Y]\textsc{)}$ \ est \ un \ successeur \ \textsc{(}resp. prédécesseur\textsc{)} de \ $[Y$] \ \textsc{(}resp. $[X]\textsc{)}$ \ ssi \ $\textit{Supp}\textsc{(}X\textsc{)} \ < \ \textit{Supp}\textsc{(}Y\textsc{)}$ \ et \ $\textit{Supp}\textsc{(}X\textsc{)} = \textit{Supp}\textsc{(}X \cup Y\textsc{)}$.
\end{prop}

\begin{demo}
\mbox{}
\begin{enumerate}
{\em \item \mbox{}

\textsc{(}a\textsc{)} \quad $X \subseteq \textsc{(}X \cup Y\textsc{)} \wedge \textit{Supp}\textsc{(}X\textsc{)} = \textit{Supp}\textsc{(}X \cup Y\textsc{)}  \Rightarrow \gamma\textsc{(}X\textsc{)} = \gamma\textsc{(}X \cup Y\textsc{)}$ \textsc{(}d'après le Lemme \ref{lemmeTitanics-support}\textsc{)}.

\textsc{(}b\textsc{)} \quad $Y \subseteq \textsc{(}X \cup Y\textsc{)} \wedge \textit{Supp}\textsc{(}Y\textsc{)} = \textit{Supp}\textsc{(}X \cup Y\textsc{)} \Rightarrow \gamma\textsc{(}Y\textsc{)} = \gamma \textsc{(}X \cup Y\textsc{)}$ \textsc{(}d'après le Lemme \ref{lemmeTitanics-support}\textsc{)}. D'après \textsc{(}a\textsc{)} et \textsc{(}b\textsc{)}, $\gamma\textsc{(}X\textsc{)} = \gamma\textsc{(}Y\textsc{)}$ et donc $X$ et $Y$ appartiennent à la même classe d'équivalence c'est-à-dire $[X]$ et $[Y]$ sont identiques.}

{\em \item \mbox{}

\textsc{(}a\textsc{)} \quad $X \subseteq \textsc{(}X \cup Y\textsc{)} \wedge \textit{Supp}\textsc{(}X\textsc{)} = \textit{Supp}\textsc{(}X \cup Y\textsc{)} \Rightarrow \gamma\textsc{(}X\textsc{)} = \gamma\textsc{(}X \cup Y\textsc{)}$ \textsc{(}d'après le Lemme \ref{lemmeTitanics-support}\textsc{)}.

\textsc{(}b\textsc{)} \quad $Y \subseteq \textsc{(}X \cup Y\textsc{)} \wedge \textit{Supp}\textsc{(}Y\textsc{)} \ne \textit{Supp}\textsc{(}X \cup Y\textsc{)} \Rightarrow \gamma\textsc{(}Y\textsc{)} \subset\gamma\textsc{(}X \cup Y\textsc{)}$ or, d'après \textsc{(}a\textsc{)}, $\gamma\textsc{(}X\textsc{)} = \gamma\textsc{(}X \cup Y\textsc{)}$ et donc $\gamma\textsc{(}Y\textsc{)} \subset\gamma\textsc{(}X\textsc{)}$ d'où $[X]$ \textsc{(}resp. $[Y]$\textsc{)} est un successeur \textsc{(}resp. prédécesseur\textsc{)} de $[Y]$ \textsc{(}resp. $[X]$\textsc{)}. \quad $\diamondsuit$}
\end{enumerate}
\end{demo}

Dans tous les autres cas, $[X]$ et $[Y]$ sont dits {\em incomparables}. Le lemme suivant présente ces cas.

\begin{lemme}\label{lemme_CEq_incomparables} $[X]$ et $[Y]$ sont incomparables {\em ssi} $\textit{Supp}\textsc{(}X\textsc{)} > \textit{Supp}\textsc{(}X \cup Y\textsc{)}$ et $\textit{Supp}\textsc{(}Y\textsc{)} > \textit{Supp}\textsc{(}X \cup Y\textsc{)}$.
\end{lemme}

\begin{demo}
{\em \mbox{}

\textsc{(}CS\textsc{)} \quad si $\textsc{(}Supp\textsc{(}X\textsc{)} > \textit{Supp}\textsc{(}X \cup Y\textsc{)}$ et $\textit{Supp}\textsc{(}Y\textsc{)} >  \textit{Supp}\textsc{(}X Y\textsc{)}\textsc{)}$ alors $[X]$ et $[Y]$ ne peuvent vérifier aucune des deux clauses précédentes \textsc{(}cf. Proposition \ref{proprelation}\textsc{)} ; ils sont donc incomparables.

\textsc{(}CN\textsc{)} \quad soient $[X]$ et $[Y]$ incomparables. Ils vérifient donc la conjonction \textsc{(}A\textsc{)} ci-dessous des négations des deux premières clauses :

\vspace{1.5mm}

\noindent \textsc{(}A\textsc{)} \quad $\textsc{(}\textsc{(}Supp\textsc{(}X\textsc{)} \neq \textit{Supp}\textsc{(}Y\textsc{)}\textsc{)}$ ou $\textsc{(}Supp\textsc{(}X\textsc{)} \neq \textit{Supp}\textsc{(}X \cup Y\textsc{)}\textsc{)}$ ou $\textsc{(}Supp\textsc{(}Y\textsc{)}\neq \textit{Supp}\textsc{(}X \cup Y\textsc{)}\textsc{)}\textsc{)}$

\hspace{0.5cm} et

\hspace{0.5cm} $\textsc{(}\textsc{(}Supp\textsc{(}X\textsc{)} \geq \textit{Supp}\textsc{(}Y\textsc{)}\textsc{)}$ ou \textsc{(}$\textit{Supp}\textsc{(}X\textsc{)} \neq \textit{Supp}\textsc{(}X \cup Y\textsc{)}\textsc{)}\textsc{)}$

\hspace{0.5cm} et

\hspace{0.5cm} $\textsc{(}\textsc{(}Supp\textsc{(}Y\textsc{)} \geq \textit{Supp}\textsc{(}X\textsc{)}\textsc{)}$ ou  \textsc{(}$\textit{Supp}\textsc{(}Y\textsc{)} \neq \textit{Supp}\textsc{(}X \cup Y\textsc{)}\textsc{)}\textsc{)}$.\\

En remarquant que si $\textit{Supp}\textsc{(}T\textsc{)} \neq \textit{Supp}\textsc{(}T \cup S\textsc{)}$, alors $\textit{Supp}\textsc{(}T\textsc{)} > \textit{Supp}\textsc{(}T \cup S\textsc{)}$, et en
notant $x = \textit{Supp}\textsc{(}X\textsc{)}, y = \textit{Supp}\textsc{(}Y\textsc{)}$ et $u = \textit{Supp}\textsc{(}X \cup Y\textsc{)}$, la formule \textsc{(}A\textsc{)} se reécrit :

\noindent\textsc{(}B\textsc{)} \quad $\textsc{(}\textsc{(}x \neq y\textsc{)}$ ou $\textsc{(}x > u\textsc{)}$ ou $\textsc{(}y > u\textsc{)}\textsc{)}$

\hspace{0.5cm} et

\hspace{0.5cm} $\textsc{(}\textsc{(}x \geq y\textsc{)}$ ou$ \textsc{(}x > u\textsc{)}\textsc{)}$

\hspace{0.5cm} et

\hspace{0.5cm} $\textsc{(}\textsc{(}y \geq x\textsc{)}$ ou $\textsc{(}y > u\textsc{)}\textsc{)}$.\\

On cherche à montrer qu'elle implique $\textsc{(}x > u\textsc{)}$ et $\textsc{(}y > u\textsc{)}$. En développant \textsc{(}B\textsc{)}, on obtient la disjonction de 12 formules :

\begin{description}
  \item[] \textsc{(}1\textsc{)} \quad \ \ $\textsc{(}x \neq y\textsc{)}$ et $\textsc{(}x \geq y\textsc{)}$ et $\textsc{(}y \geq x\textsc{)}$
  \item[] \textsc{(}2\textsc{)} \quad \ \ $\textsc{(}x \neq y\textsc{)}$ et $\textsc{(}x \geq y\textsc{)}$ et $\textsc{(}y > u\textsc{)}$
  \item[] \textsc{(}3\textsc{)} \quad \ \ $\textsc{(}x \neq y\textsc{)}$ et $\textsc{(}x > u\textsc{)}$ et $\textsc{(}y \geq x\textsc{)}$
  \item[] \textsc{(}4\textsc{)} \quad \ \ $\textsc{(}x \neq y\textsc{)}$ et $\textsc{(}x > u\textsc{)}$ et $\textsc{(}y > u\textsc{)}$
  \item[] \textsc{(}5\textsc{)} \quad \ \ $\textsc{(}x > u\textsc{)}$ et $\textsc{(}x \geq y\textsc{)}$ et $\textsc{(}y \geq x\textsc{)}$
  \item[] \textsc{(}6\textsc{)} \quad \ \ $\textsc{(}x > u\textsc{)}$ et $\textsc{(}x \geq y\textsc{)}$ et $\textsc{(}y > u\textsc{)}$
  \item[] \textsc{(}7\textsc{)} \quad \ \ $\textsc{(}x > u\textsc{)}$ et $\textsc{(}x > u\textsc{)}$ et $\textsc{(}y \geq x\textsc{)}$
  \item[] \textsc{(}8\textsc{)} \quad \ \ $\textsc{(}x > u\textsc{)}$ et $\textsc{(}x > u\textsc{)}$ et $\textsc{(}y > u\textsc{)}$
  \item[] \textsc{(}9\textsc{)} \quad \ \ $\textsc{(}y > u\textsc{)}$ et $\textsc{(}x \geq y\textsc{)}$ et $\textsc{(}y \geq x\textsc{)}$
  \item[] \textsc{(}10\textsc{)} \quad $\textsc{(}y > u\textsc{)}$ et $\textsc{(}x \geq y\textsc{)}$ et $\textsc{(}y > u\textsc{)}$
  \item[] \textsc{(}11\textsc{)} \quad $\textsc{(}y > u\textsc{)}$ et $\textsc{(}x > u\textsc{)}$ et $\textsc{(}y \geq x\textsc{)}$
  \item[] \textsc{(}12\textsc{)} \quad $\textsc{(}y > u\textsc{)}$ et $\textsc{(}x > u\textsc{)}$ et $\textsc{(}y > u\textsc{)}$
\end{description}

\vspace{2mm}

Il est simple de montrer que chacune des formules 1 à 12 implique la formule $\textsc{(}\textsc{(}x > u\textsc{)}$ et $\textsc{(}y > u\textsc{)}\textsc{)}$. \quad $\diamondsuit$}
\end{demo}

\'Etant donné le tri imposé dans l'ensemble $\mathcal{GMF}_{\mathcal{K}}$ par rapport au support de ses éléments, la classe d'équivalence de chaque générateur minimal fréquent en cours de traitement ne peut qu'être successeur des classes d'équivalence déjà présentes dans le treillis partiellement construit \textsc{(}cf. Proposition \ref{proprelation}\textsc{)}. Les traitements associés à chaque générateur minimal fréquent sont détaillés dans la sous-section suivante.

\subsubsection{Traitements associés à chaque générateur minimal fréquent}

\noindent Chaque générateur minimal fréquent $g$ de taille $k$ \textsc{(}$k$ $\geq$ $1$\textsc{)} est inséré dans le {\em treillis des générateurs minimaux} en le comparant avec les successeurs immédiats de ses sous-ensembles de taille \textsc{(}$k$ - $1$\textsc{)}. Ceci est basé sur la propriété d'{\em isotonie} de l'opérateur de fermeture [Ganter, Wille, 1999]. En effet, si $g_{1}$ est inclus dans $g$ tel que $|g_{1}|$ = \textsc{(}$k$ - $1$\textsc{)} alors la fermeture de $g_{1}, \gamma\textsc{(}g_{1}\textsc{)}$, est incluse dans la fermeture de $g$, $\gamma\textsc{(}g\textsc{)}$. Ainsi, la classe d'équivalence [$g$] de $g$ est un successeur -- {\em pas forcément immédiat} -- de la classe d'équivalence de $g_{1}, [g_{1}]$.

\vspace{2mm}

En comparant $g$ à la liste des successeurs immédiats de $g_{1}$, disons $L$, deux cas sont à distinguer. Si $L$ est vide, alors $g$ sera simplement ajouté à $L$, sinon, $g$ sera comparé aux éléments appartenant à $L$. Dans ce dernier cas, la Proposition \ref{proprelation} est utilisée, en remplaçant $X$ par $g$ et $Y$ par les éléments de $L$. Soit $g_2$ un des éléments de $L$. Nous distinguons alors deux cas lors du calcul du support de $\mathcal{U} = \textsc{(}g \cup g_2\textsc{)}$ :

\begin{enumerate}
\item Le support de $\mathcal{U}$ est directement dérivé si cet motif fait partie de la représentation extraite lors de la première étape. Le motif $\mathcal{U}$ est alors un générateur minimal. Dans ce cas, [$g$] et [$g_{2}$] sont incomparables étant donné que $\textit{Supp}\textsc{(}\mathcal{U}\textsc{)}$ est nécessairement strictement inférieur à celui de $g$ et celui de $g_{2}$ car sinon il ne serait pas un générateur minimal.
\item Dans le cas où $\mathcal{U}$ ne fait pas partie de la représentation, la Proposition \ref{proptitanic} \textsc{(}cf. page \pageref{proptitanic}\textsc{)} est appliquée. La recherche du support s'arrête alors du moment qu'un générateur minimal inclus dans $\mathcal{U}$ et ayant un support strictement inférieur à celui de $g$ et à celui de $g_{2}$ est trouvé. Dans ce cas, [$g$] et [$g_{2}$] sont incomparables. Si un tel générateur minimal n'a pas été trouvé, nous nous trouvons alors dans un des deux cas explicités par la Proposition \ref{proprelation} et il suffit alors de comparer les supports de $g$, $g_{2}$ et $\mathcal{U}$ pour savoir la relation réelle entre [$g$] et [$g_{2}$].
\end{enumerate}

Dans la suite, nous dénotons par $\tau\textsc{(}g\textsc{)}$ l'ensemble des sous-ensembles immédiats de $g$, c'est-à-dire ceux de taille $| g | - 1$.

\subsubsection{Gestion efficace des classes d'équivalence}

\noindent Lors de ces comparaisons et afin d'éviter une des lacunes des algorithmes adoptant la stratégie {\em Générer-et-tester}, à savoir le calcul redondant des fermetures, \textsc{Prince} utilise des traitements qui se complètent. Ces derniers permettent de maintenir la notion de classe d'équivalence tout au long du traitement. \`{A} cet effet, chaque classe d'équivalence sera caractérisée par un {\em représentant}, qui est le {\em premier} générateur minimal fréquent inséré dans le {\em treillis des générateurs minimaux}. Tout générateur minimal fréquent $g$ est initialement considéré comme représentant de [$g$] et le restera tant qu'il n'est pas comparé à un générateur minimal fréquent précédemment ajouté dans le \textit{treillis des générateurs minimaux} et appartenant à [$g$].

\vspace{2mm}

Chaque générateur minimal $g$, de taille $k$, est comparé avec les listes des successeurs $L_1$, $L_2$, \ldots, et $L_k$ associées respectivement à ses sous-ensembles immédiats $g_1$, $g_2$, \ldots, et $g_k$. Lors de la comparaison d'un générateur minimal fréquent, disons $g$, avec les éléments d'une liste $L$ de successeurs immédiats d'un autre générateur minimal fréquent, des traitements dédiés à la gestion efficace des classes d'équivalence seront réalisés dans le cas où $g$ serait comparé au représentant de sa classe d'équivalence, disons $\mathcal{R}$. Ils sont décrits comme suit :

\begin{enumerate}
\item Toutes les occurrences de $g$ seront remplacées par $\mathcal{R}$ dans les listes des successeurs immédiats où $g$ a été ajouté.

\item Les comparaisons de $g$ avec le reste des éléments de $L$ s'arrêtent car elles ont été effectuées avec $\mathcal{R}$. Ceci permet de n'avoir que des représentants dans les listes des successeurs immédiats et n'affecte en rien le résultat de la deuxième étape, étant donné que $g$ et $\mathcal{R}$ appartiennent à la même classe d'équivalence.
\item Soit $g_i$ le sous-ensemble immédiat de $g$ ayant permis sa comparaison avec $\mathcal{R}$. Le générateur $g$ doit être aussi comparé aux listes des successeurs du reste de ses sous-ensembles immédiats $g_j$ \textsc{(}$i$ $<$ $j$ $\leq$ $k$\textsc{)}. Ces comparaisons seront alors effectuées moyennant $\mathcal{R}$ et non $g$. Le but de poursuivre les comparaisons avec $\mathcal{R}$ est de ne maintenir dans la liste des successeurs immédiats d'une classe d'équivalence donnée que les générateurs minimaux fréquents représentants de leurs classes respectives. Par ailleurs, si $\mathcal{R}$ a été déjà comparé à la liste des successeurs d'un des sous-ensembles immédiats de $g$, la comparaison de $\mathcal{R}$ avec cette liste ne sera pas effectuée. Ceci permet d'éviter de comparer un générateur minimal à une liste des successeurs immédiats plus d'une fois étant donné que les comparaisons qui en découlent ne vont pas donner lieu à de nouveaux arcs de succession immédiate.
\end{enumerate}

Ainsi, pour chaque classe d'équivalence, seul son représentant figure dans les listes des successeurs immédiats. Ceci permet d'optimiser la gestion des classes d'équivalence en minimisant les comparaisons inutiles entre générateurs minimaux fréquents. Par ailleurs, un traitement dédié permet de trouver, pour chaque générateur minimal fréquent $g$, le représentant $\mathcal{R}$ de sa classe d'équivalence. Ceci permettra de compléter la liste des successeurs immédiats de [$g$], égale à [$\mathcal{R}$], et qui est stockée au niveau du représentant $\mathcal{R}$. Ceci permet de n'avoir à gérer qu'une seule liste de successeurs immédiats pour chaque classe d'équivalence.

\vspace{2mm}

Ainsi, lors des comparaisons d'un générateur minimal fréquent $g$ avec les listes des successeurs de ses sous-ensembles immédiats, c'est-à-dire les éléments de $\tau\textsc{(}g\textsc{)}$, deux traitements complémentaires sont réalisés : un premier traitement leur est appliqué tant que $g$ n'est pas comparé au représentant de sa classe d'équivalence. Ensuite, un deuxième traitement leur est appliqué une fois la comparaison effectuée. La définition suivante présente un sous-ensemble des représentants des classes d'équivalence et qui résulte de la comparaison d'un générateur minimal $g$ avec la liste des successeurs de $g_1 \in \tau\textsc{(}g\textsc{)}$.

\begin{definitio} Soit $g$ un générateur minimal de support $n \geq 0$ et $g_1$ un sous-ensemble immédiat de $g$ \textsc{(}$g_1 \in \tau\textsc{(}g\textsc{)}$\textsc{)}. $\phi\textsc{(}n, g, g_1\textsc{)}$ est l'ensemble des représentants des classes $h$ tels que :

\vspace{1.5mm}

soit :
\vspace{0.5mm}

\mbox{}\\\textsc{(}1a\textsc{)} \quad $h$ est de support $n$ et \\\textsc{(}1b\textsc{)} \quad $h$ est un successeur de $\rho\textsc{(}g_1\textsc{)}$ et \\\textsc{(}1c\textsc{)} \quad  $h$ est équivalent à $g$ c'est-à-dire appartiennent à la même classe d'équivalence et\\\textsc{(}1d\textsc{)} \quad $h \prec g$ ;\\

\vspace{1.5mm}

soit :
\vspace{0.5mm}

\mbox{}\\\textsc{(}2a\textsc{)} \quad $h$ est de support $>$ $n$ et \\\textsc{(}2b\textsc{)} \quad $\textsc{(}\textsc{(}h = \rho\textsc{(}g_1\textsc{)}\textsc{)}$ ou $\textsc{(}h$ est un successeur de $\rho\textsc{(}g_1\textsc{)}\textsc{)}$ et\\\textsc{(}2c\textsc{)} \quad $g$ est successeur de $h$ et\\\textsc{(}2d\textsc{)} \quad pour tout représentant $h'$ de support $>$ $n$ tel que $h'$ est successeur de $h, g$ n'est pas successeur de $h'$.
\end{definitio}

Il est important de noter que l'ensemble $\phi\textsc{(}n, g, g_1\textsc{)}$ s'il contient un élément de support $n$, il n'en contient qu'un à savoir le représentant de la classe de $g$. Si $g$ est le représentant de sa classe, alors $\phi\textsc{(}n, g, g_1\textsc{)}$ ne contient aucun élément de support $n$, d'après \textsc{(}1c\textsc{)}, puisque il est faux que $g \prec g$. Tout élément $h$ de $\phi\textsc{(}n, g, g_1\textsc{)}$, autre que le représentant de la classe de $g$ s'il existe dans cet ensemble, correspond à un prédécesseurs immédiat de $g$ résultant de la comparaison de $g$ avec la liste des successeurs de $g_1$ \textsc{(}ou inversement, $g$ est un successeur immédiat de $h$\textsc{)}. Soit $\xi_{g_1} = \phi\textsc{(}n, g, g_1\textsc{)} \backslash \rho\textsc{(}g\textsc{)}$. Il est à noter que $\xi_{g_1}$ n'est jamais vide puisque $h$ peut être confondu avec $\rho\textsc{(}g_1\textsc{)}$ \textsc{(}condition \textsc{(}2b\textsc{)}\textsc{)} et que $g$, sur-ensemble de $g_1$, est successeur de $\rho\textsc{(}g_1\textsc{)}$.

\vspace{2mm}

Les deux lemmes suivants montrent la relation entre $g$ et les éléments de $\xi_{{g}_{1}}$ pour tout $g_1 \in \tau\textsc{(}g\textsc{)}$.

\begin{lemme} \textsc{(}{\em de validité}\textsc{)}. Soit $\xi = \textsc{(}\cup_{\textsc{(}g_1 \in \tau\textsc{(}g\textsc{)}\textsc{)}} \phi\textsc{(}n, g, g_1\textsc{)}\textsc{)} \backslash \rho\textsc{(}g\textsc{)}$. Tous les couples du produit $\xi \times \{g\}$ sont des arcs de succession immédiate, autrement dit, pour tout élément $h$ de $\xi$, $g$ est successeur immédiat de $h$.
\end{lemme}

\begin{lemme} \textsc{(}{\em de complétude}\textsc{)}. L'ensemble des arcs de succession immédiate qui \guillemotleft~aboutissent \guillemotright \ sur $g$ est égal au produit $\xi \times \{g\}$.
\end{lemme}

Rappelons maintenant que chaque classe d'équivalence est représentée dans le {\em treillis des générateurs minimaux} à travers son représentant. Ce dernier représente donc l'ensemble des générateurs minimaux fréquents de sa classe dans le treillis et seul lui, parmi les générateurs de sa classe, figure dans les listes de succession immédiate. \`A cet égard, afin que la construction du treillis soit valide, deux conditions doivent être vérifiées. La première consiste dans le fait que chaque classe d'équivalence doit être connectée à sa couverture inférieure une fois ses générateurs minimaux associés traités. La deuxième condition impose que tous les générateurs minimaux fréquents appartenant à la même classe d'équivalence doivent y être inclus une fois leur traitement effectué. Le lemme suivant traite de la première condition et celui qui le suit de la seconde.

\vspace{2mm}

Ainsi, le Lemme \ref{completude_lien_precedence_entre_CEq} montre que tout lien de précédence entre deux classes d'équivalence tel que l'une est successeur immédiat de l'autre sera construit une fois les générateurs minimaux associés introduits dans le {\em treillis des générateurs minimaux}.

\begin{lemme}\label{completude_lien_precedence_entre_CEq} Soient $\mathcal{R}$ et $\mathcal{R}'$ les représentants de leur classe d'équivalence. Si $\mathcal{R}$ est successeur immédiat de $\mathcal{R}'$, alors il existe un générateur minimal $g \approx \mathcal{R}$, et $h$ dans $\tau\textsc{(}g\textsc{)}$ tels que $\textsc{(}\textsc{(}h \approx \mathcal{R}'\textsc{)}$ ou $\textsc{(}\mathcal{R}'$ est un successeur de $\rho\textsc{(}h\textsc{)}\textsc{)}\textsc{)}$.
\end{lemme}

\begin{demo}
{\em Soit $w = \mathcal{R}' \cup \{i\}$ avec $i \in \textsc{(}\mathcal{R} \backslash \gamma \textsc{(}\mathcal{R}'\textsc{)}\textsc{)}$. L'item $i$ existe nécessairement car sinon $\mathcal{R}$ serait inclus dans la fermeture de $\mathcal{R}'$ ce qui est en contradiction avec le fait que $\mathcal{R}$ est un des générateurs minimaux d'une classe d'équivalence successeur immédiat à celle de $\mathcal{R}'$.

\vspace{1mm}

\'Etant donné que $\mathcal{R}$ est un successeur immédiat de $\mathcal{R}', \gamma\textsc{(}\mathcal{R}'\textsc{)} \subset \gamma\textsc{(}w\textsc{)} = \gamma\textsc{(}\mathcal{R}' \cup \{i\}\textsc{)} = \gamma\textsc{(}\gamma\textsc{(}\mathcal{R}'\textsc{)} \cup \{i\}\textsc{)} \subseteq\gamma\textsc{(}\mathcal{R}\textsc{)}$. Il en découle que $\gamma\textsc{(}w\textsc{)} = \gamma\textsc{(}\mathcal{R}\textsc{)}$.

\vspace{2mm}

Soit $g \subseteq w$ un générateur minimal inclus dans $w$ et ayant la même fermeture que $w : \gamma\textsc{(}g\textsc{)} = \gamma\textsc{(}w\textsc{)} = \gamma\textsc{(}\mathcal{R}\textsc{)}$. $g$ contient forcément l'item $i$ car sinon ce serait en contradiction avec le fait que $\mathcal{R}'$ soit un générateur minimal d'un fermé strictement inclus dans $\gamma\textsc{(}\mathcal{R}\textsc{)}$.

\vspace{2mm}

Soit $h$ le sous-ensemble immédiat de $g$ égal à \textsc{(}$g \backslash \{i\}$\textsc{)}. $h$ est un générateur minimal puisque $g$ l'est aussi. Par ailleurs, $h \subseteq \mathcal{R}'$.

\vspace{1mm}

Deux cas sont alors à distinguer :

\noindent -- si $h = \mathcal{R}'$, alors $\mathcal{R}'$ est égal à un des sous-ensembles immédiats de $g$, à savoir $h$. Il est ainsi évident que $h \approx \mathcal{R}'$.

\noindent -- si $h \subset \mathcal{R}'$, alors $\mathcal{R}'$ est un des successeurs de $h$. Par conséquent, $\mathcal{R}'$ est aussi un successeur de $\rho\textsc{(}h\textsc{)}$ puisque $h \approx \rho\textsc{(}h\textsc{)}$.

\vspace{1mm}

Par conséquent, le lien de précédence entre la classe d'équivalence de $\mathcal{R}'$ et celle de $g$ \textsc{(}elle-même égale à la classe d'équivalence de $\mathcal{R}$ puisque $g \approx \mathcal{R}\textsc{)}$ est construit, en comparant $g$ aux successeurs de son sous-ensemble immédiat $h$. \quad $\diamondsuit$}
\end{demo}

Le lemme suivant prouve que si un générateur minimal fréquent $g$ n'est pas le représentant de sa classe d'équivalence c'est-à-dire $g \neq \rho\textsc{(}g\textsc{)}$, ou d'une manière équivalente $\rho\textsc{(}g\textsc{)} \prec g\textsc{)}$, alors $g$ sera comparé au représentant de sa classe d'équivalence lors de ses comparaisons avec les successeurs de ses sous-ensembles immédiats. Il est à noter que si $g$ est l'unique générateur minimal de sa classe d'équivalence ou est son représentant, le problème ne se pose pas car il sera le premier générateur minimal de la classe à être traité. Ce lemme permet de montrer donc qu'une fois l'ensemble des générateurs minimaux de même support insérés dans le treillis, chaque classe d'équivalence associée contient tous ses générateurs minimaux.

\begin{lemme}\label{lemme_comparaison_entre_GMFs_meme_CEq} Si $g$ est un générateur minimal fréquent tel que $\rho\textsc{(}g\textsc{)} \prec g$, alors il existe un générateur minimal $X$ équivalent à $g$ \textsc{(}$X$ $\approx$ $g$\textsc{)} et $X \prec g$ pour lequel il existe $Y \in \tau\textsc{(}X\textsc{)}$ tel que :

\begin{itemize}
\item \textsc{(}$\rho\textsc{(}g\textsc{)}$ est un successeur de $\rho\textsc{(}Y\textsc{)}$\textsc{)}\\
et
\item \textsc{(}$\rho\textsc{(}Y\textsc{)}$ est un successeur de $\rho\textsc{(}h\textsc{)}$\textsc{)} si $h$ désigne le plus petit des éléments de $\tau\textsc{(}g\textsc{)}$ au sens de l'ordre $\prec$.
\end{itemize}
\end{lemme}

\begin{demo} {\em La condition $\rho\textsc{(}g\textsc{)} \prec g$ du lemme impose à la classe de $g$ d'avoir au moins deux éléments. Soit $g$ un générateur minimal fréquent dont la classe d'équivalence à au moins deux éléments. Elle ne peut donc pas être la classe singleton contenant le seul générateur minimal fréquent vide. On pose $z = \rho\textsc{(}g\textsc{)}$ le représentant de $g$. On a $z \prec g$. Puisque $g$ et $z$ sont minimaux et différents, on a : \textsc{(}1\textsc{)} \  $\textsc{(}z \not\subset g\textsc{)}$ et $\textsc{(}g \not\subset z\textsc{)}$. Soient $\mu\textsc{(}z\textsc{)}$ et $\mu\textsc{(}g\textsc{)}$ les mots associés à $z$ et à $g$ selon l'ordre $\prec \footnote{\ Si $F$ est un ensemble fini muni d'un ordre total $\prec$, le mot $\mu\textsc{(}G\textsc{)}$ associé à tout sous-ensemble $G$ de $F$ est l'unique bijection de l'intervalle $[1, | G |]$ dans $G$ qui respecte l'ordre $\prec$. Les mots associés permettent d'ordonner lexicographiquement les parties de $F$.} : \mu\textsc{(}z\textsc{)} = z_1 z_2 \ldots z_p, p > 0$, et $\mu\textsc{(}g\textsc{)} = g_1 g_2 \ldots g_q, q > 0$. D'après la propriété \textsc{(}1\textsc{)} ci-dessus, il existe un unique indice $i, 1 \leq i \leq min\textsc{(}p, q\textsc{)}$, tel que $z_i \prec g_i$ et pour tout $k, 1 \leq k < i, z_k = g_k$. On pose $u = z_i \prec g_i$. On a $u \notin g$, sinon on aurait $u \prec g_i \prec u$. On considère le plus petit élément $h$ de $\tau\textsc{(}g\textsc{)}$ au sens de l'ordre $\prec$. Il est obtenu en enlevant à $g$ son plus grand élément au sens de l'ordre $\prec$.
On a donc $\mu\textsc{(}h\textsc{)} = g_1 g_2 \ldots g_{q-1}$. L'ensemble $h$, qui peut être vide, est un générateur minimal fréquent puisqu'il est inclus dans le générateur minimal fréquent $g$ et son support est strictement supérieur à $n$.

\vspace{2mm}

On a $\gamma\textsc{(}h \cup z\textsc{)} = \gamma\textsc{(}h \cup \gamma\textsc{(}z\textsc{)}\textsc{)}$ \ \textsc{(}propriété des chemins indépendants\textsc{)}

\hspace*{2.5cm}$= \gamma\textsc{(}h \cup \gamma\textsc{(}g\textsc{)}\textsc{)}$ \ \textsc{(}$z$ et $g$ sont équivalents\textsc{)}

\hspace*{2.5cm}$= \gamma\textsc{(}\gamma\textsc{(}g\textsc{)}\textsc{)}\textsc{(}h \subset g \subseteq \gamma\textsc{(}g\textsc{)}\textsc{)}$

\hspace*{2.5cm}$= \gamma\textsc{(}g\textsc{)}$ \textsc{(}idempotence de $\gamma\textsc{)} = \gamma\textsc{(}z\textsc{)}$.

\vspace{2mm}

Puisque $u \in z, h \cup z = h \cup \{u\} \cup\textsc{(}z \backslash \{u\}\textsc{)}$. L'ensemble $h \cup \{u\} \cup\textsc{(}z\backslash \{u\}\textsc{)}$ est équivalent à $z$ mais n'est pas un générateur minimal fréquent puisqu'il contient le générateur minimal fréquent $z$. Il existe donc un sous-ensemble $z'$ de $z \backslash \{u\}$ tel que l'ensemble $X = h \cup \{u\} \cup z'$ \textsc{(}qui est contenu dans $h \cup z$\textsc{)} est un générateur minimal fréquent de support $n$, donc équivalent à $z$. Il est à noter que l'ensemble $z'$ est vide lorsque $X = h \cup \{u\}$ est de support $n$. Soit $X_1 X_2 \ldots X_r, r >$ 0, le mot associé au générateur minimal fréquent $X$. Par construction, on a, pour tout $k, 1 \leq k < i, X_k = g_k = z_k$ et $X_i = u$. Il en résulte que $X \prec g$. L'ensemble $Y = h \cup z'$ est un élément de $\tau\textsc{(}X\textsc{)}$. $Y$ est donc un générateur minimal fréquent de support $> n$. Il contient le générateur minimal fréquent $h$, donc $\rho\textsc{(}Y\textsc{)}$ est un successeur de $\rho\textsc{(}h\textsc{)}$. Par ailleurs, le générateur minimal fréquent $X$ contenant le générateur minimal fréquent $Y, z = \rho\textsc{(}g\textsc{)} = \rho\textsc{(}X\textsc{)}$ est un successeur de $\rho\textsc{(}Y\textsc{)}$. \quad $\diamondsuit$}
\end{demo}

\subsubsection{Pseudo-code de la deuxième étape de l'algorithme \textsc{Prince}}

\noindent Le pseudo-code de la deuxième étape de l'algorithme \textsc{Prince} est donné par la procédure \textsc{Gen-Ordre} \textsc{(}cf. Algorithme \ref{algogenordre}, page \pageref{algogenordre}\textsc{)}. Dans ce pseudo-code, {\em gmf} est l'abréviation de générateur minimal fréquent. \`{A} la fin de l'exécution de la procédure \textsc{Gen-Ordre}, le {\em treillis des générateurs minimaux} est construit et est égal à la relation d'ordre sur l'ensemble des représentants des classes d'équivalence, c'est-à-dire $\{\rho\textsc{(}g\textsc{)} | g\in \mathcal{GMF}_{\mathcal{K}}\}$. Le champ \texttt{\textit{succs-immédiats}} associé à un générateur minimal fréquent $g$ sera alors vide si ce dernier n'est pas le représentant de sa classe d'équivalence ou si $g$ appartient à une classe d'équivalence n'ayant pas de successeurs. Sinon, cette liste ne contiendra que des représentants. La procédure \textsc{Gen-Ordre} insère un générateur minimal fréquent $g$ dans le {\em treillis des générateurs minimaux} en le comparant aux listes des successeurs immédiats de ses sous-ensembles de taille \textsc{(}$k$ - $1$\textsc{)}. Cette procédure implante alors la coupure en deux parties complémentaires du traitement de l'énumération de $\tau\textsc{(}g\textsc{)}$, c'est-à-dire avant que $g$ ne soit comparé au représentant de sa classe et après.

\vspace{2mm}

Les notations utilisées dans le pseudo-code de cette procédure sont comme suit sachant que le symbole $\$$ permet de distinguer les objets de programmation des objets mathématiques auxquels ils sont liés.

\newpage

\incmargin{1em}
\restylealgo{algoruled}
\linesnumbered
\begin{algorithm}[!t]\small{
 \caption{\textsc{Gen-Ordre}}  \label{algogenordre}
   \SetVline
   \setnlskip{-3pt}
   \Donnees{         - L'ensemble $\mathcal{GMF}$$_{\mathcal{K}}$ des générateurs minimaux fréquents.}
\Res{       -  Le \textit{treillis des générateurs minimaux}.}}
 \end{algorithm}
\decmargin{1em}
\begin{figure}[!t]
\parbox{17.cm}{
\vspace{-0.54cm}\hspace{0.12cm}
\parbox{6cm}{
\includegraphics[scale = 0.94]{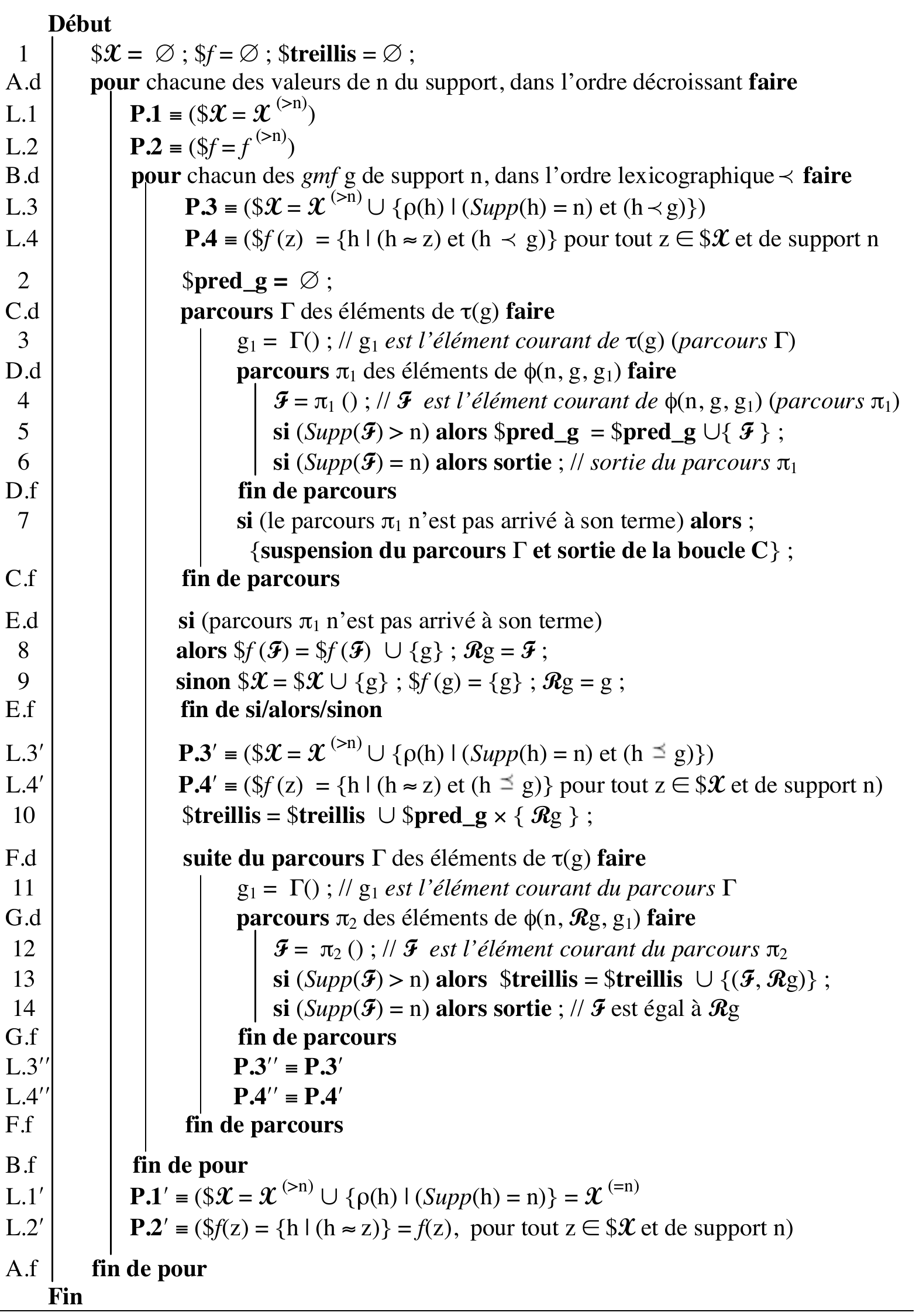}
}
}
\end{figure}

\noindent -- La primitive {\sc parcours} permet de parcourir {\em tout ou une partie} des éléments d'un ensemble.\\
\noindent -- $\mathcal{X}$ : est l'ensemble des représentants des classes.\\
\noindent -- $f$ : est une application permettant d'associer à chaque générateur minimal son représentant, c'est-à-dire l'élément de $\mathcal{X}$ qui lui est équivalent par $\approx$.\\
\noindent -- $\textsc{(}\mathcal{X}, f\textsc{)}$ : est égal à la forme fonctionnelle de la partition des générateurs minimaux fréquents par la relation d'équivalence $\approx$.\\
\noindent -- $\textsc{(}\mathcal{X}^{\textsc{(}>n\textsc{)}}$, $f^{\textsc{(}>n\textsc{)}}\textsc{)}$ : est égal à la forme fonctionnelle de la partition des générateurs minimaux fréquents de support $>$ n par $\approx$.\\
\noindent -- $\textsc{(}\mathcal{X}^{\textsc{(}\geq n\textsc{)}}$, $f^{\textsc{(}\geq n\textsc{)}}\textsc{)}$ : stocke la forme fonctionnelle de la partition des générateurs minimaux fréquents de support $\geq n$ par $\approx$.\\
\noindent -- $treillis$ : est égal au treillis des générateurs minimaux fréquents, c'est-à-dire, le treillis sur l'ensemble $\mathcal{X}$ des représentants.\\
\noindent -- $pred$\_$g$ : permet de stocker les prédécesseurs immédiats de $g$.\\
\noindent -- $g_1$ : permet de stocker l'élément courant de $\tau\textsc{(}g\textsc{)}$.\\
\noindent -- $\mathcal{R}_g$ : permet de stocker le représentant de la classe de $g$.\\
\noindent -- $\mathcal{F}$ : permet de stocker l'élément courant de $\phi\textsc{(}n, g, g_1\textsc{)}$, puis de $\phi\textsc{(}n, \mathcal{R}_g, g_1\textsc{)}$.

\vspace{2mm}

\bigskip

\noindent Dans ce qui suit, nous décrivons en détail le déroulement de \textsc{Gen-Ordre} :

\begin{enumerate}

\item Dans la colonne de gauche, \textsc{A}.d signifie début du bloc \textsc{A} et \textsc{A}.f signifie fin du bloc \textsc{A}. Il en est de même pour les autres lettres.

 \item La procédure \textsc{Gen-Ordre} traite l'ensemble des générateurs minimaux fréquents par support décroissant \textsc{(}cf. bloc \textsc{A}\textsc{)} et pour un ensemble de générateurs minimaux de même support, elle les traite par ordre lexicographique $\prec$ \textsc{(}cf. bloc \textsc{B}\textsc{)}. Pour chaque générateur minimal $g$, les traitements qui lui sont associés sont divisés en deux parties complémentaires. La première s'intéresse aux liens de succession immédiate obtenus avant la comparaison de $g$ avec le représentant de sa classe $\mathcal{R}_g$ \textsc{(}cf. bloc \textsc{C}\textsc{)}. La seconde parcours le reste des éléments de $\tau\textsc{(}g\textsc{)}$ en utilisant $\mathcal{R}_g$ pour faire les comparaisons avec les listes de successeurs.

 \item La procédure \textsc{Gen-Ordre} a pour effet de calculer les variables \$$\mathcal{X}$, \$$f$ et {\em \$treillis}. Elles sont initialisées à vide \textsc{(}ligne 1\textsc{)}. Chaque itération de la boucle \textsc{A} aura pour effet de prendre en compte les générateurs minimaux fréquent de support égal à $n$ et de calculer les nouvelles valeurs des variables \$$\mathcal{X}$, \$$f$ et {\$\em treillis}. Les supports sont énumérés dans l'ordre décroissant.

\item L'intérieur de l'énumération \textsc{B}, est en trois parties :

 \begin{itemize}
   \item ligne 2 + bloc \textsc{C} : début du parcours des éléments de $\tau\textsc{(}g\textsc{)}$ et actions associées. Le parcours s'arrête soit parce qu'il est arrivé à son terme, soit parce qu'il a rencontré le représentant de la classe de $g$ \textsc{(}ligne 6\textsc{)} ;
   \item bloc \textsc{E} + ligne 10 : actions conclusives de la partie précédente ;
   \item bloc \textsc{F} : suite et fin du parcours des éléments de $\tau\textsc{(}g\textsc{)}$ et actions associées.
  \end{itemize}

\item L'énumération des éléments $g_1$ de $\tau\textsc{(}g\textsc{)}$ est en deux parties distinguées : bloc \textsc{C} et bloc \textsc{F}. Dans le bloc \textsc{C}, l'algorithme ne connaît pas le représentant de $g$ et travaille avec l'ensemble $\phi\textsc{(}n, g, g_\textsc{)}$ tandis que, dans le bloc \textsc{F}, il connaît le représentant $\mathcal{R}_g$ de $g$ et travaille avec l'ensemble $\phi\textsc{(}n, \mathcal{R}_g, g_1\textsc{)}$.

\item \`A l'intérieur du bloc \textsc{C}, le bloc \textsc{D} parcourt les éléments $\mathcal{F}$ de $\phi\textsc{(}n, g, g_1$\textsc{)} \textsc{(}bloc \textsc{D}\textsc{)}. Si $\mathcal{F}$ est de support $> n$, il s'agit d'un prédécesseur immédiat de $g$. Il est provisoirement stocké comme tel dans la variable {\$\em pred\_g} \textsc{(}ligne 5\textsc{)}. Si $\mathcal{F}$ est de support = $n$, il s'agit du représentant de $g$. Dès lors le parcours \textsc{D} s'arrête avant son terme \textsc{(}ligne 6\textsc{)} et le parcours \textsc{C} sera alors suspendu \textsc{(}ligne 7\textsc{)} et reprendra plus tard \textsc{(}bloc \textsc{F}\textsc{)}.

\item bloc conclusif : le parcours \textsc{C} s'arrête avant son terme lorsque l'ensemble $\phi\textsc{(}n, g, g_1\textsc{)}$ contient le représentant de $g$. La variable $\mathcal{F}$ est alors égale à ce représentant. L'ensemble \$$f \textsc{(}\mathcal{F}\textsc{)}$ des éléments équivalents à $\mathcal{F}$ est donc augmenté de $g$ \textsc{(}ligne 8\textsc{)}. Dans le cas où le parcours \textsc{C} s'arrête à son terme, $g$ n'a pas de représentant : il devient le représentant de sa classe et est donc ajouté comme tel à l'ensemble \$$\mathcal{X}$ des représentants \textsc{(}ligne 9\textsc{)}. Enfin, les affectations $\mathcal{R}_g = \mathcal{F}$ à la fin de la ligne 8 et $\mathcal{R}_g = g$ à la fin de la ligne 9, font qu'en sortie du bloc \textsc{E}, avant la ligne 10, la variable $R_g$ est égale au représentant de $g$. La ligne 10 met alors à jour la variable {\$\em treillis}.

\item le bloc \textsc{F} \textsc{(}partie 3\textsc{)} réalise ce qui reste à réaliser du parcours des éléments de $\tau\textsc{(}g\textsc{)}$. \`A l'intérieur du bloc \textsc{F}, le bloc $G$ parcourt les éléments de $\phi\textsc{(}n, \mathcal{R}_g, g_1\textsc{)}$. Si le support de l'élément courant $\mathcal{F}$ est $> n$, il s'agit d'un prédécesseur immédiat de $g$. La variable {\$\em treillis} est alors mise à jour \textsc{(}ligne 13\textsc{)}. Si son support est égal à $n$, alors $\mathcal{F} = \mathcal{R}_g$. La ligne 14 arrête le parcours \textsc{G} puisque la comparaison de $\mathcal{R}_g$ avec la liste des successeurs a été déjà effectuée et ne va donc pas donner lieu à de nouveaux liens de succession immédiate.
\end{enumerate}

\bigskip

Il est important de noter le lien entre les objets mathématiques $\phi\textsc{(}n, g, g_1\textsc{)}$ et $\phi\textsc{(}n, \mathcal{R}_g, g_1\textsc{)}$ et les objets de programmation \$$\phi\textsc{(}n, g, g_1\textsc{)}$ et \$$\phi\textsc{(}n, \mathcal{R}_g, g_1\textsc{)}$, qui leurs sont respectivement associés. En effet, les ensembles $\phi\textsc{(}n, g, g_1\textsc{)}$ et $\phi\textsc{(}n, \mathcal{R}_g, g_1\textsc{)}$ dépendent du treillis tout entier. Or, de ce treillis, la procédure \textsc{Gen-Ordre} n'a que la partie stockée par la variable {\$\em treillis}. L'idée qui sous-tend cette procédure est que la partie du treillis stockée par la variable {\$\em treillis} suffit au calcul des ensembles $\phi$. \`A cet égard, nous introduisons les ensembles \$$\phi\textsc{(}n, g, g_1$\textsc{)}. Leur définition est calquée sur celle des ensembles $\phi\textsc{(}n, g, g_1\textsc{)}$. La différence est qu'elle utilise l'ordre stocké dans {\$\em treillis}. Dans la définition ci-dessous, \guillemotleft~successeur \guillemotright \ signifie \guillemotleft~successeur au sens de la Proposition \ref{proprelation} \guillemotright, tandis que \guillemotleft~\$successeur \guillemotright \ signifie \guillemotleft~successeur selon l'ordre stocké dans la variable \$treillis \guillemotright. La définition de \$$\phi\textsc{(}n, g, g_1\textsc{)}$ est ainsi comme suit :

\begin{definitio} Soient $g$ un générateur minimal de support $n$ et $g_1$ un sous-ensemble de $g$ de cardinal $| g | - 1 \ \textsc{(}g_1 \in \tau\textsc{(}g\textsc{)}\textsc{)}$. Soit \$$\phi\textsc{(}n, g, g_1\textsc{)}$ l'ensemble des $h$ de \$$\mathcal{X}$ tels que :
\vspace{1.5mm}

soit :
\vspace{0.5mm}

\mbox{}\\\textsc{(}1a\textsc{)} \quad $h$ est de support $n$ et\\\textsc{(}1b\textsc{)} \quad $h$ est un \$successeur de $\rho\textsc{(}g_1\textsc{)}$ et\\\textsc{(}1c\textsc{)} \quad $h$ est équivalent à $g$ \textsc{(}selon la Proposition \ref{proprelation}\textsc{)} et\\\textsc{(}1d\textsc{)} \quad $h \prec g$ ;\\

\vspace{1.5mm}

soit :
\vspace{0.5mm}

\mbox{}\\\textsc{(}2a\textsc{)} \quad $h$ est de support $> n$ et\\\textsc{(}2b\textsc{)} \quad $\textsc{(}\textsc{(}h = \rho\textsc{(}g_1\textsc{)}\textsc{)}$ ou $\textsc{(}h$ est un \$successeur de $\rho\textsc{(}g_1$\textsc{)}\textsc{)}\textsc{)} et\\\textsc{(}2c\textsc{)} \quad $g$ est successeur de $h$ \textsc{(}selon la Proposition \ref{proprelation}\textsc{)} et\\\textsc{(}2d\textsc{)} \quad pour tout représentant $h'$ de support $> n$ tel que $h'$ est \$successeur de $h, g$ n'est pas successeur de $h'$ \textsc{(}selon la Proposition \ref{proprelation}\textsc{)}.
\end{definitio}

Il en résulte qu'à la ligne \textsc{D}.d de la procédure \textsc{Gen-Ordre}, $\phi\textsc{(}n, g, g_1\textsc{)}= \$\phi\textsc{(}n, g, g_1\textsc{)}$. Par ailleurs, à la ligne \textsc{G}.d de la procédure \textsc{Gen-Ordre}, $\phi\textsc{(}n, \mathcal{R}_g, g_1\textsc{)} = \$\phi\textsc{(}n, \mathcal{R}_g, g_1\textsc{)}$. On peut donc remplacer dans le pseudo-code, l'occurrence de $\phi\textsc{(}n, g, g_1\textsc{)}$ par \$$\phi\textsc{(}n, g, g_1\textsc{)}$ et celle $\phi\textsc{(}n, \mathcal{R}_g, g_1\textsc{)}$ par \$$\phi\textsc{(}n, \mathcal{R}_g, g_1\textsc{)}$.

\vspace{2mm}

\begin{rem}
\mbox{}

{\em -- Si nous avions opté pour {\em n'importe quel autre ordre} dans le tri de $\mathcal{GMF}_{\mathcal{K}}$ \textsc{(}par exemple, tri par ordre {\em croissant} par rapport aux supports\textsc{)} et si la classe d'équivalence de $g$ est incomparable avec celles des éléments de $L$, les comparaisons de $g$ avec les listes des successeurs immédiats des éléments de $L$ seraient dans ce cas obligatoires. En effet, deux classes d'équivalence incomparables \textsc{(}celle de $g$ et celle d'un représentant appartenant à $L$\textsc{)} peuvent avoir des successeurs en commun, existants déjà dans le \textit{treillis des générateurs minimaux}. Ainsi, tout autre choix de tri augmenterait considérablement le nombre, et par conséquent le coût, des comparaisons pour construire le {\em treillis des générateurs minimaux}.

-- Lors de la première étape, \textsc{Prince} adopte l'optimisation introduite par \textsc{A-Close} pour détecter le niveau à partir duquel les générateurs minimaux fréquents ne sont plus forcément des fermés. Ceci permettra, lors de la seconde étape, d'optimiser encore plus la construction de la relation d'ordre : une partie \textsc{(}ou quasiment la totalité\textsc{)} du treillis pouvant être déjà construite dès la première étape. En effet, lorsque le motif fermé fréquent se confond avec son générateur c'est-à-dire la classe d'équivalence associée ne contient qu'un seul motif\textsc{)}, ses prédécesseurs immédiats seront ses sous-ensembles immédiats. Les liens vers ces derniers étant stockés dans le champ \texttt{\textit{sous-ens-directs}}, nous avons ainsi les prémisses nécessaires pour extraire les règles approximatives valides. Par ailleurs, aucune règle exacte ne peut être extraite dans ce cas.}
\end{rem}

\vspace{2mm}

Nous allons maintenant décrire les invariants de la procédure \textsc{Gen-Ordre}. Rappelons qu'un invariant d'un algorithme est un couple \textsc{(}$P, L$\textsc{)} composé d'une propriété $P$ liant certaines variables de l'algorithme et d'un endroit, un lieu, $L$ dans ledit algorithme tel que la propriété $P$ est vraie chaque fois que l'algorithme passe à l'endroit $L$. Dans ce sens, l'invariant au début de la boucle \textsc{B}.d est donné par le lemme suivant.

\begin{lemme} Soit $\textsc{max}\textsc{(}\mathcal{R}_1, \mathcal{R}_2$\textsc{)} l'ensemble des prédécesseurs immédiats de $\mathcal{R}_2$ qui sont soit des successeurs de $\mathcal{R}_1$, soit égaux à $\mathcal{R}_1$, avec $\mathcal{R}_1$ et $\mathcal{R}_2$ deux représentants de leur classe d'équivalence $\textsc{(}\mathcal{R}_1, \mathcal{R}_2 \in \mathcal{X}$\textsc{)}. Soit $\textit{Supp}^{\textsc{(}=n\textsc{)}}$ l'ensemble des générateurs minimaux de support = $n$ et {\em treillis}$^{\textsc{(}>n\textsc{)}}$ la restriction du treillis {\em treillis} à l'ensemble $\mathcal{X}^{\textsc{(}>n\textsc{)}}$ des représentants de support $> n$.\\L'égalité suivante : ${\$treillis} = {treillis}^{\textsc{(}>n\textsc{)}} \cup \textsc{(}\cup_{\textsc{(}g_1 \ \in \ \textit{Supp}^{\textsc{(}=n\textsc{)}} \ et \ g_1 \ \prec \ g\textsc{)}} \textsc{(}\cup_{\textsc{(}h \ \in \ \tau\textsc{(}g_1\textsc{)}\textsc{)}}$ $\textsc{max}\textsc{(}$ $\rho\textsc{(}h\textsc{)},\mathcal{R}\textsc{)} \times \mathcal{R}\textsc{)}\textsc{)}$ avec $\mathcal{R}$ = $\rho$\textsc{(}$g_1$\textsc{)} est un invariant au début de la boucle \textsc{B} de la procédure \textsc{Gen-Ordre}.
\end{lemme}

Les invariants qui régissent la relation d'équivalence entre générateurs minimaux fréquents, à savoir l'évolution des variables \$$\mathcal{X}$ et \$$f$ sont indiqués dans le pseudo-code sous forme de couples \textsc{(}$L.i$, $P.i$\textsc{)}, dans lesquels $L.i$ identifie un lieu dans l'algorithme et $P.i$ une propriété. Dans ce cadre, les invariants \textsc{(}$L.3$, $P.3$\textsc{)} et \textsc{(}$L.4$, $P.4$\textsc{)} signifient que tous les générateurs minimaux traités avant le générateur minimal courant $g$ ont été correctement placés dans leur classe. Le lemme suivant prouve la validité de ces invariants.

\begin{lemme} Tous les couples \textsc{(}$L.i$, $P.i$\textsc{)}, $i$ = $1$, $2$, $3$, $4$ sont des invariants de la procédure \textsc{Gen-Ordre}.
\end{lemme}

\begin{demo}{\em \textsc{(}partielle\textsc{)}} {\em La preuve du lemme se construit à partir de deux récurrences imbriquées à l'image de l'imbrication des boucles \textsc{A} et \textsc{B}. La récurrence extérieure établit que les couples \textsc{(}$L.i$, $P.i$\textsc{)}, $i$ = $1$, $2$ sont des invariants ; la récurrence intérieure établit que, pour une valeur donnée du support $n$, les couples \textsc{(}$L.i$, $P.i$\textsc{)}, $i$ = $3$, $4$ sont des invariants. La seule difficulté de la démonstration consiste à montrer que si les propriétés $P.i$ sont vraies en $L.i$, $i$ = $3$, $4$, alors les propriétés $P.i'$ sont vraies en $L.i'$. Notons aussi que la différence entre $P.i$ et $P.i', i = 3,4$ est la substitution du signe $\preceq$ au signe $\prec$. Nous développons ici uniquement cette preuve. Elle utilise le Lemme \ref{lemme_comparaison_entre_GMFs_meme_CEq}. Soit $g$ le générateur minimal fréquent courant de la boucle \textsc{B}. On suppose que les propriétés $P.i$ sont vraies en $L.i$, $i$ = $3$, $4$. Deux cas se présentent selon que $g$ est ou n'est pas le représentant de sa classe.

\vspace{1.5mm}

\noindent {\sc cas 1 :} $g$ est le représentant de sa classe, $g = \rho\textsc{(}g\textsc{)}$. Dans ce cas, quel que soit $g_1$ dans $\tau\textsc{(}g\textsc{)}$, l'ensemble $\phi\textsc{(}n, g, g_1\textsc{)}$ ne contient pas $g$. L'instruction 6 n'est donc jamais exécutée et le parcours \textsc{D} \textsc{(}variable $\pi_1\textsc{)}$ se termine à son terme. Il en résulte que l'instruction 7 n'est jamais exécutée et que le parcours \textsc{C} \textsc{(}variable $\Gamma\textsc{)}$ se termine lui aussi à son terme. Dans le bloc conclusif \textsc{E}, l'instruction 9 est alors exécutée. Elle augmente la variable \$$\mathcal{X}$ du générateur minimal fréquent $g$ et initialise la variable \$$f\textsc{(}g\textsc{)}$ avec le singleton $\{g\}$. Ces modifications conduisent, à partir de \textsc{(}$L.i$, $P.i$\textsc{)}, $i$ = $3$, $4$ à \textsc{(}$L.i'$, $P.i'$\textsc{)}, $i$ = $3$, $4$.

\vspace{1.5mm}

\noindent {\sc cas 2 :} $g$ n'est pas le représentant de sa classe. On a $\rho\textsc{(}g\textsc{)} \prec g$, donc $\rho\textsc{(}g\textsc{)}$ est stocké dans la variable \$$\mathcal{X}$, d'après l'hypothèse $P.3$. Il résulte du Lemme \ref{lemme_comparaison_entre_GMFs_meme_CEq} que $g$ sera comparé à $\rho\textsc{(}g\textsc{)}$. Le parcours \textsc{D} sera donc arrêté avant son terme par l'instruction 6 et le parcours \textsc{C} sera suspendu par l'instruction 7. Dès lors, l'instruction 8 du bloc conclusif \textsc{E} sera exécutée. Elle intègre $g$ dans sa classe. Il en résulte que la propriété $P.4'$ est vérifiée en $L.4'$. Par ailleurs, la propriété $P.3'$ est \textsc{(}trivialement\textsc{)} vérifiée en $L.3'$. En effet, la variable \$$\mathcal{X}$ n'étant pas modifiée, la propriété $P.3$ est vérifiée, après le bloc conclusif \textsc{E}, en $L.3'$. Or, \textsc{(}$P.3$ et \textsc{(}$\rho$\textsc{(}$g$\textsc{)} $\prec$ $g$\textsc{)}\textsc{)} $\Rightarrow$ $P.3'$. \quad $\diamondsuit$}
\end{demo}

\subsection{extraction des bases génériques de règles}

\noindent Dans cette dernière étape, \textsc{Prince} extrait les règles génériques informatives valides formées par l'union de la base générique de règles exactes et de la réduction transitive de la base informative de règles approximatives.

\subsubsection{Dérivation des motifs fermés fréquents et des règles informatives}

\noindent Pour chaque classe d'équivalence du treillis d'Iceberg, \textsc{Prince} dérive simplement le motif fermé fréquent correspondant {\em via} l'application de la proposition donnée ci-dessous, dont la preuve utilise le théorème suivant :

\begin{theoreme} \label{theopfaltz}
Soient $f \in \mathcal{IFF}_{\mathcal{K}}$ et $GM_{f}$ l'ensemble de ses générateurs minimaux. Si $f_{1} \in \mathcal{IFF}_{\mathcal{K}}$ tel que $f$ couvre $f_{1}$ dans le treillis d'Iceberg $\textsc{(}\hat{\mathcal{L}}, \subseteq$\textsc{)} alors $face\textsc{(}f | f_{1}\textsc{)}$ est un bloqueur minimal de $GM_{f}$ {\em [Pfaltz, Taylor, 2002]}.
\end{theoreme}

\begin{prop} \label{propiff} Soient $f$ et $f_{1} \in \mathcal{IFF}_{\mathcal{K}}$ tels que $f$ couvre $f_{1}$ dans le treillis d'Iceberg $\textsc{(}\hat{\mathcal{L}}, \subseteq$\textsc{)}. Soit $GM_{f}$ l'ensemble des générateurs minimaux de $f$. Alors, le motif fermé fréquent $f$ est égal à :
$$f = \cup \{g | g \in GM_{f}\} \cup f_{1}$$
\end{prop}

\begin{demo}
{\em \'{E}tant donné que l'union des éléments de $GM_{f}$ est un bloqueur de $GM_{f}$, la face $\textsc{(}f |f_{1}\textsc{)}$, qui est un bloqueur minimal pour $GM_{f}$ d'après Théorème \ref{theopfaltz}, est incluse dans l'union des éléments de $GM_{f}$. Ainsi, il suffit de calculer l'union de $f_{1}$ avec les éléments de $GM_{f}$ pour dériver $f$. \quad $\diamondsuit$}
\end{demo}

Il est à noter que la Proposition \ref{propiff} a pour avantage d'assurer l'extraction {\em sans redondance} de l'ensemble des motifs fermés fréquents. En effet, chaque motif fermé n'est déterminé qu'une seule fois.

\vspace{2mm}

Pour chaque classe d'équivalence, une fois son motif fermé fréquent déterminé et étant donnés les liens de précédence établis lors de l'étape précédente, la dérivation des règles informatives qui lui sont associées se fait d'une manière immédiate \textsc{(}cf. Définition \ref{defbase-exact} et Définition \ref{defbase-trans}\textsc{)}.

\subsubsection{Pseudo-code de la troisième étape de l'algorithme \textsc{Prince}}

\noindent Le pseudo-code de cette étape est donné par la procédure \textsc{Gen-BGRs} \textsc{(}cf. Algorithme \ref{algogenbgr}\textsc{)}\footnote{\ BGR est l'acronyme de Base Générique de Règles.}. Dans la procédure \textsc{Gen-BGRs}, $L_{1}$ désigne la liste des classes d'équivalence à partir desquelles sont extraites les règles d'association informatives. Par $L_{2}$, nous désignons la liste des classes d'équivalence qui couvrent celles formant $L_{1}$.

\incmargin{1em}
\restylealgo{algoruled}
\linesnumbered
\begin{algorithm}[!h]\small{
 \caption{\textsc{Gen-BGRs}}}  \label{algogenbgr}
   \SetVline
   \setnlskip{-3pt}
    \Donnees { - le \textit{treillis des générateurs minimaux}, et le seuil \textit{minconf}.}

\Res{\begin{enumerate} \item Le motif fermé fréquent de
chaque classe d'équivalence.
    \item La base générique de règles exactes $\mathcal{BG}$.
    \item La réduction transitive des règles
    approximatives $\mathcal{RI}$.
\end{enumerate}}

    \Deb {
  $\mathcal{BG}$ = $\emptyset$\;  $\mathcal{RI}$ = $\emptyset$\;
 $L_{1}$ = $\{\emptyset\}$\;
 $L_{2}$ = $\emptyset$\;

 \Tq{\textsc{(}$L_{1}$ $\neq$ $\emptyset$\textsc{)}}{\PourCh{\textsc{(}$g$ $\in$
$L_{1}$\textsc{)}}{\Si {\textsc{(}$g$.\texttt{\textit{iff}}
$\ne$ $g$\textsc{)}}{$\mathcal{BG}$ = $\mathcal{BG}$ $\cup$
{\{}\textsc{(}t $\Rightarrow$ \textsc{(}$g$.\texttt{\textit{iff}}
$\backslash$ $t$\textsc{)}, $g$.\texttt{\textit{support}}\textsc{)} $\vert$
$t$ $\in$ $\mathcal{GMF}_{\mathcal{K}}$ et $t$ $\in$
[$g$]{\}};}

\PourCh{$g_{1}$ $\in$ $g$.\texttt{\textit{succs-immédiats}}}{\Si{\textsc{(}$g_{1}$.\texttt{\textit{iff}} = $\emptyset$\textsc{)}}{$g_{1}$.\texttt{\textit{iff}} = $\cup$
{\{}$t$ $\in$ $\mathcal{GMF}_{\mathcal{K}}$ $|$ $t$ $\in$
[$g_{1}$]{\}} $\cup$ $g$.\texttt{\textit{iff}}\;$L_{2}$ = $L_{2}$ $\cup$ $\{$$g_{1}$$\}$\;}

\Si{\textsc{(}$\frac{\displaystyle
g_{1}.\texttt{\textit{support}}}{\displaystyle
g.\texttt{\textit{support}}}$ $\ge$
\textit{minconf}\textsc{)}}{$\mathcal{RI}$ = $\mathcal{RI}$ $\cup$
{\{}\textsc{(}$t$ $ \Rightarrow $
\textsc{(}$g_{1}$.\texttt{\textit{iff}} $\backslash$ $t$\textsc{)},
$g_{1}$.\texttt{\textit{support}}, $\frac{\displaystyle
g_{1}.\texttt{\textit{support}}}{\displaystyle
g.\texttt{\textit{support}}}$\textsc{)} $\vert $ $t$ $\in$
$\mathcal{GMF}_{\mathcal{K}}$ et $t$ $\in$
[$g$]{\}}\;}}}

  $L_{1}$ = $L_{2}$\;
  $L_{2}$ = $\emptyset$\;
}
 }
 \end{algorithm}
\decmargin{1em}

L'ensemble des règles informatives exactes $\mathcal{BG}$ est initialement vide\textsc{(}ligne 2\textsc{)}. Il en est de même pour l'ensemble des règles informatives approximatives $\mathcal{RI}$ \textsc{(}ligne 3\textsc{)}. Le parcours du {\em treillis des générateurs minimaux} s'effectue d'une manière ascendante en partant de la classe d'équivalence dont le générateur est l'ensemble vide \textsc{(}notée [$\emptyset$]\textsc{)}. Ainsi, $L_{1}$ est initialisée par ce générateur minimal \textsc{(}ligne 4\textsc{)}. Rappelons que la fermeture de l'ensemble vide a été déjà calculée dès la première étape en collectant les items qui se répètent dans tous les objets \textsc{(}cf. lignes 7-8 de l'algorithme \ref{algoGen-GMs}, page \pageref{algoGen-GMs}\textsc{)}. La liste $L_{2}$ est initialement vide \textsc{(}ligne 5\textsc{)}. Les traitements de cette étape s'arrêtent lorsqu'il n'y a plus de classes d'équivalence à partir desquelles seront extraites des règles génériques \textsc{(}ligne 6\textsc{)}. Si la fermeture de l'ensemble vide n'est pas nulle, la règle {\em exacte informative} mettant en jeu l'ensemble vide et sa fermeture sera extraite \textsc{(}lignes 8-9\textsc{)}. Ayant l'ordre partiel construit, \textsc{Gen-BGRs} extrait les règles {\em approximatives informatives valides} mettant en jeu l'ensemble vide et les motifs fermés fréquents de la
couverture supérieure de [$\emptyset$] \textsc{(}lignes 10-15\textsc{)}. Ces fermetures sont dérivées en appliquant la Proposition \ref{propiff} aux générateurs minimaux fréquents de chaque classe d'équivalence et la fermeture de l'ensemble vide \textsc{(}ligne 10\textsc{)}. Cette couverture supérieure est stockée afin que le même traitement soit réalisé pour les classes d'équivalence la composant \textsc{(}ligne 11\textsc{)}. Une fois les traitements relatifs à la classe d'équivalence de l'ensemble vide terminés, $L_{1}$ prendra pour valeur le contenu de $L_{2}$ \textsc{(}ligne 16\textsc{)} afin d'appliquer le même processus aux classes d'équivalence qui sont successeurs immédiats de [$\emptyset$]. $L_{2}$ est initialisée de nouveau au vide \textsc{(}ligne 17\textsc{)} et contiendra les successeurs immédiats des classes d'équivalence contenues dans $L_{1}$. \'{E}tant donné qu'une classe d'équivalence peut avoir plusieurs prédécesseurs immédiats, un test est réalisé pour vérifier qu'elle n'a pas été déjà insérée dans $L_{2}$. Ce test consiste à vérifier si le motif fermé fréquent correspondant a été déjà calculé \textsc{(}ligne 11\textsc{)}. De la même manière, \textsc{Gen-BGRs} traite les niveaux supérieurs du {\em treillis des générateurs minimaux} jusqu'à atteindre ses sommets \textsc{(}c'est-à-dire les classes d'équivalence n'ayant pas de successeurs\textsc{)}. $L_{1}$ serait alors vide et la condition de la ligne 6 ne sera plus vérifiée. Ainsi, la troisième étape de l'algorithme \textsc{Prince} prend fin et toutes les règles génériques sont extraites.

\vspace{2mm}

\begin{rem} {\em Il est important de noter que dans [Zaki, 2004], l'auteur montre que les bases génériques offrent un facteur de réduction du nombre total de règles pouvant atteindre $O\textsc{(}\frac{\displaystyle 2^n}{\displaystyle n}\textsc{)}$, où $n$ étant la taille de le motif fréquent le plus long \textsc{(}par rapport au nombre d'items\textsc{)}.}
\end{rem}

\begin{figure}[!t]
\begin{center}
\parbox{4.5cm}{
\small{
\includegraphics[scale = .6]{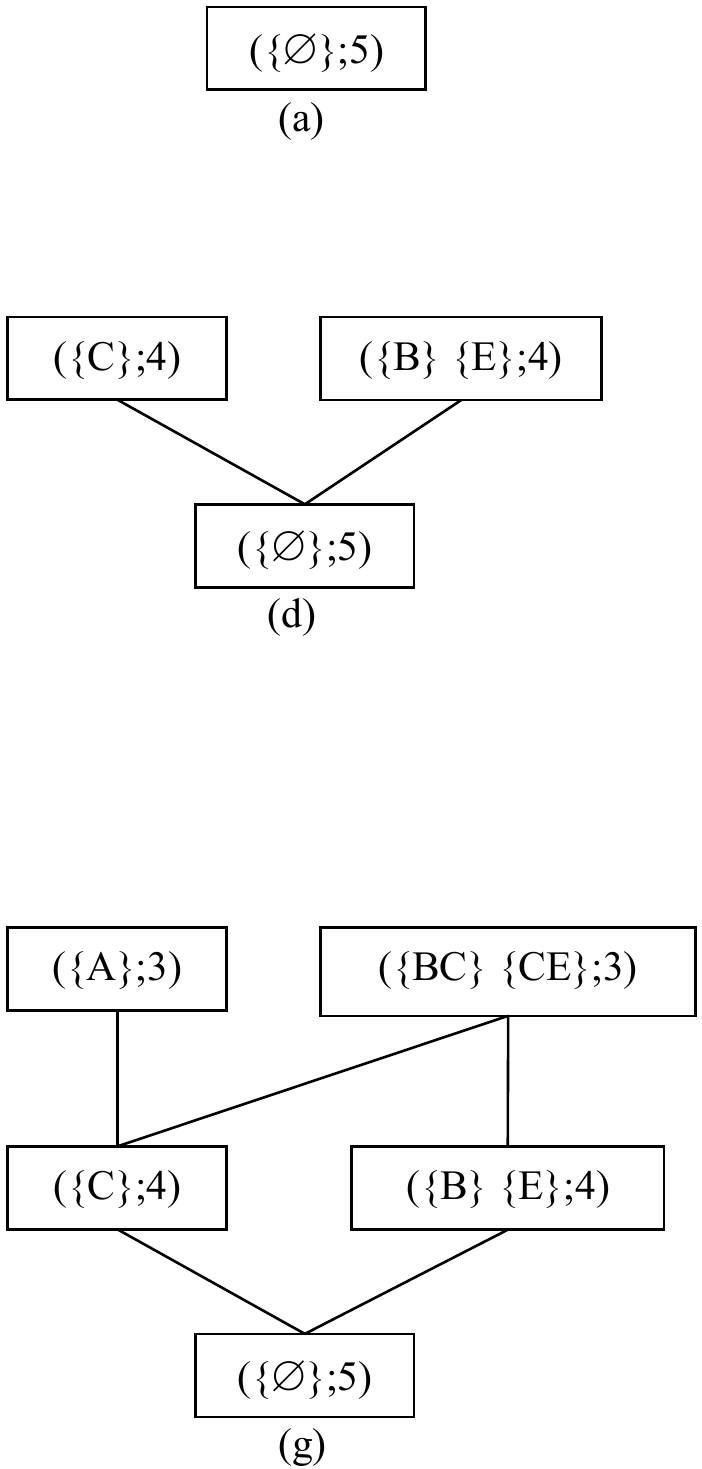}}}
\hspace{0.4cm}
\parbox{4.5cm}{
\small{
\includegraphics[scale = .6]{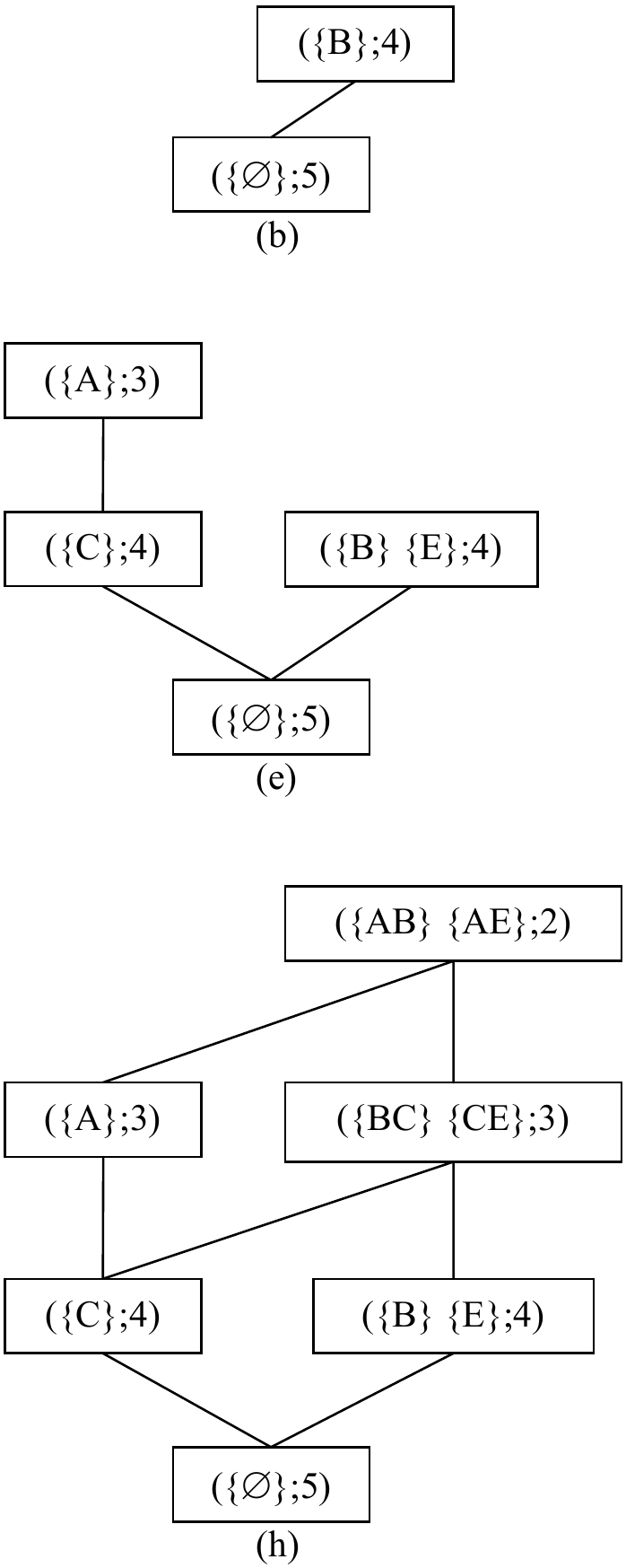}}}
\hspace{0.4cm}
\parbox{4.5cm}{
\small{
\includegraphics[scale = .6]{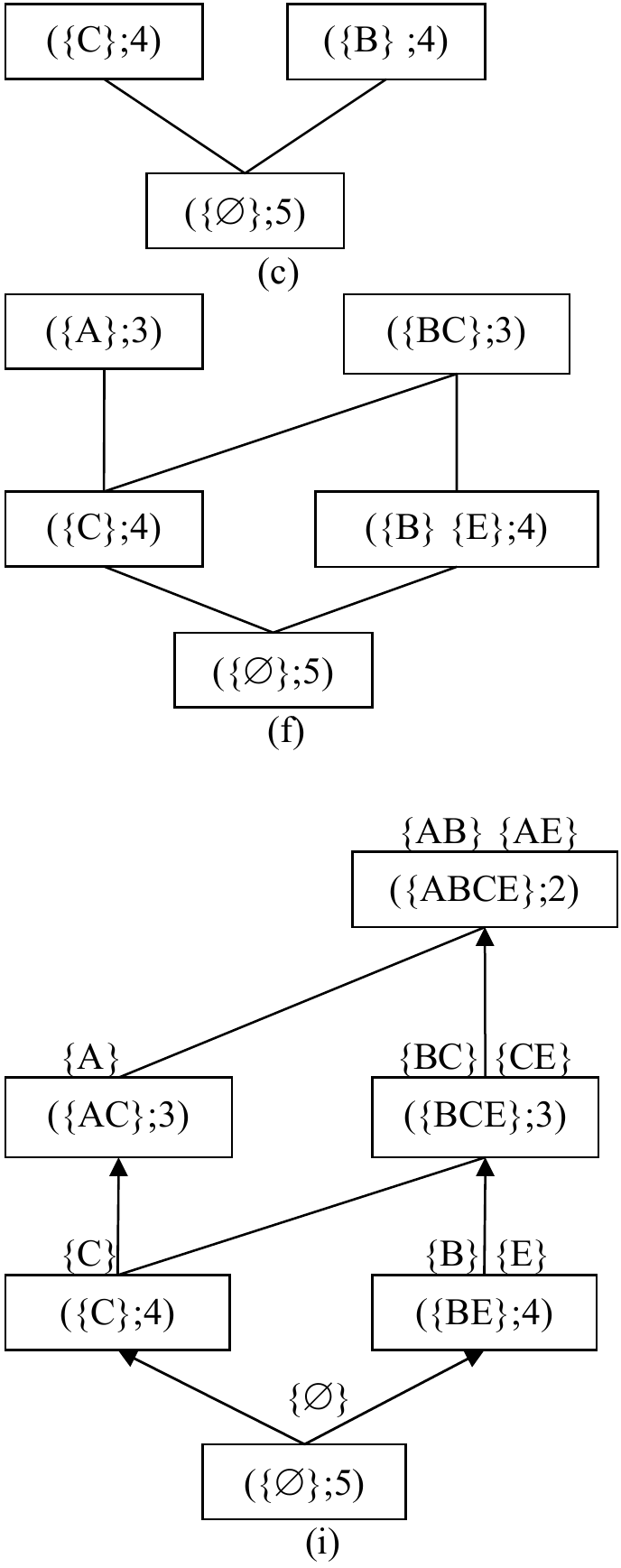}}}
\caption{\'Etapes de la construction du \textit{treillis des générateurs minimaux} et du treillis d'Iceberg
associés au contexte d'extraction $\mathcal{K}$ pour \textit{minsupp} = 2.} \label{figexemple}
\end{center}
%\end{figure}

\bigskip

\begin{center}
\parbox{4cm}{\small
    \begin{tabular}{|l|}
   \hline
\multicolumn{1}{|c|}{Règles informatives exactes}\\\hline
\hline
  $R_{_1}$ : \texttt{E}$\Rightarrow$\texttt{B} \textsc{(}{\xmplbx4}\textsc{)} \\\hline
  $R_{_2}$ : \texttt{B}$\Rightarrow$\texttt{E} \textsc{(}{\xmplbx4}\textsc{)}\\\hline
  $R_{_3}$ : \texttt{A}$\Rightarrow$\texttt{C} \textsc{(}{\xmplbx3}\textsc{)}\\\hline
  $R_{_4}$ : \texttt{BC}$\Rightarrow$\texttt{E} \textsc{(}{\xmplbx3}\textsc{)}\\\hline
  $R_{_5}$ : \texttt{CE}$\Rightarrow$\texttt{B} \textsc{(}{\xmplbx3}\textsc{)}\\\hline
  $R_{_6}$ : \texttt{AB}$\Rightarrow$\texttt{CE} \textsc{(}{\xmplbx2}\textsc{)}\\\hline
  $R_{_7}$ : \texttt{AE}$\Rightarrow$\texttt{BC} \textsc{(}{\xmplbx2}\textsc{)}\\\hline
   \end{tabular}}
\hspace{1.cm}
\parbox{7cm}
{\small
   \begin{tabular}{|l|l|}
      \hline
     \multicolumn{2}{|c|}{Règles informatives approximatives}      \\\hline\hline
     $R_{_8}$ : $\emptyset$$\overset{0,80}{\Rightarrow}$\texttt{BE} \textsc{(}{\xmplbx4}\textsc{)}& $R_{_{13}}$ : \texttt{E}$\overset{0,75}{\Rightarrow}$\texttt{BC} \textsc{(}{\xmplbx3}\textsc{)} \\\hline
     $R_{_9}$ : $\emptyset$$\overset{0,80}{\Rightarrow}$\texttt{C} \textsc{(}{\xmplbx4}\textsc{)} & $R_{_{14}}$ : \texttt{A}$\overset{0,66}{\Rightarrow}$\texttt{BCE} \textsc{(}{\xmplbx2}\textsc{)}\\\hline
     $R_{_{10}}$ : \texttt{C}$\overset{0,75}{\Rightarrow}$\texttt{A} \textsc{(}{\xmplbx3}\textsc{)}& $R_{_{15}}$ : \texttt{BC}$\overset{0,66}{\Rightarrow}$\texttt{AE} \textsc{(}{\xmplbx2}\textsc{)} \\\hline
     $R_{_{11}}$ : \texttt{C}$\overset{0,75}{\Rightarrow}$\texttt{BE} \textsc{(}{\xmplbx3}\textsc{)}&  $R_{_{16}}$ : \texttt{CE}$\overset{0,66}{\Rightarrow}$\texttt{AB} \textsc{(}{\xmplbx2}\textsc{)}\\\hline
 $R_{_{12}}$ : \texttt{B}$\overset{0,75}{\Rightarrow}$\texttt{CE} \textsc{(}{\xmplbx3}\textsc{)} &\\\hline
\end{tabular}}
\caption{{\sc Gauche} : La base générique de règles exactes $\mathcal{BG}$. {\sc Droite} : La
réduction transitive des règles approximatives $\mathcal{RI}$. Le support de chaque règle est mis entre parenthèses.} \label{rulescontext}
\end{center}
\end{figure}

\begin{exemple}\label{exemple_deroulement_algo}{\em Afin d'illustrer le déroulement de l'algorithme \textsc{Prince}, considérons le contexte d'extraction $\mathcal{K}$ donné par la figure \ref{runfig} \textsc{(}Gauche\textsc{)} pour \textit{minsupp} = {\xmplbx2} et \textit{minconf} = {\xmplbx0,50}. La première étape permet de déterminer l'ensemble des générateurs minimaux $\mathcal{GMF}\mathcal{_{K}}$ trié, ainsi que la bordure $\mathcal{GB}$d$^{-}$. $\mathcal{GMF}_{\mathcal{K}}$ = $\{$\textsc{(}$\emptyset$, {\xmplbx5}\textsc{)}, \textsc{(}\texttt{B}, {\xmplbx4}\textsc{)}, \textsc{(}\texttt{C}, {\xmplbx4}\textsc{)}, \textsc{(}\texttt{E}, {\xmplbx4}\textsc{)}, \textsc{(}\texttt{A}, {\xmplbx3}\textsc{)}, \textsc{(}\texttt{BC}, {\xmplbx3}\textsc{)}, \textsc{(}\texttt{CE}, {\xmplbx3}\textsc{)}, \textsc{(}\texttt{AB}, {\xmplbx2}\textsc{)}, \textsc{(}\texttt{AE}, {\xmplbx2}\textsc{)}$\}$ et $\mathcal{GB}$d$^{-}$ = $\{$\textsc{(}\texttt{D}, {\xmplbx1}\textsc{)}$\}$.  Dans la deuxième étape, \textsc{Prince} parcourt $\mathcal{GMF}_{\mathcal{K}}$ en comparant chaque générateur minimal fréquent $g$ de taille $k$ $\textsc{(}k \geq 1\textsc{)}$ aux listes des successeurs immédiats de ses sous-ensembles de taille $\textsc{(}k - 1\textsc{)}$. L'ensemble vide, n'ayant aucun sous-ensemble, est inséré directement dans le {\em treillis des générateurs minimaux} \textsc{(}cf. Figure \ref{figexemple}.a\textsc{)}. Ensuite, \texttt{B} est ajouté à $\emptyset$.\texttt{\textit{succs-immédiats}} \textsc{(}cf. Figure \ref{figexemple}.b\textsc{)}, la liste des successeurs immédiats du $\emptyset$, initialement vide. Ensuite, \texttt{C} sera comparé à \texttt{B}. Le motif \texttt{BC} étant un générateur minimal, $[$\texttt{B}$]$ et $[$\texttt{C}$]$ sont alors incomparables et \texttt{C} est ajouté à $\emptyset$.\texttt{\textit{succs-immédiats}} \textsc{(}cf. Figure \ref{figexemple}.c\textsc{)}. Le générateur minimal fréquent \texttt{E} est alors comparé à cette liste. En comparant \texttt{E} à \texttt{B}, nous avons \texttt{E}.\texttt{\textit{support}} = \texttt{B}.\texttt{\textit{support}} = \textit{Supp}\textsc{(}\texttt{BE}\textsc{)}. Ainsi, \texttt{E} $\in$ $[$\texttt{B}$]$, dont \texttt{B} est le représentant \textsc{(}cf. Figure \ref{figexemple}.d\textsc{)}. Afin de ne maintenir que des représentants dans les listes des successeurs immédiats, les occurrences de \texttt{E} s'il y en a seront ainsi à remplacer par \texttt{B} dans des listes des successeurs immédiats \textsc{(}dans ce cas, il n'y a aucune occurrence\textsc{)} et les comparaisons seront à poursuivre avec \texttt{B} au lieu de \texttt{E} \textsc{(}dans ce cas, il n'y a plus de comparaisons à faire \textit{via} \texttt{E}\textsc{)}. Les traitements s'arrêtent alors pour \texttt{E}. \`{A} ce moment du traitement, $\emptyset$.\texttt{\textit{succs-immédiats}} = $\{$\texttt{B}, \texttt{C}$\}$. Le générateur minimal fréquent \texttt{A} est alors comparé à \texttt{B}. Comme \texttt{AB} $\in$ $\mathcal{GMF}\mathcal{_{K}}$, $[$\texttt{B}$]$ et $[$\texttt{A}$]$ sont incomparables. Par contre, en comparant \texttt{A} et \texttt{C}, \texttt{A}.\texttt{\textit{support}} $<$ \texttt{C}.\texttt{\textit{support}} et \texttt{A}.\texttt{\textit{support}} = \textit{Supp}\textsc{(}\texttt{AC}\textsc{)} et donc $[$\texttt{A}$]$ est un successeur de $[$\texttt{C}$]$. Le générateur minimal fréquent \texttt{A} est tout simplement ajouté à \texttt{C}.\texttt{\textit{succs-immédiats}} étant donné qu'elle est encore vide \textsc{(}cf. Figure \ref{figexemple}.e\textsc{)}. Le motif \texttt{BC} est comparé aux listes des successeurs immédiats de \texttt{B} et de \texttt{C}. La liste des successeurs immédiats de \texttt{B} est vide, \texttt{BC} est alors ajouté. La liste des successeurs immédiats de \texttt{C} contient \texttt{A}. Le générateur minimal fréquent \texttt{BC} est alors comparé à \texttt{A} et comme \texttt{BC}.\texttt{\textit{support}} = \texttt{A}.\texttt{\textit{support}} mais \texttt{BC}.\texttt{\textit{support}} $\neq$ \textit{Supp}\textsc{(}\texttt{ABC}\textsc{)}, $[$\texttt{BC}$]$ et $[$\texttt{A}$]$ sont incomparables et \texttt{BC} est donc ajouté à \texttt{C}.\texttt{\textit{succs-immédiats}} \textsc{(}cf. Figure \ref{figexemple}.f\textsc{)}. Le motif \texttt{CE} est comparé aux listes des successeurs immédiats de \texttt{C} et de \texttt{E}. Celle de \texttt{C} contient \texttt{A} et \texttt{BC}. Les classes d'équivalence $[$\texttt{CE}$]$ et $[$\texttt{A}$]$ sont incomparables, puisque \texttt{CE}.\texttt{\textit{support}} = \texttt{A}.\texttt{\textit{support}} mais \texttt{CE}.\texttt{\textit{support}} $\neq$ \textit{Supp}\textsc{(}\texttt{ACE}\textsc{)}. En comparant \texttt{CE} à \texttt{BC}, \texttt{CE}.\texttt{\textit{support}} = \texttt{BC}.\texttt{\textit{support}} = \textit{Supp}\textsc{(}\texttt{BCE}\textsc{)} alors le motif \texttt{CE} va être affecté à la classe d'équivalence de \texttt{BC} et les traitements dédiés à la gestion efficace des classes d'équivalence sont invoquées \textsc{(}cf. Figure \ref{figexemple}.g\textsc{)}. En particulier, les comparaisons de \texttt{CE} aux successeurs immédiats de $[$\texttt{E}$]$ seront faites avec \texttt{BC} qui est le représentant de la classe associée. Comme $[$\texttt{E}$]$ a pour représentant \texttt{B}, \texttt{BC} est donc comparé aux éléments de \texttt{B}.\texttt{\textit{succs-immédiats}}. Cependant, comme \texttt{B}.\texttt{\textit{succs-immédiats}} ne contient que \texttt{BC} alors les comparaisons se terminent. Le même traitement est appliqué pour \texttt{AB} et \texttt{AE}. Ainsi, la procédure de construction de l'ordre partiel prend fin. Le \textit{treillis des générateurs minimaux} obtenu est donné par la Figure \ref{figexemple}.h. Pour la dérivation des règles génériques, le \textit{treillis des générateurs minimaux} est parcouru d'une manière ascendante à partir de $[$$\emptyset$$]$. Comme $\gamma\textsc{(}\emptyset\textsc{)} = \emptyset$, il n'y a donc pas de règle exacte informative relative à $[$$\emptyset$$]$. Nous avons $\emptyset$.\texttt{\textit{succs-immédiats}} = $\{$\texttt{B}, \texttt{C}$\}$. Le motif fermé fréquent correspondant à $[$\texttt{B}$]$ est alors dérivé et est égal à \texttt{BE} \textsc{(}cf. Figure \ref{figexemple}.i\textsc{)}. La règle informative approximative valide $\emptyset\Rightarrow \texttt{BE}$ de support {\xmplbx4} et de confiance {\xmplbx0,80} sera alors extraite. Il en est de même pour la règle $\emptyset\Rightarrow \texttt{C}$, ayant les mêmes valeurs de support et de confiance que la précédente. De la même manière et à partir de $[$\texttt{B}$]$ et $[$\texttt{C}$]$, le parcours du treillis se fait d'une façon ascendante jusqu'à extraire toutes les règles d'association informatives valides.

\vspace{2mm}

\`{A} la fin de l'exécution de l'algorithme, nous obtenons le treillis d'Iceberg associé au contexte d'extraction $\mathcal{K}$ \textsc{(}cf. Figure \ref{figexemple}.i\textsc{)}\footnote{\ Dans la Figure \ref{figexemple}.i, les flèches indiquent les liens de précédence utilisés pour dériver les motifs fermés fréquents.} ainsi que la liste les règles informatives valides, donnée par la Figure \ref{rulescontext}.}
\end{exemple}

\section{PROPRI\'ET\'ES DE L'ALGORITHME \textsc{Prince}}\label{section_proprietes_Prince}

\noindent Dans ce qui suit, nous allons prouver différentes propriétés associées à l'algorithme \textsc{Prince}.

\subsection{\sc preuves de validité, de terminaison et de complétude}

\noindent Dans cette sous-section, nous allons montrer la validité de l'algorithme \textsc{Prince}, puis sa terminaison et sa complétude.

\begin{prop}
L'algorithme \textsc{Prince} est valide. En effet, il permet de déterminer tous les motifs fermés fréquents, leurs générateurs minimaux associés ainsi que toutes les règles génériques informatives valides.
\end{prop}

\begin{demo} {\em Nous allons montrer la validité de \textsc{Prince} en montrant la validité de chacune des étapes le constituant.

\begin{enumerate}
    \item {\sc première étape} : \textit{Détermination des générateurs minimaux}

     \textsc{Prince} détermine tous les générateurs minimaux fréquents ainsi que la bordure négative non fréquente. En effet, \textsc{Prince} parcourt l'espace de recherche par niveau \textsc{(}et donc par taille croissante des candidats générateurs minimaux\textsc{)}. Tout au long de ce parcours, \textsc{Prince} élimine tout candidat $g$ ne pouvant pas être un générateur minimal. L'élagage d'un tel candidat est basé sur la Proposition \ref{propOI} et le Lemme \ref{lemmeTitanics-support}.

    \item {\sc deuxième étape} : \textit{Construction du \textit{treillis des générateurs minimaux}}

    Chaque générateur minimal $g$ est inséré dans le \textit{treillis des générateurs minimaux} \textit{via} sa comparaison avec les listes des successeurs immédiats \textit{des classes d'équivalence auxquelles appartiennent ses sous-ensembles de taille \textsc{(}$k$ - $1$\textsc{)}}. Au moment de l'introduction de $g$, les classes d'équivalence de ses sous-ensembles de taille \textsc{(}$k$ - $1$\textsc{)} ont été déjà insérées dans le \textit{treillis des générateurs minimaux}. En effet, le support des générateurs minimaux fréquents composant chacune de ces classes d'équivalence est strictement supérieur à celui de $g$. Par ailleurs, lors des comparaisons de $g$ avec les éléments des listes des successeurs de l'ensemble $\tau$\textsc{(}$g$\textsc{)}, les différents cas possibles sont gérés moyennant la Proposition \ref{proprelation} \textsc{(}cf. page \pageref{proprelation}\textsc{)}.

Ainsi, chaque liste des successeurs immédiats est comparée à tous les générateurs minimaux fréquents susceptibles d'y appartenir. Les traitements s'arrêtent lorsqu'il n'y a plus de générateurs minimaux fréquents à introduire dans le \textit{treillis des générateurs minimaux}.

    \item {\sc troisième étape} : \textit{Extraction des bases génériques informatives de règles}

    Dans cette étape, le parcours du \textit{treillis des générateurs minimaux} se fait en
    partant de [$\emptyset$]. Toute classe d'équivalence $\mathcal{CE}$, autre que [$\emptyset$], admet au moins un prédécesseur immédiat. La classe $\mathcal{CE}$ sera donc incluse dans au moins une couverture supérieure d'une autre classe d'équivalence. Ceci permettra de l'inclure dans la liste des classes d'équivalence à traiter lors de la prochaine itération \textsc{(}c'est-à-dire dans la liste L$_{2}$ de l'Algorithme \ref{algogenbgr}, page \pageref{algogenbgr}\textsc{)}. Les traitements s'arrêtent lorsque les couvertures supérieures des classes d'équivalence traitées en dernier sont vides. Ainsi, toutes les classes d'équivalence du treillis d'Iceberg seront traitées. En appliquant la Proposition \ref{propiff} \textsc{(}cf. page \pageref{propiff}\textsc{)}, tous les motifs fermés fréquents sont dérivés au fur et à mesure de ce parcours ascendant. Les règles génériques informatives valides sont alors extraites d'une manière directe. \quad $\diamondsuit$\\
\end{enumerate}}
\end{demo}

\begin{prop}
L'algorithme \textsc{Prince} termine dans tous les cas et son résultat est complet.
\end{prop}

\begin{demo} {\em Montrons que l'algorithme \ \textsc{Prince} \ termine quel que soit le \ contexte d'extraction $\mathcal{K}$ donné en entrée et quelles que soient les valeurs de \textit{minsupp} et \textit{minconf}. \'{E}tant donné que le nombre de couples du contexte d'extraction est fini, le nombre de générateurs minimaux fréquents extraits du contexte $\mathcal{K}$ lors de la première étape est fini. Afin de construire le treillis d'Iceberg, les comparaisons entre générateurs minimaux sont de nombre fini étant donné que le nombre d'éléments de chaque liste de successeurs immédiats est fini. Il en est de même pour le nombre de générateurs minimaux fréquents déjà insérés dans le treillis partiellement construit lors du traitement d'un générateur donné. De même le treillis d'Iceberg parcouru lors de la troisième étape a une taille finie, égale au nombre de ses classes d'équivalence.

Le résultat de l'algorithme \textsc{Prince} est complet étant donné que la construction du treillis d'Iceberg assure l'exhaustivité des éléments \textsc{(}les générateurs minimaux fréquents\textsc{)} de chaque classe d'équivalence. De même, elle assure l'exhaustivité des éléments de la liste des successeurs immédiats de chaque classe d'équivalence. Les traitements effectués lors de la troisième étape prennent ainsi en compte toutes les classes d'équivalence. \quad $\diamondsuit$}
\end{demo}

\subsection{\sc complexité de l'algorithme \textsc{Prince}}

\noindent Dans cette sous-section, nous allons étudier la complexité théorique au \textit{pire des cas} de l'algorithme \textsc{Prince}.

\begin{prop}
Dans le pire des cas, la complexité de l'algorithme \textsc{Prince} est $O\textsc{(}\textsc{(}n^{3} + m\textsc{)} \times 2^{n}\textsc{)}$, où $n$ \textsc{(}{\em resp.} $m$\textsc{)} est le nombre d'items \textsc{(}{\em resp. objets}\textsc{)} du contexte d'extraction.
\end{prop}

\begin{demo} {\em Soit un contexte d'extraction $\mathcal{K}$ = \textsc{(}$\mathcal{O}$, $\mathcal{I}$, $\mathcal{M}$\textsc{)}, le pire des cas est obtenu quand le nombre de générateurs minimaux fréquents est égal au nombre de motifs fréquents et est égal au nombre de motifs fermés fréquents \textsc{(}c'est-à-dire 2$^{|\mathcal{I}|}$\textsc{)}. En d'autres termes, chaque générateur minimal fréquent est aussi un fermé et le \textit{treillis des générateurs minimaux} se confond avec le treillis des motifs fréquents.

Soient $m$ = $|$$\mathcal{O}$$|$ et $n$ = $|$$\mathcal{I}|$. La taille maximale d'un objet est égale au nombre maximum d'items distincts, c'est-à-dire $n$. Nous allons considérer que tous les objets ont pour longueur $n$. Pour chaque objet et dans le pire des cas, nous déterminons ses 2$^{n}$ sous-ensembles afin de calculer les supports des motifs. Nous allons calculer la complexité théorique au pire des cas de l'algorithme \textsc{Prince} en calculant la complexité au pire des cas de chacune des trois étapes le constituant.\\

\begin{itemize}
    \item {\sc  première étape} : \textit{Détermination des générateurs minimaux}
        \begin{enumerate}
        \item L'initialisation de l'ensemble des items candidats se fait en $O\textsc{(}1\textsc{)}$ \textsc{(}ligne 2 dans Algorithme \ref{algoGen-GMs}, page \pageref{algoGen-GMs}\textsc{)}.
        \item Le coût du calcul des supports des items est de l'ordre de $O\textsc{(}m \times n\textsc{)}$ \textsc{(}ligne 3\textsc{)}.
        \item Les affectations des lignes 4-5 se font en $O\textsc{(}1\textsc{)}$.
        \item Le coût de l'élagage par rapport aux supports des items est de l'ordre de $O\textsc{(}n\textsc{)}$ \textsc{(}lignes 6-14\textsc{)}.
        \item Le coût de la détermination des générateurs minimaux fréquents de tailles supérieures ou égales à {\xmplbx2} \textsc{(}lignes 15-16\textsc{)} est égal à la somme des coûts suivants :
\begin{enumerate}
    \item \textsc{Apriori-Gen} : il y a \textsc{(}2$^{n}$ - $n$ - 1\textsc{)} candidats à générer. Ainsi, le coût de cette phase est de l'ordre de $O\textsc{(}2^{n} - n\textsc{)}$ \textsc{(}lignes 2-3 dans Algorithme \ref{algoGen-GMs-suivants}, page \pageref{algoGen-GMs-suivants}\textsc{)} ;
    \item Le coût de la vérification de l'idéal d'ordre, et en même temps, le calcul du support estimé et le stockage des liens vers les sous-ensembles adéquats est de l'ordre de $O\textsc{(}n^{2} \times
\textsc{(}2^{n} - n\textsc{)}\textsc{)}$ \textsc{(}lignes 4-14\textsc{)} ;
        \item Le coût du calcul des supports des candidats de tailles supérieures ou égales à
    {\xmplbx2} est de l'ordre de $O\textsc{(}m \times \textsc{(}2^{n} - n\textsc{)}\textsc{)}$ \textsc{(}ligne 16\textsc{)} ;
    \item Le coût de l'élagage par rapport aux supports des candidats est de l'ordre de $O\textsc{(}2^{n} - n\textsc{)}$ \textsc{(}lignes 17-22\textsc{)}.
\end{enumerate}
\end{enumerate}
La complexité $\mathcal{C}_{1}$ de cette étape est alors : $\mathcal{C}_{1}$ = $O\textsc{(}m \times n + n +
\textsc{(}2^{n} - n\textsc{)} + n^{2} \times \textsc{(}2^{n} - n\textsc{)} + m \times \textsc{(}2^{n} - n\textsc{)} + 2^{n} - n\textsc{)}$ = $O\textsc{(}m \times 2^{n}+2^{n}+\textsc{(}n^{2}+1\textsc{)} \times \textsc{(}2^{n} - n\textsc{)}\textsc{)}$.

\'{E}tant donné que 2$^{n}$ est largement supérieur à $n$, alors $\mathcal{C}_{1}$ = $O\textsc{(}m \times
2^{n}+2^{n}+\textsc{(}n^{2}+1\textsc{)} \times 2^{n}\textsc{)}$  = $O\textsc{(}m \times 2^{n}+\textsc{(}n^{2}+2\textsc{)} \times 2^{n}\textsc{)}$ = $O\textsc{(}m \times 2^{n}+n^{2} \times 2^{n}\textsc{)}$ = $O\textsc{(}\textsc{(}n^{2}+m\textsc{)} \times 2^{n}\textsc{)}$.

D'où, $\mathcal{C}_{1}$ = $O\textsc{(}\textsc{(}n^{2}+m\textsc{)} \times 2^{n}\textsc{)}$.\\

    \item {\sc deuxième étape} : \textit{Construction du treillis des générateurs minimaux}

\begin{enumerate}
    \item La boucle {\em pour} du bloc \textsc{A} \textsc{(}dans Algorithme \ref{algogenordre} page
    \pageref{algogenordre}\textsc{)} se répète $2^{n}$ fois étant donné que dans le pire des cas, nous avons $2^{n}$ générateurs minimaux fréquents.

    \item Pour chaque générateur minimal fréquent $g$ de support $n$ et de taille $k$, la boucle {\em parcours} du bloc \textsc{C} se répète $k$ fois étant donné que $g$ admet $k$ sous-ensembles de taille \textsc{(}$k$ - $1$\textsc{)} \textsc{(}$k$, le nombre de sous-ensembles, sera largement majoré par $n$\textsc{)}. Pour chaque sous-ensemble de $g$ de taille \textsc{(}$k$ - $1$\textsc{)}, disons $g_{1}$, les traitements suivants sont réalisés :
        \begin{enumerate}
            \item Le coût du repérage du représentant de la classe de $g_{1}$ est en $O\textsc{(}1\textsc{)}$ étant donné que dans le pire des cas, $g_{1}$ est le seul générateur minimal fréquent de sa classe d'équivalence et est donc son représentant \textsc{(}$g_{1}$ est égal à $\rho$\textsc{(}$g_{1}$\textsc{)}\textsc{)}.
            \item Le motif $g_{1}$, étant de taille égale à \textsc{(}$k$ - $1$\textsc{)}, possède au maximum $n$ - \textsc{(}$k$ - $1$\textsc{)} - $1$ successeurs immédiats, au moment de l'intégration de $g$ dans le \textit{treillis des générateurs minimaux} \textsc{(}$n$ - $k$\textsc{)}, le nombre maximal de successeurs immédiats, sera largement majoré par $n$\textsc{)}. Afin de déterminer l'ensemble $\phi$\textsc{(}$n$, $g$, $g_1$\textsc{)}, $g$ est ainsi comparé à chaque élément $g_{2}$ de la liste des successeurs immédiats de $g_{1}$, c'est-à-dire, $g_{1}$.\texttt{\textit{succs-immédiats}}. Le coût de la comparaison de $g$ avec $g_{2}$ est la somme des deux coûts suivants :
\begin{enumerate}
    \item Le premier coût est relatif à l'union de $g$ et $g_{2}$ et qui est, au pire des cas, de l'ordre de $O\textsc{(}n\textsc{)}$. Soit $q$, le motif résultat de cette union.
    \item Le second coût est relatif à la recherche du support adéquat de $q$ et qui est, au pire des cas, de l'ordre de $O\textsc{(}n\textsc{)}$. Notons que le motif $q$ est un générateur minimal fréquent. En effet, $q$ est inclus dans le motif contenant tous les items, c'est-à-dire le motif $\mathcal{I}$. Or, ce dernier -- dans le pire des cas -- est un générateur minimal fréquent et d'après la propriété de l'idéal d'ordre des générateurs minimaux fréquents \textsc{(}cf. Proposition \ref{propOI}, page \pageref{propOI}\textsc{)}, $q$ est un générateur minimal fréquent.
\end{enumerate}
        \'{E}tant \ \ donné \ \ que \ \ l'union \ \ de \ \ $g$ \ \ avec \ \ tout \ \ élément \ \ de \ \ $g_{1}$.\texttt{\textit{succs-immédiats}} donne toujours un générateur minimal, aucune des deux conditions spécifiées par la Proposition \ref{proprelation} \textsc{(}cf. page \pageref{proprelation}\textsc{)} n'est vérifiée. Ainsi, [$g$] est incomparable avec toutes les classes d'équivalence des représentants successeurs immédiats de $g_{1}$. Par conséquent, $\phi$\textsc{(}$n$, $g$, $g_1$\textsc{)} contient seulement $g_1$ \textsc{(}étant représentant de sa classe\textsc{)}. Le motif $g$ est donc ajouté en tant que successeur immédiat de $g_1$ en $O\textsc{(}1\textsc{)}$.
        \end{enumerate}
\end{enumerate}

Ainsi, la complexité $\mathcal{C}_{2}$ de cette étape est de l'ordre de : $\mathcal{C}_{2}$ = $O\textsc{(}2^{n}
\times n \times \textsc{(}1 + \textsc{(}n \times\textsc{(}n+n\textsc{)}\textsc{)} + 1\textsc{)}\textsc{)}$ = $O\textsc{(}2^{n} \times n \times \textsc{(}n \times \textsc{(}n+n\textsc{)}\textsc{)}\textsc{)}$.

D'où, $\mathcal{C}_{2}$ = $O\textsc{(}n^{3} \times 2^{n}\textsc{)}$.\\

\item {\sc troisième étape} : {\em Extraction des bases génériques informatives de règles}
\begin{enumerate}
    \item Le coût des initialisations des lignes 2-5 \textsc{(}dans Algorithme \ref{algogenbgr}, page \pageref{algogenbgr}\textsc{)} se fait en $O\textsc{(}1\textsc{)}$.
    \item La condition de la boucle {\em tant que} est vérifiée tant qu'il y a des classes d'équivalence à traiter \textsc{(}ligne 6\textsc{)}. Dans le pire des cas, chaque générateur minimal fréquent fermé forme une classe d'équivalence. Ainsi, $2^{n}$ classes d'équivalence sont traitées. Pour chaque classe d'équivalence $\mathcal{CE}$ \textsc{(}ligne 7\textsc{)}, le coût total de cette étape est égal à la somme des deux coûts suivants :
\begin{enumerate}
    \item Le premier coût consiste à dériver la fermeture correspondante à $\mathcal{CE}$. Cette dernière est le résultat de l'union du générateur minimal fréquent de $\mathcal{CE}$ avec la fermeture d'un de ses prédécesseurs immédiats \textsc{(}lignes 11-13\textsc{)}. Ceci est réalisé, au pire des cas, en $O\textsc{(}n\textsc{)}$.
    \item Le deuxième coût est relatif à l'extraction des règles d'association informatives.
    \begin{enumerate}
        \item \'{E}tant donné que, le générateur minimal fréquent de $\mathcal{CE}$ est égal à sa fermeture, la condition de la ligne 8, testée en $O\textsc{(}1\textsc{)}$, n'est pas vérifiée. Ainsi, aucune règle exacte informative n'est extraite \textsc{(}ligne 9\textsc{)}.
        \item Soit $k$ la taille de le motif fermé fréquent de $\mathcal{CE}$. La classe $\mathcal{CE}$ possède alors $k$ prédécesseurs immédiats. Ainsi, pour une valeur de \textit{minconf} égale à {\xmplbx0}, $k$ règles approximatives informatives valides sont extraites \textsc{(}lignes 14-15\textsc{)}. Le coût de l'extraction de chacune des
règles approximatives est égal au coût du calcul de la différence entre la prémisse et la conclusion et qui est, au pire des cas, de l'ordre de $O\textsc{(}n\textsc{)}$.
    \end{enumerate}
Le coût total, pour chaque classe d'équivalence, est alors de l'ordre de $O\textsc{(}n + 1 + n \times n\textsc{)}$ \textsc{(}la taille $k$ d'un motif fermé fréquent étant largement majorée par $n$\textsc{)}.
\end{enumerate}
\end{enumerate}

La complexité $\mathcal{C}_{3}$ de cette étape est alors de l'ordre de : $\mathcal{C}_{3}$ = $O\textsc{(}1 +
\textsc{(}n + 1+ n^{2}\textsc{)} \times 2^{n}$\textsc{)}.

D'où, $\mathcal{C}_{3}$ = $O\textsc{(}n^{2} \times 2^{n}\textsc{)}$.\\

\end{itemize}

Au pire des cas, la complexité totale de l'algorithme \textsc{Prince} est de l'ordre de : $\mathcal{C}_{pire}$ = $\mathcal{C}_{1}$ + $\mathcal{C}_{2}$ + $\mathcal{C}_{3}$ = $O\textsc{(}\textsc{(}n^{2}+m\textsc{)} \times 2^{n}\textsc{)}$ +
$O\textsc{(}n^{3} \times 2^{n}\textsc{)}$ + $O\textsc{(}n^{2} \times 2^{n}\textsc{)}$ = $O\textsc{(}\textsc{(}n^{3} + 2 \times n^{2} + m\textsc{)}2^{n}\textsc{)}$ = $O\textsc{(}\textsc{(}n^{3} + m\textsc{)} \times 2^{n}\textsc{)}$. D'où, $\mathcal{C}_{pire}$ = $O\textsc{(}\textsc{(}n^{3} + m\textsc{)} \times 2^{n}\textsc{)}$. \quad $\diamondsuit$}
\end{demo}

Ainsi, la complexité de \textsc{Prince} est de même ordre de grandeur que celle des algorithmes dédiés à l'extraction des motifs fermés fréquents [Godin {\em \& al.}, 1995 ; Pasquier, 2000 ; Stumme {\em \& al.}, 2002], bien qu'il détermine les
trois composantes nécessaires à l'extraction des bases génériques de règles d'association.

\section{\'EVALUATION EXP\'{E}RIMENTALE}\label{section_XPs}

\noindent Dans cette section, nous présentons les performances de \textsc{Prince} comparées à celles des algorithmes \textsc{Close}, \textsc{A-Close} et \textsc{Titanic}. \textsc{Prince} est implanté en langage C \footnote{\ Une version de l'algorithme \textsc{Prince} est disponible à l'adresse suivante : \\
http://www.cck.rnu.tn/sbenyahia/software\_release.htm.} tandis que les trois autres algorithmes le sont en C++. Toutes les expérimentations ont été réalisées sur un PC muni d'un processeur Pentium IV ayant une fréquence d'horloge de {\xmplbx2,40} GHz et {\xmplbx512} Mo de mémoire RAM \textsc{(}avec {\xmplbx2} Go de swap\textsc{)} tournant sous la plate-forme \textsc{S.u.S.E} Linux {\xmplbx9.0.} Les programmes ont été compilés avec le compilateur gcc {\xmplbx3.3.1}.

\vspace{2mm}

Dans le cadre de nos expérimentations, nous avons comptabilisé le temps d'exécution total des algorithmes sur des contextes de référence denses et épars\footnote{\ L'ensemble de ces contextes est
disponible à [Goethals, 2004].} ainsi que sur des contextes \guillemotleft~pire des cas \guillemotright \ [Hamrouni {\em \& al.}, \`a para\^{i}tre]. La définition d'un contexte \guillemotleft~pire des cas \guillemotright \ est comme suit :

\begin{definitio}\label{def_base_pire} Un contexte \guillemotleft~pire des cas \guillemotright \ est un contexte où la taille de l'ensemble des items $\mathcal{I}$ est égale à $n$ et celle de l'ensemble des objets $\mathcal{O}$ est égale à \textsc{(}$n$ + $1$\textsc{)}. Dans cette base, chaque item est vérifié par $n$ objets
différents. Chaque objet, parmi les $n$ premiers, est vérifié par \textsc{(}$n$ - $1$\textsc{)} items différents. Le dernier objet est vérifié par tous les items. Les objets sont tous différents deux à deux. Les items sont tous différents deux à deux.
\end{definitio}

Les caractéristiques des contextes testés sont résumées par le Tableau ~\ref{Caractéristiques BDTs}. Typiquement, les quatre premiers contextes sont considérés comme denses, c'est-à-dire qu'ils produisent plusieurs motifs fréquents longs même pour des valeurs de supports élevées [Bayardo, 1998]. Le reste des contextes testés sont considérés comme épars. \textsc{T10I4D100K} et \textsc{T40I10D100K} sont deux contextes synthétiques générées par un programme développé dans le cadre du projet db\textsc{Quest}\footnote{\ Le générateur des bases synthétiques est disponible à l'adresse suivante :\\
 http://www.almaden.ibm.com/software/quest/resources/datasets/data/.}. Les autres contextes de référence sont réels.

\begin{table}[h]
\begin{center} \parbox{4.2cm}{
\small{
\begin{tabular}{|p{10pt}|p{170pt}|}
\hline
 $|$T$|$ &   Taille moyenne des objets \\\hline $|$I$|$ &
  Taille moyenne des motifs maximaux potentiellement fréquents\\\hline$|$D$|$&   Nombre d'objets générés \\
\hline
\end{tabular}
}}\hspace{3.5cm}\parbox{3.cm}{ \small{
\begin{tabular}{|p{5pt}|c|c|c|c|}
\hline  & $i_{1}$ & $i_{2}$ & $i_{3}$ & $i_{4}$\\
\hline $o_{1}$ &  & $\times$ & $\times$ & $\times$ \\ \hline
$o_{2}$ & $\times$ &  &$\times$ & $\times$\\ \hline  $o_{3}$ &
$\times$ & $\times$ &   &$\times$\\ \hline  $o_{4}$ & $\times$&
$\times$ &
$\times$&\\ \hline $o_{5}$ & $\times$& $\times$ & $\times$&$\times$\\
\hline
\end{tabular}
 }}
 \vspace{0.2cm}
\small{
    \begin{tabular}{|l|r|r|r|r|}
  \hline
\textsc{Contextes de référence}& \textsc{Type du}&\textsc{Nombre}&\textsc{Nombre}& \textsc{Taille moyenne}\\
&\textsc{contexte} &\textsc{d'items} &\textsc{d'objets} &\textsc{d'un objet}\\
  \hline   \hline   \textsc{\textbf{Pumsb}} &Dense&            7 117      &   49 046           & 74        \\
    \hline
       \textsc{\textbf{Connect}} & Dense&      129         &   67 557           & 43        \\
  \hline
    \textsc{\textbf{Chess}}& Dense  &          75          &   3 196            & 37        \\
  \hline
   \textsc{\textbf{Mushroom}}& Dense        &  119         &   8 124            & 23        \\
    \hline \hline
  \textsc{\textbf{T10I4D100K}}& \'{E}pars&    1 000      &   100 000          & 10        \\
   \hline
   \textsc{\textbf{T40I10D100K}}& \'{E}pars&  1 000      &   100 000          & 40        \\
    \hline
      \textsc{\textbf{Retail}}& \'{E}pars&    16 470     &   88 162           & 10        \\
  \hline
  \textsc{\textbf{Accidents}}& \'{E}pars &    469         &   340 183          &   33      \\
\hline \hline \textbf{Contexte \guillemotleft~pire des cas \guillemotright} &\'{E}pars& $n$  &
$n$ + $1$  & $\frac{\displaystyle n^{2}}{\displaystyle n\ +\ 1}$
\\\hline

\end{tabular}}
\caption{\textsc{(}{\sc haut à gauche}\textsc{)} Paramètres des contextes synthétiques. \textsc{(}{\sc haut à droite}\textsc{)} Exemple d'un contexte \guillemotleft~pire des cas \guillemotright \ pour $n$ = {\xmplbx4}. \textsc{(}{\sc bas}\textsc{)} Caractéristiques des contextes de test. } \label{Caractéristiques BDTs}
\end{center}
\end{table}

Pour l'algorithme \textsc{Prince} et dans toutes les figures qui vont suivre, nous utilisons un \textit{minconf} égal à {\xmplbx0}\ \%, c'est-à-dire que nous avons considéré, pour une valeur de \textit{minsupp} fixée, le pire des cas par rapport au nombre de règles générées. En outre, tout au long de cette section, les figures présentées ont une échelle logarithmique. Dans les tableaux comparatifs des temps d'exécution, nous avons arrondi les valeurs à l'entier le plus proche pour les avoir toutes en seconde étant donné les écarts dans les temps d'exécution des algorithmes. Par ailleurs, \guillemotleft~/ \guillemotright \ signifie que l'exécution ne s'est pas terminée correctement et que le ratio correspondant ne peut être calculé.

\vspace{2mm}

Avant de nous intéresser aux performances de \textsc{Prince}, nous donnons à titre indicatif un exemple montrant l'utilité du couple de bases \textsc{(}$\mathcal{BG}$, $\mathcal{RI}$\textsc{)} dans la réduction sans perte
d'information du nombre de règles offertes à l'utilisateur. Par exemple, pour le contexte \textsc{Mushroom} et pour un \textit{minsupp} = {\xmplbx10}\ \%, la taille de \textsc{(}$\mathcal{BG}$, $\mathcal{RI}$\textsc{)} est égale à {\xmplbx33 762} alors que celle de l'ensemble de toutes les règles valides\footnote{\ Fourni par l'implémentation de Bart Goethals disponible à : \\
http ://www.adrem.ua.ac.be/$^{\thicksim}$goethals/software/.} est égale à {\xmplbx380 791 946}, ce qui constitue un taux de réduction égal à {\xmplbx1 127,88} fois. Ceci montre clairement que l'extraction des bases génériques est plus avantageuse que l'extraction de toutes les règles et confirme les résultats obtenus dans d'autres travaux tels que [Ashrafi {\em \& al.}, 2007 ; Ceglar, Roddick, 2006 ; Kryszkiewicz, 2002 ; Zaki, 2004].

\subsection{performances de \textsc{Prince} \textit{versus} \textsc{Close}, \textsc{A-Close} et \textsc{Titanic}}

\subsubsection{Expérimentations sur les contextes denses}

\begin{figure}[h!]
\begin{center}
\parbox{7cm}{\includegraphics[scale = 0.9]{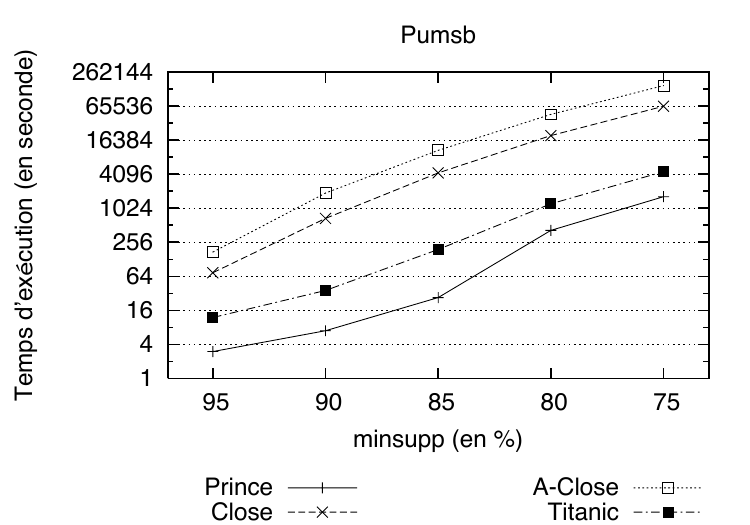}}
\parbox{7cm}{\includegraphics[scale = 0.9]{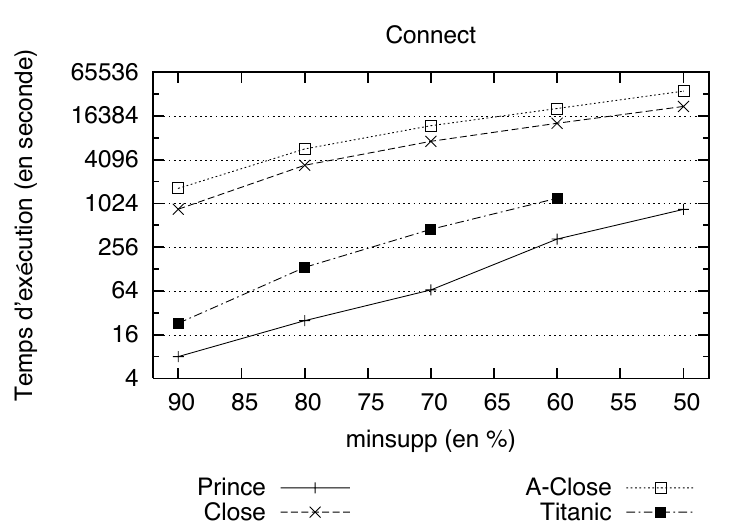}}\\
\parbox{7cm}{\includegraphics[scale = 0.9]{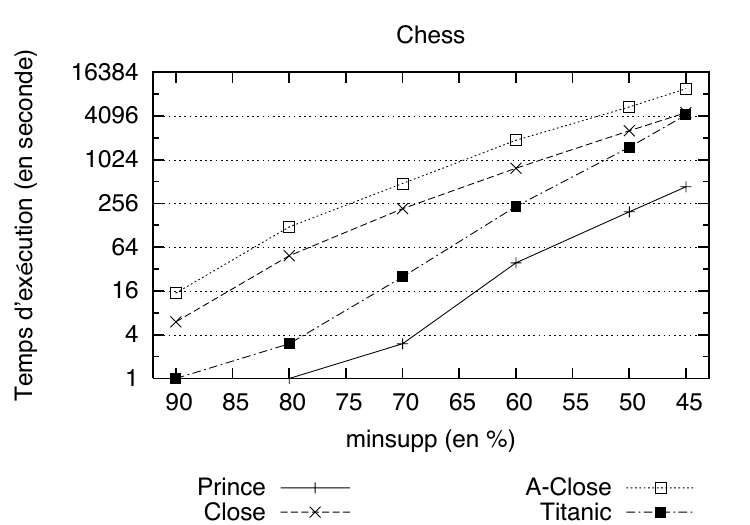}}
\parbox{7cm}{\includegraphics[scale = 0.9]{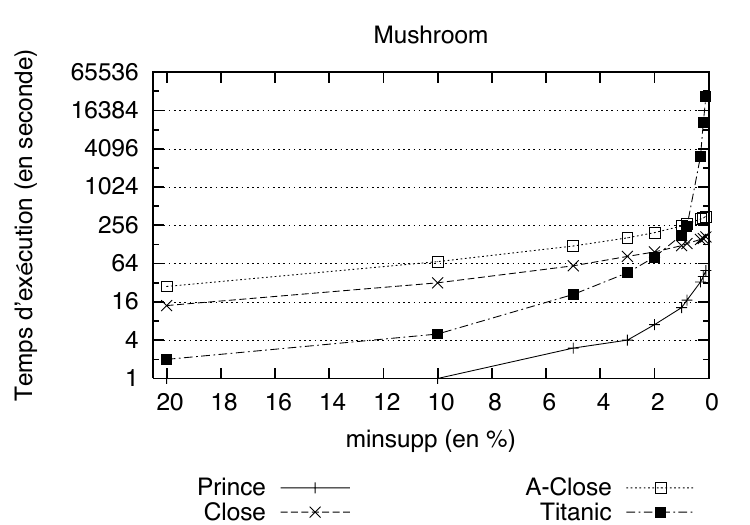}}
\end{center}
\caption{Les performances de \textsc{Prince}
\textit{versus} \textsc{Close}, \textsc{A-Close} et \textsc{Titanic}
pour les contextes denses.} \label{fig_tps_exec_denses}
\end{figure}

\noindent Les temps d'exécution de l'algorithme \textsc{Prince} comparés respectivement aux algorithmes \textsc{Close}, \textsc{A-Close} et \textsc{Titanic} sur les contextes denses sont
présentés par la Figure \ref{fig_tps_exec_denses} et le Tableau \ref{ratiosdense}.

- \textsc{\textbf{Pumsb}} : pour cette base, les performances de \textsc{Prince} sont meilleures que celles de \textsc{Close}, \textsc{A-Close} et \textsc{Titanic} pour toutes les valeurs de \textit{minsupp}. Les performances de \textsc{Close} et \textsc{A-Close} se dégradent considérablement étant donné qu'ils effectuent des intersections sur un grand nombre d'objets de taille élevée. Le mécanisme de comptage
par inférence adopté par \textsc{Titanic} s'avère plus efficace que le calcul des intersections adopté par
\textsc{Close} et \textsc{A-Close}. Ceci peut être expliqué par le nombre réduit d'items fréquents à
prendre en considération lors du calcul de la fermeture d'un générateur minimal fréquent. Les performances de \textsc{Prince} peuvent être expliquées par le fait que les traitements à effectuer pour un générateur minimal fréquent sont beaucoup moins coûteux que ceux effectués pour les trois
autres algorithmes. Les traitements de gestion des classes d'équivalence permettent aussi à \textsc{Prince} d'éviter le calcul redondant des fermetures. En effet, pour une valeur de \textit{minsupp} égale à
{\xmplbx75}\ \%, le nombre de générateurs minimaux fréquents \textsc{(}égal à {\xmplbx248 406}\textsc{)} est presque égal à {\xmplbx2,46} fois le nombre de motifs fermés fréquents \textsc{(}égal à {\xmplbx101 083}\textsc{)}.

- \textsc{\textbf{Connect}} : pour cette base, et bien qu'il n'y ait aucun calcul redondant des fermetures pour les trois algorithmes \textsc{Close}, \textsc{A-Close} et \textsc{Titanic} \textsc{(}le nombre de générateurs minimaux fréquents étant égal à celui des motifs fermés fréquents\textsc{)}, les performances de notre algorithme restent les meilleures. En effet, tout comme \textsc{Pumsb}, \textsc{Connect} est caractérisé par un grand nombre d'objets de taille élevée. Ces caractéristiques compliquent la tâche des algorithmes
\textsc{Close} et \textsc{A-Close}. \textsc{Prince} est alors avantagé par un coût réduit des comparaisons effectuées pour un générateur minimal fréquent comparé aux tentatives d'extension exécutées pour un générateur minimal fréquent dans le cas de \textsc{Titanic}. Il est à noter que pour \textit{minsupp} = {\xmplbx50}\ \%, l'exécution de \textsc{Titanic} n'a pu se terminer pour manque d'espace mémoire.

- \textsc{\textbf{Chess}} : pour cette base, les performances de \textsc{Prince} sont largement meilleures que celles de \textsc{Close}, \textsc{A-Close} et \textsc{Titanic} pour toutes les valeurs de \textit{minsupp}. Le mécanisme de comptage par inférence adopté par \textsc{Titanic} s'avère plus efficace que le calcul des intersections adopté respectivement par les algorithmes \textsc{Close} et \textsc{A-Close}. Cependant, cette efficacité tend à diminuer avec la baisse de la valeur de \textit{minsupp}. En effet, pour calculer les motifs fermés fréquents, \textsc{Titanic} nécessite des recherches de supports dont le nombre augmente considérablement avec l'augmentation du nombre d'items fréquents, au fur et à mesure que la valeur de \textit{minsupp} diminue.

- \textsc{\textbf{Mushroom}} : dans le cas du contexte \textsc{Mushroom}, \textsc{Prince} est encore une fois meilleur que \textsc{Close}, \textsc{A-Close} et \textsc{Titanic}. \textsc{Prince} bénéficie du rôle important joué par les traitements de gestion des classes d'équivalence. En effet, pour une valeur de \textit{minsupp} égale à {\xmplbx0,10}\ \%, le nombre de générateurs minimaux fréquents \textsc{(}égal à {\xmplbx360 166}\textsc{)} est presque égal à {\xmplbx2,20} fois le nombre de motifs fermés fréquents \textsc{(}égal à {\xmplbx164 117}\textsc{)}. Ainsi, bien que \textsc{Close} et \textsc{A-Close} bénéficiaient d'un nombre réduit d'objets ainsi qu'une taille moyenne des objets relativement petite sur lesquelles ils exécutent des intersections, ces deux algorithmes sont handicapés par un calcul redondant des fermetures. De son côté, \textsc{Titanic} est handicapé au fur et à mesure que la valeur de \textit{minsupp} diminue par le nombre de recherches des items nécessaires au calcul des fermetures. En effet, pour \textit{minsupp} = {\xmplbx0,10}\ \%, {\xmplbx116} items sont fréquents parmi {\xmplbx119} possibles alors que la taille maximale d'un générateur minimal fréquent est égale à {\xmplbx10} items seulement.

\begin{table}[h!]
\begin{center}
  \parbox{16cm}{\hspace{0.2cm}
  \scriptsize{
  \begin{tabular}{|l||r||r|r|r|r||r|r|r|}
   \hline
    Contexte & \textit{minsupp} \textsc{(}\%\textsc{)} & \textsc{Prince}& \textsc{Close} & \textsc{A-Close} & \textsc{Titanic}& $\frac{\displaystyle \textsc{Close}}{\displaystyle \textsc{Prince}}$ & $\frac{\displaystyle \textsc{A-Close}}{\displaystyle \textsc{Prince}}$ & $\frac{\displaystyle \textsc{Titanic}}{\displaystyle \textsc{Prince}}$ \\
&&\textsc{(}sec.\textsc{)}&\textsc{(}sec.\textsc{)}&\textsc{(}sec.\textsc{)}&\textsc{(}sec.\textsc{)}&&&\\\hline
\hline \textsc{Pumsb}
&95,00&      3&      74&     172&      12    &    24,67 &    57,33   &4,00\\
&90,00&      7&      679&    1 914&     36    &    97,00   &      273,43 &5,14\\
&85,00&      27&     4 332&   10 875&    192   &    \textbf{160,44} &      \textbf{402,78} &\textbf{7,11}\\
&80,00&      416&    19 848&  47 167&    1 233  &    47,71   &    113,38   &2,96\\
&75,00&      1 641&   65 690& 152 050&    4 563  &     40,03  &   92,66    &2,78\\

\hline\textsc{Connect}
&90,00 &     8      &853    &1 652   &23&106,62&206,50&2,87\\
&80,00 &     25     &3 428   &5 771   &135&\textbf{137,12}&\textbf{230,84}&5,40\\
&70,00 &     66    &7 358   &12 028  &452&111,48&182,24&\textbf{6,85}\\
&60,00 &     332    &13 024  &20 747  &1 205&39,23&62,49&3,63\\
&50,00 &     847    &22 164 &35 903  &/&26,17&42,39&/\\

\hline \textsc{Chess}
&90,00       &1         &6    &15     &1  &6,00&15,00&1,00    \\
&80,00       &1         &49   &122    &3  &49,00&122,00&3,00    \\
&70,00       &3        &217  &481    &25   &\textbf{72,33}&\textbf{160,33}&8,33  \\
&60,00       &39       &784  &1 896   &233   &20,10&48,61&5,97 \\
&50,00       &197       &2 560 &5 451   &1 520 &12,99&27,67&7,71  \\
&45,00       &435      &4 550 &9 719   &4 237  &10,46&22,34&\textbf{9,74} \\

\hline\textsc{Mushroom}
&20,00&  1 &      14 &  28&   2 &14,00&28,00&2,00\\
&10,00&  1 &      32&   69&   5&\textbf{32,00}&\textbf{69,00}&5,00\\
&5,00&   3 &      59&   121&  21&19,67&40,33&7,00\\
&3,00&   4&      83 &  163&  46&20,75&40,75&11,50\\
& 2,00 & 7 &     98&   197&  81&14,00&28,14&11,57\\
& 1,00 & 13&       123&  250&  180&9,46&19,23&13,85\\
&0,80 &17&      132&  270 & 246&7,76&15,88&14,47\\
& 0,30 &33 &   154&  322 & 3 127&4,67&9,76&94,76\\
& 0,20 &40 &   159&  336 & 10 518&3,97&8,40&262,95\\
& 0,10 &50&    168&  352 & 26 877&3,36&7,04&\textbf{537,54}\\
\hline
\end{tabular}
  }}
\caption{Tableau comparatif des temps d'exécution de \textsc{Prince} \textit{versus} \textsc{A-Close}, \textsc{Close} et \textsc{Titanic} pour les contextes denses.}\label{ratiosdense}
\end{center}
\end{table}

L'algorithme \textsc{Prince} s'avère performant sur les contextes denses et pour toutes les valeurs de \textit{minsupp}. La différence entre les performances de \textsc{Prince} et celles de \textsc{Close} et \textsc{A-Close} atteint son maximum pour le contexte \textsc{Pumsb}. En effet, \textsc{Prince} est environ {\xmplbx160} \textsc{(}resp. {\xmplbx403}\textsc{)} fois plus rapide que \textsc{Close} \textsc{(}resp. \textsc{A-Close}\textsc{)} pour un support de {\xmplbx85}\ \%. De même, \textsc{Prince} est environ {\xmplbx538} fois plus
rapide que \textsc{Titanic} pour le contexte \textsc{Mushroom} et pour un support égal à {\xmplbx0,01}\ \%. Pour ces contextes denses, \textsc{Prince} est en moyenne {\xmplbx42} \textsc{(}resp. {\xmplbx89} et {\xmplbx41}\textsc{)} fois plus rapide que \textsc{Close} \textsc{(}resp. \textsc{A-Close} et \textsc{Titanic}\textsc{)}.

\subsubsection{Expérimentations sur les contextes épars}

\begin{figure}[h!]
\begin{center}
\parbox{7cm}{\includegraphics[scale = 0.9]{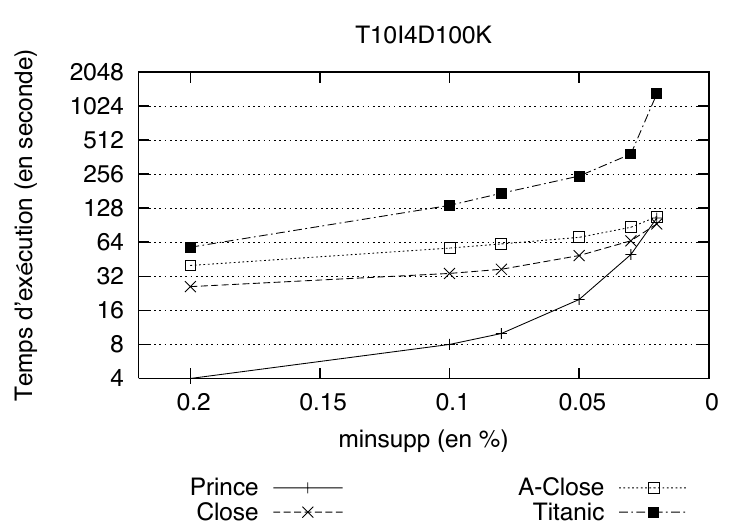}}
\parbox{7cm}{\includegraphics[scale = 0.9]{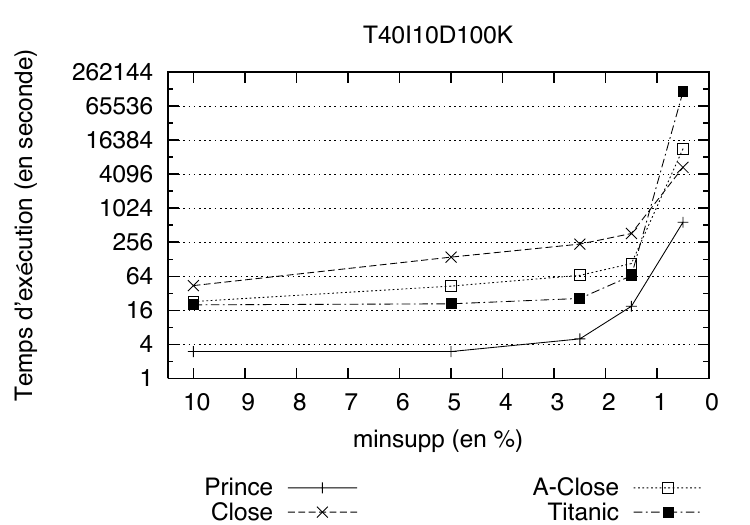}}\\
\parbox{7cm}{\includegraphics[scale = 0.9]{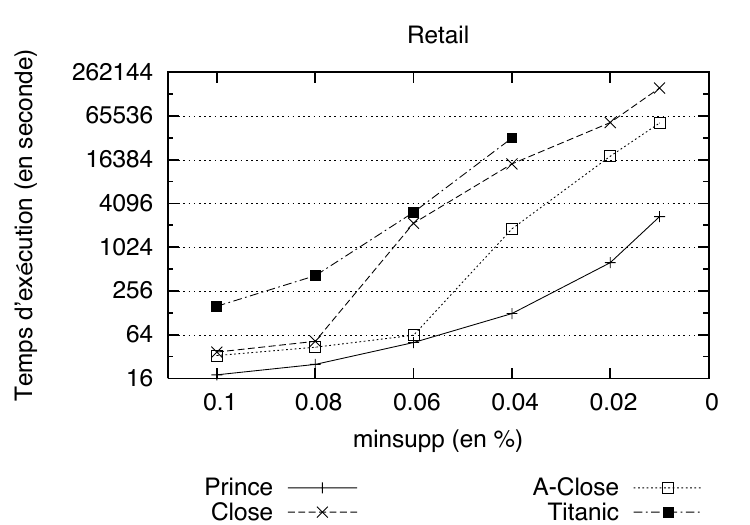}}
\parbox{7cm}{\includegraphics[scale = 0.9]{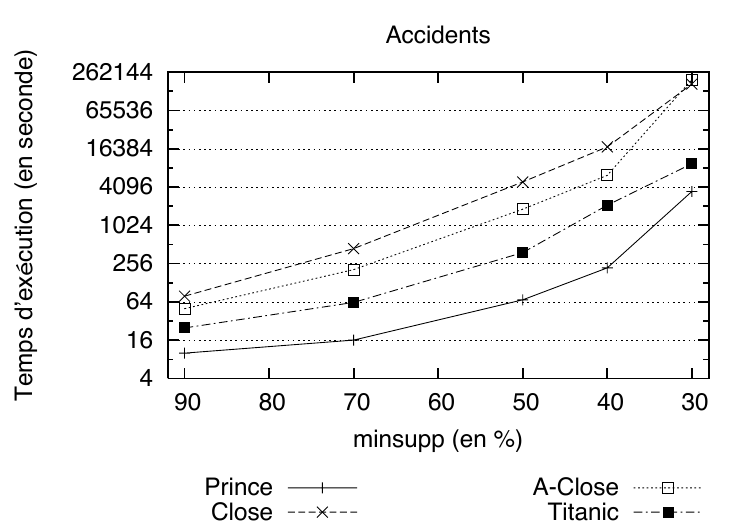}}
\end{center}
\caption{Les performances de \textsc{Prince}
\textit{versus} \textsc{Close}, \textsc{A-Close} et \textsc{Titanic}
pour les contextes épars.} \label{fig_tps_exec_sparses}
\end{figure}

\noindent Les temps d'exécution de l'algorithme \textsc{Prince} comparés respectivement aux algorithmes \textsc{Close}, \textsc{A-Close} et \textsc{Titanic} sur les contextes épars sont présentés par la Figure \ref{fig_tps_exec_sparses} et le Tableau \ref{ratiossparse}.

- \textsc{\textbf{T10I4D100K}} : pour cette base, \textsc{Prince} fait mieux que \textsc{A-Close} et \textsc{Titanic} pour toutes les valeurs de \textit{minsupp}. En comparant les performances de \textsc{Prince} à celles de \textsc{Close}, \textsc{Prince} fait mieux que cet algorithme pour des valeurs de \textit{minsupp}
supérieures ou égales à {\xmplbx0,03}\ \%. Alors que c'est l'inverse pour des supports inférieurs à
{\xmplbx0,03}\ \%. Ceci peut être expliqué par le fait que \textsc{Prince} est pénalisé par le coût de la
construction du treillis d'Iceberg pour de très basses valeurs de \textit{minsupp}. Cependant, \textsc{Close} est avantagé par une taille moyenne des objets relativement faible \textsc{(}{\xmplbx10} items\textsc{)}. \textsc{A-Close} se voit moins rapide en raison du balayage supplémentaire de l'ensemble des générateurs minimaux fréquents de taille \textsc{(}$k$ - $1$\textsc{)} effectué pour chaque candidat $g$ de taille $k$ afin de savoir si $g$ est minimal ou non. Au fur et à mesure que la valeur de \textit{minsupp} diminue, les performances de \textsc{Titanic} se dégradent considérablement. En effet, pour \textit{minsupp} = {\xmplbx0,02}\ \%, {\xmplbx859} items sont fréquents parmi {\xmplbx1 000} possibles alors que la taille maximale d'un générateur minimal fréquent est égale à {\xmplbx10} items seulement.

- \textsc{\textbf{T40I10D100K}} : les performances de \textsc{Prince} dans ce contexte sont largement meilleures que celles de \textsc{Close}, \textsc{A-Close} et \textsc{Titanic} quelle que soit la valeur de \textit{minsupp}. \textsc{Close} et \textsc{A-Close} sont handicapés par une taille moyenne élevée des objets
\textsc{(}{\xmplbx40} items\textsc{)} sachant que \textsc{Close} prend le dessus sur \textsc{A-Close}. De même, les
performances de \textsc{Titanic} se dégradent d'une manière considérable pour les mêmes raisons
évoquées auparavant. Le coût des comparaisons, dans le cas de \textsc{Prince}, est nettement plus réduit que le coût du calcul des intersections \textsc{(}resp. des tentatives d'extension\textsc{)} pour calculer les fermetures dans le cas de \textsc{Close} et \textsc{A-Close} \textsc{(}resp. \textsc{Titanic}\textsc{)}. Ceci explique l'écart dans les performances entre \textsc{Prince} et les trois autres algorithmes. Il est aussi à noter l'augmentation surprenante du nombre de classes d'équivalence en passant de {\xmplbx1,50}\ \% à {\xmplbx0,50}\ \% comme valeurs de \textit{minsupp}. En effet, nous passons de {\xmplbx6 540} à {\xmplbx1 275 940} classes d'équivalence, c'est-à-dire plus de {\xmplbx195} fois. Les performances de \textsc{Prince} demeurent intéressantes même pour un tel nombre de classes d'équivalence.

- \textsc{\textbf{Retail}} : pour cette base, notre algorithme réalise des temps d'exécution beaucoup moins importants que ceux réalisés respectivement par \textsc{Close}, \textsc{A-Close} et \textsc{Titanic}. Les performances réalisées peuvent être expliquées par l'influence énorme du nombre élevé d'items dans \textsc{Retail}. En effet, \textsc{Close} est handicapé par un nombre énorme de candidats pour lesquels il est obligé de calculer la fermeture alors qu'un grand nombre d'entre eux s'avèrera non fréquent. Le nombre de candidats affecte aussi les performances de \textsc{A-Close} à cause des balayages supplémentaires ainsi que le coût du calcul des fermetures. De son côté, \textsc{Titanic} est
considérablement alourdi par un grand nombre d'items fréquents à considérer lors du calcul de la fermeture d'un générateur minimal fréquent \textsc{(}pour \textit{minsupp} = {\xmplbx0,04}\ \%, {\xmplbx4 463} items sont fréquents alors que la taille maximale d'un générateur minimal fréquent est égale à {\xmplbx6} items seulement\textsc{)} seulement. L'exécution de \textsc{Titanic}, pour des supports inférieurs à {\xmplbx0,04}\ \%, s'arrête pour manque d'espace mémoire.

- \textsc{\textbf{Accidents}} : dans cette base, les performances de \textsc{Prince} sont meilleures que celles de \textsc{Close}, \textsc{A-Close} et \textsc{Titanic} pour toutes les valeurs de \textit{minsupp}. Notons que pour des valeurs de \textit{minsupp} supérieures ou égales à {\xmplbx40}\ \%, chaque générateur minimal fréquent est un fermé. \textsc{Prince} ne nécessite alors pas l'exécution de sa procédure \textsc{Gen-Ordre} étant donné que le treillis d'Iceberg est déjà construit à la fin de sa première étape.
Les performances de \textsc{Close} et \textsc{A-Close} sont énormément désavantagées par un nombre ainsi qu'une taille moyenne des objets relativement élevés, ce qui induit des coûts énormes pour le calcul des fermetures. Le mécanisme de comptage par inférence adopté par \textsc{Titanic} s'avère plus efficace que le calcul des intersections. Ceci peut être expliqué par le nombre réduit d'items fréquents à prendre en considération lors du calcul des fermetures. En effet, seuls {\xmplbx32} items parmi {\xmplbx468} sont fréquents pour \textit{minsupp} égal à {\xmplbx30}\ \% et la taille maximale d'un générateur minimal fréquent est égale à {\xmplbx12} items.

\begin{table}[h!]
\begin{center}
  \parbox{16cm}{\hspace{0.1cm}
    \scriptsize{
  \begin{tabular}{|l||r||r|r|r|r||r|r|r|}
   \hline
    Contexte & \textit{minsupp} \textsc{(}\%\textsc{)} & \textsc{Prince}& \textsc{Close} & \textsc{A-Close} & \textsc{Titanic}& $\frac{\displaystyle \textsc{Close}}{\displaystyle \textsc{Prince}}$ & $\frac{\displaystyle \textsc{A-Close}}{\displaystyle \textsc{Prince}}$ & $\frac{\displaystyle \textsc{Titanic}}{\displaystyle \textsc{Prince}}$ \\
&&\textsc{(}sec.\textsc{)}&\textsc{(}sec.\textsc{)}&\textsc{(}sec.\textsc{)}&\textsc{(}sec.\textsc{)}&&&\\\hline
\hline \textsc{T10I4D100K}
&0,50&    3          &17  &9    & 7 &5,67&3,00&2,33\\
&0,20      &4         &26  &40   &58&\textbf{6,50}&\textbf{10,00}&14,50\\
&0,10      &8       &34  &57   &136&4,25&7,12&17,00\\
&0,08     &10       &37  &62   &174&3,70&6,20&\textbf{17,40}\\
& 0,05    &20       &49  &71   &247&2,45&3,55&12,35\\
& 0,03    &50      &66  &87   &387&1,32&1,74&7,74\\
& 0,02    &105  &93  &108  &1 316&0,88&1,03&12,53\\

\hline\textsc{T40I10D100K}
& 10,00 &3       &44   &23    &20 &14,67&7,67&6,67\\
& 5,00 &3        &140  &43    &21&46,67&14,33&7,00\\
&2,50  &5      &238  &67    &26&\textbf{47,60}&13,40&5,20\\
& 1,50 &19     &366  &108   &66&19,26&5,68&3,47\\
& 0,50 &582   &5 420 &11 564 &117 636&9,31&\textbf{19,87}&\textbf{202,12}\\

\hline\textsc{Retail}
& 0,10     &18&       37&      33&      158&2,05&1,83&8,78\\
&0,08     &25&       52&      43&      415&2,08&1,72&16,60\\
& 0,06    &50 &      2 185&    63&      3 089&43,70&1,26&61,78\\
& 0,04    &125&     14 358&    1 833&    32 663&\textbf{114,86}&14,66&\textbf{261,30}\\
&0,02     &626&    53 208&    18 269&   /&85,00&\textbf{29,18}&/\\
&0,01     &2 699&   159 217&  52 162&    /&58,99&19,33&/\\

\hline \textsc{Accidents}
& 90,00  &10         &79   &50    &25&7,90&5,00&2,50\\
&70,00  & 16         &440  &206   &63&27,50&12,87&3,94\\
& 50,00 &69          &4 918 &1 839  &381&71,27&26,65&5,52\\
&40,00  &219      &17 528    &6 253  &2 120&\textbf{80,04}&\textbf{28,55}&\textbf{9,68}\\
&30,00  &3 482     &170 540&  199 980 &9 530&48,98&57,43&2,74\\

\hline  \end{tabular}
  }}
\caption{Tableau comparatif des temps d'exécution de \textsc{Prince} \textit{versus} \textsc{A-Close}, \textsc{Close} et \textsc{Titanic} pour les contextes épars.}
\label{ratiossparse}
\end{center}
\end{table}

L'algorithme \textsc{Prince} s'avère aussi performant sur les contextes épars et pour toutes les valeurs de \textit{minsupp}. La différence entre les performances de \textsc{Prince} et celles de \textsc{Close} et \textsc{Titanic} atteint son maximum pour le contexte \textsc{Retail}. En effet, \textsc{Prince} est
environ {\xmplbx115} \textsc{(}resp. {\xmplbx261}\textsc{)} fois plus rapide que \textsc{Close} \textsc{(}resp. \textsc{Titanic}\textsc{)} pour un support de {\xmplbx0,04}\ \%. De même, \textsc{Prince} est environ {\xmplbx57} fois plus rapide que \textsc{A-Close} pour le contexte \textsc{Accidents} et pour une valeur de \textit{minsupp} égale à {\xmplbx30}\ \%. Pour ces contextes épars, \textsc{Prince} est en moyenne {\xmplbx31} \textsc{(}resp. {\xmplbx13} et {\xmplbx32}\textsc{)} fois plus rapide que \textsc{Close} \textsc{(}resp. \textsc{A-Close} et \textsc{Titanic}\textsc{)}.

\subsubsection{Expérimentations sur les contextes \guillemotleft~pire des cas \guillemotright}

\noindent Pour toutes ces expérimentations, la valeur de \textit{minsupp} est fixée à {\xmplbx1}. Nous avons testé {\xmplbx26} contextes \guillemotleft~pire des cas \guillemotright \ représentant la variation de la valeur de $n$ de {\xmplbx1} à {\xmplbx26}. \'{E}tant donné que ces contextes représentent le pire des cas, tout motif candidat est un générateur minimal fréquent et est égal à sa fermeture.

\vspace{2mm}

Les temps d'exécution des quatre algorithmes ne commencent à être distinguables qu'à partir d'une valeur de $n$ égale à {\xmplbx15}. Pour les contextes \guillemotleft~pire des cas \guillemotright, les performances de \textsc{Prince} restent meilleures que celles de \textsc{Close}, \textsc{A-Close} et \textsc{Titanic}. Chaque générateur minimal fréquent est aussi un fermé. Ainsi, \textsc{Prince} n'exécute pas sa deuxième étape, le treillis d'Iceberg étant construit dès la fin de sa première étape. De son côté, \textsc{A-Close} ne calcule pas les fermetures. D'où, ses performances sont meilleures que celles
de \textsc{Close} et celles de \textsc{Titanic}. Il est important de noter que les calculs de fermetures effectués par \textsc{Close} et \textsc{Titanic} s'avèrent sans utilité car tout générateur minimal est égal à sa fermeture. L'exécution de ces deux algorithmes n'a pu se terminer pour une valeur de $n$ égale à {\xmplbx24}, en raison du manque d'espace mémoire. Pour la même raison, l'exécution de \textsc{A-Close} n'a pu se terminer pour une valeur de $n$ égale à {\xmplbx25}. L'exécution de notre algorithme prend aussi fin, pour $n$ = {\xmplbx26}.

\vspace{2mm}

\begin{figure}[h!]
\begin{center}
\parbox{7cm}{\includegraphics[scale = 0.9]{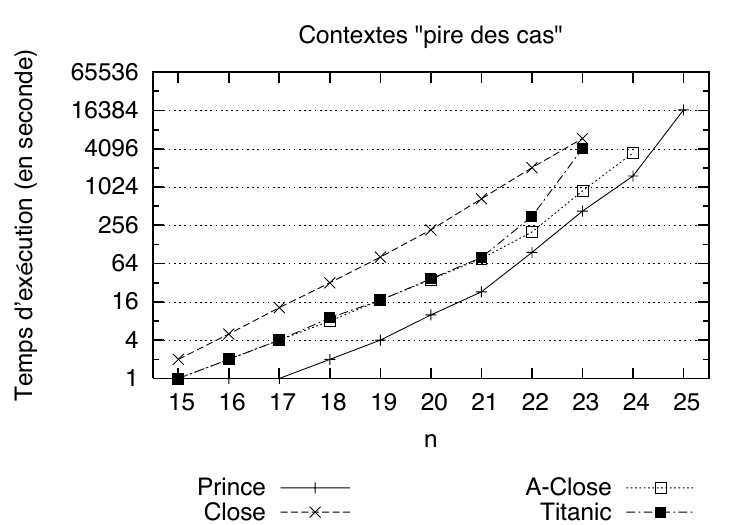}}
\caption{Test de mise à l'échelle de \textsc{Prince} \textit{versus} \textsc{Close}, \textsc{A-Close} et
\textsc{Titanic} pour les contextes \guillemotleft~pire des cas \guillemotright}
\label{fig_tps_pire_des cas}
\end{center}
\end{figure}

La Figure \ref{fig_tps_pire_des cas} et le Tableau \ref{ratiospirecas} donnent le comportement des algorithmes suivant la variation de la valeur de $n$. Dans le Tableau \ref{ratiospirecas}, les trois dernières colonnes indiquent le facteur multiplicatif des temps d'exécution de \textsc{Close}, \textsc{A-Close} et \textsc{Titanic} par rapport à l'algorithme \textsc{Prince}. L'algorithme \textsc{Prince} reste meilleur que les algorithmes \textsc{Close}, \textsc{A-Close} et \textsc{Titanic} sur les contextes \guillemotleft~pire des cas \guillemotright. La différence entre les performances de \textsc{Prince} et celles de \textsc{Close} \textsc{(}resp. \textsc{A-Close} et \textsc{Titanic}\textsc{)} est maximale pour $n$ = {\xmplbx21} \textsc{(}resp. {\xmplbx19} et {\xmplbx23}\textsc{)}. En effet, pour ces valeurs de $n$, \textsc{Prince} est environ {\xmplbx29} \textsc{(}resp. {\xmplbx4} et {\xmplbx10}\textsc{)} fois plus rapide que \textsc{Close} \textsc{(}resp. \textsc{A-Close} et \textsc{Titanic}\textsc{)}. Pour ces contextes \guillemotleft~pire des cas \guillemotright, \textsc{Prince} est en moyenne {\xmplbx16} \textsc{(}resp. {\xmplbx3} et {\xmplbx4}\textsc{)} fois plus rapide que \textsc{Close} \textsc{(}resp.
\textsc{A-Close} et \textsc{Titanic}.

\begin{table}[h!]
  \begin{center}
    \scriptsize{
  \begin{tabular}{|l||r|r|r|r||r|r|r|}
   \hline
    $n$ & \textsc{Prince}& \textsc{Close} & \textsc{A-Close} & \textsc{Titanic}& $\frac{\displaystyle \textsc{Close}}{\displaystyle \textsc{Prince}}$ & $\frac{\displaystyle \textsc{A-Close}}{\displaystyle \textsc{Prince}}$ & $\frac{\displaystyle \textsc{Titanic}}{\displaystyle \textsc{Prince}}$ \\
&\textsc{(}sec.\textsc{)}&\textsc{(}sec.\textsc{)}&\textsc{(}sec.\textsc{)}&\textsc{(}sec.\textsc{)}&&&\\
\hline
\hline 15&  1&      2&      1&      1 &2,00&1,00&1,00\\
\hline 16&  1&      5&      2&      2 &5,00&2,00&2,00\\
\hline 17&  1&      13&     4&      4 &13,00&4,00&4,00\\
\hline 18&  2&      32&     8&      9 &16,00&4,00&4,50\\
\hline 19&  4&      81&     17&     17 &20,25&\textbf{4,25}&4,25\\
\hline 20&  10&     217&    36&     37 &21,70&3,60&3,70\\
\hline 21&  23&     670&    77&     80 &\textbf{29,13}&3,35&3,48\\
\hline 22&  96&     2 058&   200&    353 &21,44&2,08&3,68\\
\hline 23&  429&    5 932&   890&    4 129 &13,83&2,10&\textbf{9,62}\\
\hline 24&  1 525&   /&      3 519&   / &/&2,31&/\\
\hline 25&  16 791&   /&      /&   / &/&/&/\\
\hline 26&  /&   /&      /&   / &/&/&/\\
\hline
\end{tabular}
  }
\caption{Tableau comparatif des temps d'exécution de \textsc{Prince} \textit{versus} \textsc{A-Close}, \textsc{Close} et \textsc{Titanic} pour les contextes \guillemotleft~pire des cas \guillemotright.}\label{ratiospirecas}
\end{center}
\end{table}

\subsection{\sc comparaisons des étapes de \textsc{Prince}}

\noindent Le but de cette sous-section est de comparer le coût des trois étapes constituant \textsc{Prince}. \`{A} titre indicatif, le Tableau \ref{princeetape} montre le coût des différentes étapes sur deux contextes denses et deux contextes épars. Suite à nos expérimentations, nous pouvons noter les
points suivants :

\begin{itemize}
  \item le coût de la deuxième étape est généralement plus réduit que le coût de la première étape. Toutefois, ceci dépend étroitement des caractéristiques du contexte. En effet, le coût du calcul du support des motifs \textsc{(}c'est-à-dire l'opération la plus coûteuse de la première étape\textsc{)} dépend du nombre d'objets que contient le contexte et de leur taille moyenne.

  \item la troisième étape est la moins coûteuse et ceci pour les différents contextes testés ainsi que pour les différentes valeurs de \textit{minsupp}. En effet, du moment que la structure contenant les générateurs minimaux fréquents a été construite, la dérivation des motifs fermés fréquents et des règles d'association génériques devient immédiate. Notons aussi que la variation de la valeur de \textit{minconf} n'a pas d'influence significative sur le temps d'exécution de \textsc{Prince}.
\end{itemize}

\vspace{2mm}

Les résultats obtenus confirment ainsi l'utilité de l'approche proposée.

\begin{table}[h!]
  \centering
     \scriptsize{
  \begin{tabular}{|l||r||r||r|r|r|}
    \hline
Contexte&\textit{minsupp} \textsc{(}\%\textsc{)}&Temps
d'exécution&1$^{i\grave{e}re}$ étape&2$^{i\grave{e}me}$
étape& 3$^{i\grave{e}me}$ étape\\&&total
\textsc{(}sec.\textsc{)}&&&\\\hline \hline\textsc{Connect}
&90,00 & 8   &8&0&0 \\
&80,00 & 25  &24&1&0  \\
&70,00 & 66  &63&3&0  \\
&60,00 & 332 &322&9&1   \\
&50,00 & 847 &822&22&3   \\

\hline\textsc{Mushroom}
&20,00&1& 1&0&0 \\
&10,00& 1 &1&0&0\\
&5,00&3  & 2&1&0\\
&3,00& 4 &2&2 &0\\
& 2,00 & 7 &3&3&1\\
& 1,00 & 13 &5&7&1\\
&0,80 & 17 &6&10&1\\
& 0,30 & 33 &9&22&2\\
& 0,20 & 40 &10&26&4\\
& 0,10 & 50 &11&35&4\\
\hline\hline

\textsc{T10I4D100K}
& 0,50 &3 &3&0&0\\
&0,20   &4 &4&0&0\\
&0,10   &8 &7&1&0\\
&0,08  &10 &9&1&0\\
& 0,05 &20 &16&4&0\\
& 0,03 &50 &35&15&0\\
& 0,02 &105 &58&47&0\\

\hline \textsc{T40I10D100K}
& 10,00 &3&3&0&0\\
& 5,00 &3&3&0&0\\
&2,50  &5&5&0&0\\
& 1,50 &19&18&0&1\\
& 0,50 &582&480&86&16\\
\hline

\end{tabular}
  }
\caption{Tableau comparatif des étapes de \textsc{Prince} pour les contextes denses \textsc{Connect} et \textsc{Mushroom} et les contextes épars \textsc{T10I4D100K} et \textsc{T40I10D100K}.}
\label{princeetape}
\end{table}

\section{CONCLUSION}

\noindent Dans ce papier, nous avons proposé un nouvel algorithme, \textsc{Prince}, permettant de
réduire le nombre de règles d'association en n'extrayant que des bases génériques composées seulement de règles informatives. \textsc{Prince} est le seul algorithme à offrir les composantes nécessaires à une telle extraction sans avoir à l'associer avec un autre algorithme. \`{A} cet effet, \textsc{Prince} extrait en premier lieu l'ensemble des générateurs minimaux fréquents et en déduit le treillis d'Iceberg. Les règles d'association informatives sont alors dérivés par un simple parcours ascendant de la structure partiellement ordonnée. Grâce à une gestion efficace des classes d'équivalence, \textsc{Prince} présente aussi comme avantages l'extraction non-redondante des motifs fermés fréquents ainsi qu'une réduction du coût de la construction du treillis d'Iceberg. Les expérimentations, que nous avons réalisées, ont montré qu'il est efficace sur les différents contextes testés couvrant les différents types de contextes. Ainsi, \textsc{Prince} offre un consensus entre la qualité de l'information extraite et l'efficacité d'extraction.

\vspace{2mm}

Les perspectives de travaux futurs concernent les pistes suivantes :

- Du point de vue de la qualité des connaissances extraites par \textsc{Prince}, l'adaptation de ce dernier pour l'exploration de l'espace disjonctif de recherche, où les items sont liés par l'opérateur de disjonction au lieu de l'opérateur de conjonction [Hamrouni {\em \& al.}, 2009], est une piste
intéressante afin d'extraire différentes formes de règles généralisées [Hamrouni {\em \& al.}, 2010]. Ces dernières sont connues pour être plus riches en information étant donné qu'elles véhiculent différents liens entre les items et ne se limitent donc pas à leur conjonction. Les métriques de support et de confiance des différentes formes sont facilement calculables puisque les supports conjonctifs et négatifs peuvent être déduits à partir des supports disjonctifs grâce aux identités d'inclusion-exclusion [Galambos, Simonelli, 2000] et à la loi de De Morgan, respectivement. Par ailleurs, l'utilisation de contraintes
utilisateur [Bonchi, Lucchese, 2006] ainsi que d'autres mesures de qualité [H\'ebert, Cr\'emilleux, 2007] dans le processus d'extraction des règles d'association informatives est une tâche importante. En effet, ceci permettra de réduire tant que faire se peut le nombre de règles tout en gardant celles qui sont les plus intéressantes pour l'utilisateur.

- Du point de vue de algorithmique, il serait d'une part intéressant d'étudier la possibilité d'intégrer
dans \textsc{Prince} le travail proposé dans [Calders, Goethals, 2007]. Ce travail s'applique à tout ensemble vérifiant la propriété d'idéal d'ordre, ce qui est le cas de l'ensemble des générateurs minimaux utilisé dans notre cas. Il permet de réduire le nombre de candidats auxquels un accès au contexte d'extraction est nécessaire pour calculer le support. \'{E}tant donné que cette étape est
la plus coûteuse de la première étape, une telle intégration permettra de réduire un tel coût. D'autre
part, la mise en place d'un choix adaptatif de la bordure à ajouter à l'ensemble des générateurs minimaux fréquents est envisagée. En effet, pour certains contextes, il est plus avantageux du point de vue cardinalité de maintenir une autre bordure que celle actuellement maintenue, à savoir $\mathcal{GB}$d$^{-}$ [Liu {\em \& al.}, 2008]. Ceci dépend étroitement des caractéristiques des contextes [Hamrouni {\em \& al.}, \`a para\^{i}tre] et qui seront donc à étudier pour bien sélectionner la bordure à maintenir.

\vspace{4mm}

\noindent {\footnotesize {\em Remerciements.} Nous tenons à remercier les relecteurs anonymes pour leurs remarques et leurs suggestions très utiles et qui ont permis d'améliorer considérablement la qualité du travail. Nous tenons aussi à remercier Yves Bastide pour avoir mis à notre disposition les codes source des algorithmes \textsc{Close}, \textsc{A-Close} et \textsc{Titanic}. Ce travail est partiellement soutenu par le projet \textsc{PHC-Utique 11G1417}, \textsc{EXQUI}}.

\vspace{0.2cm}

\begin{center}
{\small BIBLIOGRAPHIE}
\end{center}

\vspace{0.2cm}

{\small

\noindent {\sc agrawal r., imielinski t., swami a.} \textsc{(}1993\textsc{)},  ``Mining association rules between sets of items in large databases'',  \textit{Proceedings of the ACM-SIGMOD International Conference on Management of Data}, Washington \textsc{(}DC\textsc{)}, p. 207--216.

\vspace{1mm}

\noindent {\sc agrawal r., srikant r.} \textsc{(}1994\textsc{)}, ``Fast algorithms for mining association rules'', \textit{Proceedings of the 20th International Conference on Very Large Databases} \textsc{(}VLDB 1994\textsc{)}, Santiago, Chile, p.~ 478-499.

\vspace{1mm}

\noindent {\sc ashrafi m.z., taniar d., smith k.} \textsc{(}2007\textsc{)}, ``Redundant association rules reduction techniques'', {\em International Journal Business Intelligence and Data Mining} 2\textsc{(}1\textsc{)}, p. 29-63.

\vspace{1mm}

\noindent {\sc barbut m., monjardet b.} \ \textsc{(}1970\textsc{)}, \ {\em Ordre et classification. Alg\`ebre et combinatoire}, \ Paris, \ Hachette, tome II.

\vspace{1mm}

\noindent {\sc bastide y.} \textsc{(}2000\textsc{)}, {\em Data mining : Algorithmes par niveau, techniques d'implantation et   applications}, Thèse de doctorat, \'{E}cole Doctorale Sciences pour l'Ingénieur de Clermont-Ferrand, Université Blaise Pascal.

\vspace{1mm}

\noindent {\sc bastide y., pasquier n., taouil r., stumme g., lakhal l.} \textsc{(}2000\textsc{(}a\textsc{)}\textsc{)}, ``Mining minimal non-redundant association rules using frequent closed itemsets'', {\em Proceedings of the 1st International Conference on Computational Logic} \textsc{(}DOOD 2000\textsc{)}, Springer-Verlag, LNAI, volume 1861, London \textsc{(}UK\textsc{)}, p. 972-986.

\vspace{1mm}

\noindent {\sc bastide y., taouil r., pasquier n., stumme g., lakhal l.} \textsc{(}2000\textsc{(}b\textsc{)}\textsc{)}, ``Mining frequent patterns with counting inference'', {\em ACM-SIGKDD Explorations} 2\textsc{(}2\textsc{)}, p. 66-75.

\vspace{1mm}

\noindent {\sc bayardo r.j.} \textsc{(}1998\textsc{)}, ``Efficiently mining long patterns from databases'', \textit{Proceedings of the International Conference on Management of Data} \textsc{(}SIGMOD 1998\textsc{)}, Seattle \textsc{(}WA\textsc{)}, p. 85-93.

\vspace{1mm}

\noindent {\sc ben yahia s., mephu nguifo e.} \textsc{(}2004\textsc{)}, ``Revisiting generic bases of association rules'', {\em Proceedings of 6th International Conference on Data Warehousing and Knowledge Discovery} \textsc{(}DaWaK 2004\textsc{)}, Springer-Verlag, LNCS, volume 3181, Zaragoza, Spain, p. 58-67.

\vspace{1mm}

\noindent {\sc ben yahia s., hamrouni t., mephu nguifo e.} \textsc{(}2006\textsc{)}, ``Frequent closed itemset based algorithms: A thorough structural and analytical survey'', {\em ACM-SIGKDD Explorations}, 8\textsc{(}1\textsc{)}, p.~93-104.

\vspace{1mm}

\noindent {\sc bodon f., r\'oyai l.} \textsc{(}2003\textsc{)}, ``Trie: An alternative data structure for data mining
  algorithms'', \textit{Mathematical and Computer Modelling}, 38\textsc{(}7-9\textsc{)}, p. 739-751.

\vspace{1mm}

\noindent {\sc bonchi f., lucchese c.} \textsc{(}2006\textsc{)}, ``On condensed representations of constrained frequent patterns'',  \textit{Knowledge and Information Systems} 9\textsc{(}2\textsc{)}, p. 180-201.

\vspace{1mm}

\noindent {\sc boulicaut j.-f., bykowski a., rigotti c.} \textsc{(}2003\textsc{)}, ``Free-sets: A condensed representation of \textsc{B}oolean data for the approximation of frequency queries'', \textit{Data Mining and Knowledge Discovery} 7\textsc{(}1\textsc{)}, p. 5-22.

\vspace{1mm}

\noindent {\sc calders t., rigotti c., boulicaut j.-f.} \textsc{(}2005\textsc{)}, ``A survey on condensed representations for frequent sets'', {\em Constraint Based Mining and Inductive Databases}, Springer-Verlag, LNAI, volume 3848, p. 64-80

\vspace{1mm}

\noindent {\sc calders t., goethals b.} \textsc{(}2007\textsc{)}, ``Non-derivable itemset mining'', {\em Data Mining and Knowledge Discovery}, Springer}, 14\textsc{(}1\textsc{)}, p. 171-206

\vspace{1mm}

\noindent {\sc ceglar a., roddick j.f.} \textsc{(}2006\textsc{)}, ``Association mining'', {\em ACM Computing Surveys} 38\textsc{(}2\textsc{)}.

\vspace{1mm}

\noindent {\sc le floc'h a., fisette c., missaoui r., valtchev p., godin r.} \textsc{(}2003\textsc{)}, \guillemotleft~\textsc{JEN} : un algorithme efficace de construction de générateurs pour l'identification des règles d'association \guillemotright, {\em Numéro spécial de la revue des Nouvelles Technologies de l'Information} 1\textsc{(}1\textsc{)}, p. 135-146.

\vspace{1mm}

\noindent {\sc galambos j., simonelli i.} \textsc{(}2000\textsc{)}, {\em Bonferroni-type inequalities with applications}, Springer.

\vspace{1mm}

\noindent {\sc ganter b., wille r.} \textsc{(}1999\textsc{)}, {\em Formal Concept Analysis}, Springer.

\vspace{1mm}

\noindent {\sc godin r., missaoui r., alaoui h.} \textsc{(}1995\textsc{)}, `` Incremental concept formation algorithms based on Galois \textsc{(}concept\textsc{)} lattices.'', \textit{Journal of Computational Intelligence} 11\textsc{(}2\textsc{)}, p. 246-267.

\vspace{1mm}

\noindent {\sc goethals b.} \textsc{(}2004\textsc{)}, ``Frequent itemset mining implementations repository'',\\
http://fimi.cs.helsinki.fi/.

\vspace{1mm}

\noindent {\sc hamrouni t., ben yahia s., mephu nguifo e.} \textsc{(}2009\textsc{)}, ``Sweeping the disjunctive search space towards mining new exact concise representations of frequent itemsets'', {\em Data \& Knowledge Engineering} 68\textsc{(}10\textsc{)}, p. 1091-1111.

\vspace{1mm}

\noindent {\sc hamrouni t., ben yahia s., mephu nguifo e.} \textsc{(}2010\textsc{)}, ``Generalization of association rules through disjunction'', {\em Annals of Mathematics and Artificial Intelligence} 59\textsc{(}2\textsc{)}, p.~201-222.

\vspace{1mm}

\noindent {\sc hamrouni t., ben yahia s., mephu nguifo e.} \textsc{(}\`a para\^{i}tre\textsc{)}, ``Looking for a structural characterization of the sparseness measure of \textsc{(}frequent closed\textsc{)} itemset contexts'', {\em \`A paraître dans Information Sciences}.

\vspace{1mm}

\noindent {\sc h\'ebert c., cr\'emilleux b.} \textsc{(}2007\textsc{)}, ``A unified view of objective interestingness measures'', {\em Proceedings of the 5th International Conference Machine Learning and Data Mining in Pattern Recognition} \textsc{(}MLDM 2007\textsc{)}, \ Springer-Verlag, \ LNCS, \ volume 4571, \ Leipzig \textsc{(}Germany\textsc{)}, p. 533--547.

\vspace{1mm}

\noindent {\sc knuth d.e.} \textsc{(}1968\textsc{)}, {\em The art of computer programming}, volume 3, Addison-Wesley.

\vspace{1mm}

\noindent {\sc kryszkiewicz m.} \textsc{(}2002\textsc{)}, {\em Concise representation of frequent patterns and association rules}, Habilitation dissertation, Institute of Computer Science, University of Technology, Warsaw \textsc{(}Poland\textsc{)}.

\vspace{1mm}

\noindent {\sc kuznetsov s.o., obiedkov s.a.} \textsc{(}2002\textsc{)}, ``Comparing Performance of Algorithms for Generating Concept Lattices'', \ {\em Journal of Experimental and Theoretical Artificial Intelligence} 14\textsc{(}2-3\textsc{)}, p. 189-216.

\vspace{1mm}

\noindent {\sc li h., li j., wong l., feng m., tan y.} \textsc{(}2005\textsc{)}, ``Relative risk and odds ratio: a data mining perspective'', {\em Proceedings of the 24th ACM-SIGMOD International Symposium on Principles Of Database Systems} \textsc{(}PODS 2005\textsc{)}, Baltimore \textsc{(}MD\textsc{)}, p. 368-377.

\vspace{1mm}

\noindent {\sc liu g., li j., wong l.} \textsc{(}2008\textsc{)}, ``A new concise representation of frequent itemsets using generators and a positive border'', {\em Knowledge and Information Systems} 17\textsc{(}1\textsc{)}, p. 35-56.

\vspace{1mm}

\noindent {\sc lucchese c., orlando s., perego r.} \textsc{(}2006\textsc{)}, ``Fast and memory efficient mining of frequent closed itemsets'', \ {\em IEEE Transactions on Knowledge and Data Engineering}, 18\textsc{(}1\textsc{)}, p. 21-36.

\vspace{1mm}

\noindent {\sc mephu nguifo e.} \textsc{(}1994\textsc{)}, \guillemotleft~Une nouvelle approche basée sur le treillis de Galois pour l'apprentissage de concepts \guillemotright, {\em Mathématiques, Informatique et Sciences humaines} 134, p. 19-38.

\vspace{1mm}

\noindent {\sc pasquier n.} \textsc{(}2000\textsc{)}, ``{\em Datamining : Algorithmes d'extraction et de réduction des règles d'association dans les bases de données}, Thèse de doctorat, \'{E}cole Doctorale Sciences pour
  l'Ingénieur de Clermont Ferrand, Université Clermont Ferrand II.

\vspace{1mm}

\noindent {\sc pasquier n., bastide y., taouil r., lakhal l.} \textsc{(}1999\textsc{(}a\textsc{)}\textsc{)}, ``Discovering frequent closed itemsets for association rules'', {\em Proceedings of 7th International Conference on Database Theory} \textsc{(}ICDT 1999\textsc{)},  LNCS, volume 1540, Springer-Verlag, Jerusalem, Israel, p. 398-416.

\vspace{1mm}

\noindent {\sc pasquier n., bastide y., taouil r., lakhal l.} \textsc{(}1999\textsc{(}b\textsc{)}\textsc{)}, ``Efficient mining of association rules using closed itemset lattices'', {\em Journal of Information Systems} 24\textsc{(}1\textsc{)}, p.~25-46.

\vspace{1mm}

\noindent {\sc pei j., han j., mao r.} \textsc{(}2000\textsc{)}, `\textsc{Closet}: An efficient algorithm for mining frequent closed  itemsets'', {\em Proceedings of the ACM-SIGMOD International Workshop on Research Issues in Data Mining and Knowledge Discovery} \textsc{(}DMKD 2000\textsc{)}, Dallas \textsc{(}TX\textsc{)}, p. 21-30.

\vspace{1mm}

\noindent {\sc pfaltz j.l., taylor c.m.} \textsc{(}2002\textsc{)}, ``Scientific knowledge discovery through iterative transformation of concept lattices'', {\em Proceedings of Workshop on Discrete Applied Mathematics in conjunction with the 2nd SIAM International Conference on Data Mining}, Arlington \textsc{(}VA\textsc{)}, p. 65.74.

\vspace{1mm}

\noindent {\sc stumme g., taouil r., bastide y., pasquier n., lakhal l.} \textsc{(}2002\textsc{)}, ``Computing \textsc{I}ceberg concept lattices with \textsc{Titanic}'', {\em Data \& Knowledge Engineering} 2\textsc{(}42\textsc{)}, p. 189-222.

\vspace{1mm}

\noindent {\sc szathmary l., napoli a., kuznetsov s.} \textsc{(}2007\textsc{)}, ``\textsc{ZART}: A multifunctional itemset mining algorithm'', {\em Proceedings of the 5th International Conference on Concept Lattices and their Applications} \textsc{(}CLA 2007\textsc{)}, Montpellier,  p. 26-37.

\vspace{1mm}

\noindent {\sc uno t., asai t. uchida y., arimura h.} \textsc{(}2004\textsc{)}, ``An efficient algorithm for enumerating closed patterns in transaction databases'', {\em Proceedings of the 7th International Conference on Discovery
Science}, Padova, Italy, p. 16-31.

\vspace{1mm}

\noindent {\sc valtchev p., missaoui r., lebrun p.} \textsc{(}2000\textsc{)}, ``A fast algorithm for building the \textsc{H}asse diagram of a \textsc{G}alois lattice'', {\em Proceedings of the Colloque LaCIM 2000}, Montréal, Canada, p. 293-306.

\vspace{1mm}

\noindent {\sc zaki m.j.} \textsc{(}2004\textsc{)}, ``Mining non-redundant association rules'', {\em Data Mining and Knowledge Discovery} 9\textsc{(}3\textsc{)}, p. 223-248.

\vspace{1mm}

\noindent {\sc zaki m.j., hsiao c.j.} \textsc{(}2002\textsc{)}, ``\textsc{ChARM}: An efficient algorithm for closed itemset mining'', {\em Proceedings of the 2nd SIAM International Conference on Data Mining}, Arlington \textsc{(}VI\textsc{)}, p. 457-573.

\vspace{1mm}

\noindent {\sc zaki m.j., hsiao c.j.} \textsc{(}2005\textsc{)}, ``Efficient algorithms for mining closed itemsets and their lattice structure'', {\em IEEE Transactions on Knowledge and Data Engineering} 17\textsc{(}4\textsc{)}, p. 462-478.

\label{hamrounif}
\end{document}